\documentclass[11pt,a4paper]{article}
 \pdfoutput=1
\usepackage{jheppub}
\usepackage[utf8]{inputenc}
\usepackage[english]{babel}
\usepackage{enumitem}
\usepackage{subdepth}
\usepackage{graphicx}				
\usepackage{slashed}							
\usepackage{amssymb}
\usepackage{amsmath}
\usepackage{tikz}
\usetikzlibrary{decorations.pathmorphing}    
\usetikzlibrary{decorations.markings}
\usepackage{color}
\usepackage{hyperref}
\usepackage{verbatim}
\usepackage{mathrsfs}
\usepackage{subcaption}

\definecolor{orangeh}{rgb}{0.983045,0.585882,0.159875}
\definecolor{blueh}{rgb}{0.000000,0.659267,0.720834}

\usepackage{ulem}

\newcommand{\mJ}{ \mathcal{J}}
\newcommand{\mI}{ \mathcal{I}}
\newcommand{\sti}{\mathfrak{i}}

\newcommand{\Z}{{\tilde A}}
\newcommand{\q}{{\frak q}}
\newcommand{\flux}{\mathcal F}
\newcommand{\mz}{\lambda}

\newcommand{\mylabel}[2]{#2\def\@currentlabel{#2}\label{#1}}
\makeatletter

 \def\be{\begin{equation}}
\def\ee{\end{equation}}
 \def\ba{\begin{align}}
\def\ea{\end{align}}
\def\bea{\begin{eqnarray}}
\def\eea{\end{eqnarray}}

\def\m{\mu}

\def\O{{\cal O}}

\newcommand{\di}{\mathrm d}
\def\ev{\mathrm{eV}}

\def\At{\mathrm{At}}

\notoc

\title{Exploring the ultra-light to sub-MeV dark matter window with atomic clocks and co-magnetometers}

\author[a]{Rodrigo~Alonso,}
\author[a,b]{Diego~Blas,}
\author[c]{Peter Wolf,}

\affiliation[a]{\it Theoretical Physics Department, CERN, CH-1211 Geneva 23,
 Switzerland}
\affiliation[b]{\it Theoretical Particle Physics and Cosmology Group, Department of Physics,\\
King's College London, Strand, London WC2R 2LS, UK}
\affiliation[c]{\it SYRTE, Observatoire de Paris, Universit\'e PSL, CNRS, Sorbonne Universit\'e, LNE, 75014 Paris, France}

\emailAdd{rodrigo.alonso@cern.ch}
\emailAdd{diego.blas@cern.ch}
\emailAdd{peter.wolf@obspm.fr}

\abstract{
Particle dark matter could have a mass anywhere from that of ultralight candidates, $m_\chi\sim 10^{-21}\,$eV, to  scales well above the GeV. Conventional laboratory searches are sensitive to a range of masses close to the weak scale, while  new techniques are required to explore candidates outside this realm. In particular lighter candidates are difficult to detect due to their small momentum.  Here we study two experimental set-ups which {\it do not require transfer of momentum} to detect dark matter: atomic clocks and co-magnetometers. These experiments probe dark matter that couples to the spin of matter via the very precise measurement of the energy difference between atomic states of different angular momenta. 
This coupling  is possible (even natural) in most dark matter models, and we translate the current experimental sensitivity into implications for   different dark matter models. 
It is found that the constraints from current atomic clocks and co-magnetometers can be competitive in the mass range $m_\chi\sim 10^{-21}-10^3\,$eV, depending on the model. We also comment on the (negligible) effect of different astrophysical neutrino backgrounds.
}

\begin{document}
\preprint{CERN-TH-2018-209, KCL-PH-TH/2018-50}

\maketitle

\newpage
\renewcommand{\baselinestretch}{0.4} 
\small
\tableofcontents
\renewcommand{\baselinestretch}{1.2} 

\normalsize

\section{Introduction}

The search for dark matter (DM) in the laboratory is one of the most active   and potentially ground-shifting fields in experimental
particle physics.  A  positive result will be a momentous event in physics; determination of DM properties would be a step towards a more complete understanding of Nature and would   open a window into what lays beyond the Standard Model (SM).  At present, very little is known about DM and searches should be unbiased and
range far and wide in methodology and scope. Nevertheless the experimental program should use guidance from theory:  well-motivated  models often point towards particular properties of dark matter and how to find it.

The driving force in this endeavour has been the search for weakly interacting massive particles (WIMPs) with masses and cross-sections with the SM particles
related to the electroweak scale. The reason for this is three-fold: {\it i)} these candidates can be produced in the early Universe after they lost  equilibrium 
with the SM \cite{2010pdmo.book..121G}; {\it ii)} there are natural candidates for WIMPs in theories that try to   solve the hierarchy problem of the  SM \cite{Bertone:2004pz}; {\it iii)} DM candidates with WIMP properties may be discovered in nuclear recoils \cite{Goodman:1984dc,Drukier:1986tm}. 
Despite years of constant progress, the current detectors have only provided bounds to these ideas, see e.g. \cite{ji2017astroparticle,Aprile:2017iyp}. Similarly, there has been an active search for indirect signals from WIMPs that has  crystallised into bounds on its properties, see \cite{Gelmini:2015zpa,Lisanti:2016jxe} for recent reviews. 
Finally, the data from the Large Hadron Collider has not shown any signs of physics beyond the SM, let alone a DM candidate. 
As a consequence, the DM searches are currently moving towards other less explored, yet well motivated, territories. In particular, there is  growing interest in scrutinizing DM candidates with smaller masses, see \cite{Essig:2013lka,Alexander:2016aln,Knapen:2017xzo} for recent reviews.  The present work falls into this category and presents new ideas to explore the  range of masses from  sub-MeV
down to the lightest masses compatible with observations. 

Viable DM models are known with masses much lighter than $\mathrm{GeV}$. For models based on thermal production, cosmological observations set a limit   $m_\chi\gtrsim \mathrm{keV}$ \cite{Baur:2015jsy}.  Other `model-independent'  (in the sense of  independent of the production mechanism) bounds 
 arise by requiring the DM candidate to be able to form the smallest objects dominated by DM (dwarf spheroidals). 
For the case of {\it fermionic} DM the possibility of localizing enough fermionic degrees of freedom in these objects implies\footnote{These bounds can be stretched a bit in more complex models, see e.g. \cite{Randall:2016bqw}.} $m_\chi \gtrsim 100~\mathrm{eV}$ \cite{Tremaine:1979we,Boyarsky:2008ju}, the Tremaine-Gunn bound (see also \cite{DiPaolo:2017geq} for similar bounds with other methods). Concerning  bosonic DM,  it should satisfy $m_\chi\gtrsim 10^{-22}\, \mathrm{eV}$ to allow its de Broglie wavelength to be smaller than the size of  dwarf spheroidals. Other astrophysical observations
push this limit by an order of magnitude \cite{Kobayashi:2017jcf,Bar:2018acw}, still very far away from the WIMP scale.
These ultra-light candidates are motivated as pseudo-Goldstone bosons, which would explain their low mass and  also suggests a non-thermal generation  in the primordial universe. The leading candidates are axions or axion-like particles \cite{Marsh:2015xka,Hui:2016ltb}. 

This large span in masses can not be explored with the techniques developed for WIMPs. In fact, the kinetic energy in DM, $E_\chi =\frac{1}{2} m_\chi v^2$ with $v\sim 10^{-3}$ being the  typical velocity of the DM in the laboratory frame, is not enough to produce visible effects in nuclear recoil experiments for $m_\chi$  substantially below the  $\mathrm{GeV}$ \cite{Gelmini:2015zpa,Lisanti:2016jxe,Kouvaris:2016afs}. A  possibility to explore smaller masses is to use electronic instead of nuclear targets to capture a larger fraction of the DM's kinetic energy,  although these too have a lower threshold on mass of order $\mathrm{MeV}$ \cite{Alexander:2016aln,Hochberg:2016ajh,Essig:2017kqs}. 
Other alternatives using different physical phenomena to detect  sub-MeV DM are currently under study\cite{Essig:2011nj,Graham:2012su,Essig:2015cda,Lee:2015qva,Essig:2012yx,Knapen:2016cue,Hochberg:2015fth,Hochberg:2016ajh,Hochberg:2016sqx,Schutz:2016tid,Essig:2016crl,Brax:2017xho,Fichet:2017bng,Budnik:2017sbu,Riedel:2012ur,Bateman:2014lia,Riedel:2016acj}. Moving to the `ultra-light' regime requires yet a new battery of techniques (including relaxing the energy threshold by studying  absorption, and not only scattering, of DM), e.g.  \cite{Derevianko:2013oaa,Arvanitaki:2014faa,Stadnik:2015upa,Hees:2016gop,Hochberg:2016ajh,VanTilburg:2015oza,Yang:2016odu,Dev:2016hxv,Garcon:2017ixh,Blas:2016ddr}. For both light and ultra-light dark matter, a key idea is to use very precise set-ups that may be sensitive to small or even vanishing momentum transfer.

Atomic clocks stand among the most precise devices  ever built;  they are therefore prominent candidates to search for new physics. Atomic clocks  have already been used  to, or suggested to,  constrain ultra-light DM candidates \cite{Derevianko:2013oaa,Arvanitaki:2014faa,Stadnik:2015upa,Hees:2016gop} and other models beyond the SM \cite{Delaunay:2016brc}.  In \cite{Derevianko:2013oaa,Arvanitaki:2014faa,Stadnik:2015upa,Hees:2016gop}, the DM is assumed to be a massive scalar field  that couples non-minimally to the SM fields. 
In these models, the scalar field oscillates at a frequency equal to the DM mass and/or is composed of finite size topological features that pass by the Earth. For both cases one observes  the effect of the DM as time variations of the fundamental constants and thus of the clock frequency. Given that the fastest variations that can be measured by clocks are  of the order of a kHz, these methods are relevant for DM  masses $m_\chi\lesssim 10^{-13}$~eV \cite{Hees:2016gop,VanTilburg:2015oza,Wcislo:2016qng}.
In contrast  to the interactions probed by the aforementioned methods, here we consider DM interactions that differentiate between the  spin of the two atomic states used in clocks running with polarized samples.  These interactions produce an extra frequency shift between these states. This effect, as in other searches with atomic clocks, does not require momentum transfer and hence has no mass threshold, as discussed in detail in \cite{Wolf:2018xlz}.  Thus, it can be used  to explore masses from the lowest values allowed to a few keV, the upper limit arising from the fact that our expected constraints become worse than existing ones.   
Other complementary ideas to use quantum devices to detect light DM include \cite{Riedel:2012ur,Bateman:2014lia,Riedel:2016acj}. 

 Atomic magnetometers monitor the spin precession of atoms around a magnetic field to very high precision.
An interaction of their atomic spins with an external field can then be searched for if the standard electromagnetic interactions are suppressed or well determined. {\it Co-magnetometers} achieve this by using two different atomic species in specific configurations.  
We consider two different systems that have already been used in the search of deviations from Lorentz invariance  \cite{Brown:2010dt,Allmendinger:2013eya}. 
These set-ups were recently suggested to study axionic dark matter \cite{Graham:2017ivz} (see also \cite{Stadnik}).
 For these DM candidates, our study yields results similar to \cite{Graham:2017ivz}, though we give  more explicit 
 equations connecting the phenomenology to fundamental DM-SM couplings. We also extend the analysis for the other models of DM, which allows us to explore DM models of higher masses.
 The latter also include models of fermionic DM.

One final comment is in order before moving to the body of the paper:  even if our main target are models of DM,  the basic idea may also be useful to search for other backgrounds. One such background is the  
neutrino flux that criss-crosses the Earth \cite{Becker:2007sv,Formaggio:2013kya}. At low momentum, the distribution is dominated by cosmological neutrinos generated during the Big Bang  and that have not been directly detected yet \cite{Ringwald:2009bg,Baracchini:2018wwj}.
At higher energies, the flux is mainly due to Solar neutrinos. We will briefly discuss the (bad) prospects to detect these backgrounds with atomic clocks or magnetometers.

Our work is organized as follows: in sec.~\ref{sec:TH} we  describe the effective interactions of DM with the constituents of
the experiments. We do this for different DM candidates. Sec.~\ref{sec:AC} contains a description of the atomic states of atomic clocks and magnetometers and their time evolution in the presence of the DM background and the interactions of sec.~\ref{sec:TH}. We first describe the scattering by a flux of individual DM particles in  sec.~\ref{sec:scatt_effect} and elaborate on the case of large occupation numbers (low DM mass) in sec.~\ref{sec:classical}. In sec.~\ref{sec:devices} we present the main results of this work:  it is shown how the measured frequency in atomic clocks (sec.~\ref{sec:dev_ac}) and the Larmor frequency in magnetometers (sec.~\ref{sec:dev_mg}) are affected by the DM background. We use this to determine in secs.~\ref{sec:AC_DM} and \ref{sec:AC_Mag} the sensitivity of near-future set-ups to 
the spin-dependent 
couplings as a function of the DM mass.  These bounds are compared to other existing constraints in sec.~\ref{sec:comp}. The case of astrophysical neutrino backgrounds is briefly touched upon in sec.~\ref{sec:neutrinos}.
We conclude in sec.~\ref{sec:conclu}, where we also present 
future prospects. Appendices~\ref{sec:QFT}-\ref{app:LV} make explicit conventions and auxiliary computations.

\section{Spin-dependent interactions of dark matter with ordinary matter}\label{sec:TH}

As outlined in the introduction,  we will  consider  a background that interacts  with ordinary matter distinguishing atomic states with different  angular momenta\footnote{Other differences between states in atomic clocks or co-magnetometers can be exploited for ideas close to those in this work. We focus on angular momentum as the most salient difference and leave the study of other distinctions (other couplings)  for future work.}. The present section discusses the interactions of DM and ordinary matter that depend on the spin {\it of the latter}. 
We split the analysis into:
\begin{itemize}
\item sec.~\ref{sec:EFT}: {\it Contact interactions and a model completion.}  
Within an effective field theory (EFT) scheme one can be systematic and comprehensively list possible interactions whose effect we can estimate from the mass dimension of the operators producing the interaction. 
Here we restrict to spin-dependent interactions that do not vanish in the limit of no transferred momentum. The last condition is important since the main observable we describe is evaluated in this limit. 
An EFT with contact interactions is however not a complete picture  since unspecified `heavy' dynamics generate the interactions. 
We make explicit a simple class of models that complete the EFT with a dynamical mediator.

\item sec.~\ref{sec:axialB}: {\it Dark matter as an axial boson.} In contrast to EFT, these models are self-contained and potentially valid to high energies; they describe physics in terms of  a set of constants and can be tested by over-constraining their parameter space. 
However, where in EFT one could be systematic in listing interactions, there is no limit on the complexity of concrete models. A survey of models is out of scope of this work; instead we will take two popular cases: DM as an axion-like particle or DM as an axial vector boson.
\end{itemize}

\subsection{Contact interactions in effective field theory and the light mediator case}\label{sec:EFT}

DM couples very weakly to the known particles. The reason for this could be the same that makes neutrinos elusive: a heavy particle mediates the interaction making the {\it effective} coupling suppressed by the ratio of the energy of the observation to the mass of the mediator in analogy to Fermi's theory of beta decay. The building blocks of our effective couplings are the DM field, $\chi$, and the elementary fermions, $\psi$, present in the atom: electrons $e$ and up and down-type quarks ${\q}=u,d$. Assuming the DM to be uncharged under the known forces the interactions in an  effective field theory (EFT) read\footnote{We take here the theory after EWSB, otherwise $e_L$ couplings come together with $\nu's$ and $G_{u_L}=G_{d_L}$.}
\begin{align}
\label{eq:LagEFT}
L _{\rm int}=-\int \di^3 x \left( G_e^{{ \mI}} \bar e\, \Gamma^\mI e \,{\mJ}^\mI_\chi +\sum_{\q=u,d}G_{\q}^{{\mI}}  \bar \q\, \Gamma^\mI \q\, \mJ^\mI_\chi \right)\equiv {-}\int \di^3x\, \sum_{\psi} G^\mI_{\psi}\,\mJ^\mI_\psi \times \mJ^\mI_\chi,
\end{align}
where by $\mJ^\mI_\chi$ we denote the Lorentz representation built out of DM fields, $\Gamma^\mI$ are the possible 
Dirac structures that contracted with $\mJ^\mI_\chi$ make a Lorentz invariant  and $G_\psi^\mI$ are coupling constants,
 which have labels $\mI$ and $\psi$ to distinguish the different operators and the SM particle they couple to. The  current $\mJ^\mI_\chi$ itself depends on what type of particle DM is; here we take it to be a spinless, spin-$1/2$ or spin-1 particle. We look for interactions which are spin-dependent for the SM fermions, not-vanishing in the limit of zero transferred momenta and of dimension $\leq 6$. This requires a DM complex field\footnote{For example the operator $\bar\psi\gamma_\mu\gamma_5 \psi\partial^\mu\chi^2$ { is proportional to} the transferred momentum $q$ so we discard it, but $\partial_\mu\chi^2$ is the only current that can be built with a real field; in contrast for a complex scalar we have two: $\partial_\mu( \chi^\dagger \chi)$ and $(\partial_\mu\chi^\dagger)\chi-\chi^\dagger\partial_\mu\chi$. The second one is proportional to the sum of incoming and outgoing DM momenta in a scattering process.}, which therefore has naturally a conserved quantum number. The interactions are listed in table~\ref{DMOps}, totalling 6 operators for each charged SM fermion $\psi=e,u,d$. 
\begin{table}[h]
\begin{center}
\begin{tabular}{|c|cc|cccc|}
\hline
$\mathcal I$&$\psi=e,u,d$&&DM &Scalar&Fermion &Vector Boson\\ \hline
\it Ax. vector&$\mJ_{\psi}:$& $\bar \psi \gamma^\mu \gamma_5\psi $ &$\mJ_\chi:$ & $i\chi^\dagger \partial_\mu\chi$+h.c., & $\bar\chi\gamma_\mu\chi,$ & $i\chi_\nu^\dagger \partial_\mu\chi^\nu$+h.c.\\
&&& &  & $\bar\chi\gamma_\mu\gamma_5\chi,$ & 
\\ \hline
\it Tensor&$\mJ_\psi:$ & $\bar \psi \sigma^{\mu\nu} \psi $ &$\mJ_\chi:$& --
& $\bar\chi \sigma^{\mu\nu}\chi,$ & $\chi^\dagger_\alpha \mathcal (\Sigma_{\mu\nu})^{\alpha}_{\,\,\beta}\chi^\beta. $\\\hline
\end{tabular}
\caption{Leading interactions for scattering between DM and SM fermions in the form of operators $\mathcal O\equiv \mJ_\psi\times \mJ_\chi$ of dimension $\leq 6$. We only write operators that do not vanish in the limit of zero transferred momenta. The terms $\sigma^{\mu\nu}/2$ (or $(\Sigma_{\mu\nu})^{\alpha}_{\,\,\beta}$) are the Lorentz generators in spin 1/2 (or spin 1) space, $\sigma^{\mu\nu}=i/2[\gamma^\mu\,,\gamma^\nu]$ ($\Sigma_{\mu\nu}^{\alpha\beta}=i(\eta^{\alpha}_\mu\eta_\nu^\beta-\eta^\beta_\mu \eta^\alpha_\nu)$). \label{DMOps}}
\end{center}
\end{table}

\begin{figure}[h]
\begin{center}
\begin{tikzpicture}
\draw (-2,.5) node {$\chi$};
\draw [-, thick] (-2,-1)--(-1,0);
\draw [-, dashed, thick] (-2,1)--(-1,0);
\draw [-, ultra thick] (-2,-1)--(-1.5,-1) node[anchor=north east]{At};
\draw [-, ultra thick] (-.5,-1)--(0,-1)node[anchor=north]{At};
\draw (-2,-.5) node {$ e$};
\filldraw[black] (-1,0) circle (2pt);
\draw (0,.5) node {$\chi$};
\draw [-, dashed,thick] (-1,0)--(0,1);
\draw [-, thick] (-1,0)--(0,-1);
\draw [-, thick](-0.5 ,-1) arc (45:135:0.7);
\draw [-, thick](-0.25 ,-1) arc (34:146:0.9);
\draw (0,-.5) node {$e$};
\draw (2,.5) node {$\chi$};
\draw [-, thick] (2,-1)--(3,0);
\draw [-, dashed,thick] (2,1)--(3,0);
\draw [-, ultra thick] (2,-1)--(2.5,-1) node[anchor=north east]{At};
\draw [-, ultra thick] (3.5,-1)--(4,-1)node[anchor=north]{At};
\draw (2,-.5) node {$ \q$};
\filldraw[black] (3,0) circle (2pt);
\draw (4,.5) node {$\chi$};
\draw [-, dashed,thick] (3,0)--(4,1);
\draw [-, thick] (3,0)--(4,-1);
\draw [-, thick](3.5 ,-1) arc (45:135:0.7);
\draw [-, thick](3.75 ,-1) arc (34:146:0.9);
\draw (4,-.5) node {$\q$};
\end{tikzpicture}
\caption{Contact interaction of DM $\chi$ with the electron or quark components of the nucleus of an atom denoted by At. \label{Scat3}}
\end{center}
\end{figure}
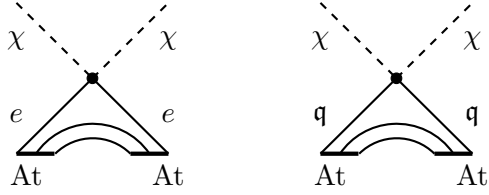
The operators in the EFT generate contact 4-particle interactions as shown diagrammatically in fig.~\ref{Scat3}. 
The actual SM degrees of freedom  in the experiments are atoms (At) where quarks are confined in the bound nucleons; the connexion between the two descriptions is given by form factors of the type $\left\langle \mathrm{At}\right| \bar \q\gamma_\mu\gamma_5 \q \left| \mathrm{At}\right\rangle$. This connection is established  step-wise; the first stage, the quarks-to-nucleons step,
can be taken by considering the RHS of eq.~\eqref{eq:LagEFT} with $\psi\to N=n,p$ and $G_{N}$ constants related to $G_\q$ (for $q^2=0$) as:
\begin{align}
{\rm Ax. \,vector}: \quad G_p=&0.897(27) G_u-0.376(27) G_d,&
G_n=&0.897(27)G_d-0.376(27) G_u ,\label{eq:q-NAx}\\ \label{eq:q-NT}
{\rm Tensor}: \quad G_p=&0.794(15)G_u-0.204(8)G_d,&
G_n=&0.794(15)G_d-0.204(8)G_u,
\end{align} 
with numerical values taken from~\cite{Bishara:2017nnn}.
The step nucleons-to-nuclei can be found in sec.~\ref{sec:contact}, after we discuss which are the atomic elements of relevance (cf. table~\ref{tab:gNNcl}).

The extension of this EFT to a model with a dynamical mediator is straight forward for the axial vector  case. We introduce an axial vector boson $\Z_\mu$ with mass $m_\Z$ and coupling to dark and ordinary matter as:
\begin{align}L_{\rm int}^{\Z{\rm-}\chi}= \int \di^3x \Z_\mu\left( g^\Z_\chi\mathcal J_\chi^\mu
+g^\Z_\psi\bar\psi\gamma^\mu\gamma_5\psi\right),  \label{ModelAVB}
\end{align} 
where $\mathcal J_\chi$ is any of the currents given in the upper block of table~\ref{DMOps}. The interaction that this mediator generates is:
\begin{align}
L_{\chi-\psi}=\int \di^3x \left[-\frac12\left( g^\Z_\chi\mathcal J_\chi^\mu+g^\Z_\psi\bar\psi\gamma^\mu\gamma_5\psi\right)\frac{g_{\mu\nu}+\partial_\mu \partial_\nu/m_\Z^2}{\partial^2+m_\Z^2}\left( g^\Z_\chi\mathcal J_\chi^\nu+g^\Z_\psi\bar\psi\gamma^\nu\gamma_5\psi\right)\right],\label{MedLagEFT}
\end{align}
 and the effective Fermi-like coupling of eq.~\eqref{eq:LagEFT} generated for heavy $m_\Z$ reads $G_\psi=g^\Z_\psi g_\chi^\Z/m_\Z^2$. For light mediators, the longitudinal mode in $\Z_\mu$ cannot be neglected in general since it couples to a non-conserved current and its Goldstone-like nature means the effective coupling scales as ($g_\psi^\Z E/m_\Z$)~\cite{Preskill:1990fr,Dror:2017ehi,Dror:2017nsg}.
One exception is the experimental set-up considered here, which is sensitive to forward scattering and hence has zero transferred momenta ($\partial\to0$). In this case actually one can extrapolate the result of the EFT, {\it even in the case where $\Z_\mu$ is light}. 
These considerations together with further phenomenology will be examined in sec.~\ref{sec:comp} where we confront  bounds on DM and its mediator with the sensitivity of atomic clocks and magnetometers.

\subsection{{Dark matter as an axial boson}}
\label{sec:axialB}

Massive axial bosons (scalars or vectors) are popular candidates for DM. This stems from their natural appearance in 
different models beyond the SM and the variety of mechanisms to generate the proper DM abundance \cite{Marsh:2015xka,Alexander:2016aln}. As discussed in the introduction, masses as low as  $m_\chi\sim 10^{-21}$~eV are compatible with observations. The couplings of the axion $a$ to SM fermions ($\psi=e,u,d,n$ or $p$) read:
\begin{align}
L_{int}^{a}=\int \di^3x \frac{i C_\psi m_\psi}{\frak f_a}\bar\psi\gamma_5\psi \,a{=} \frac{- C_\psi}{2 {\frak f}_{ a}}\int \di ^3x \,\bar\psi \gamma_\mu\gamma_5\psi \,\partial^\mu { a}, \label{LagAx}
\end{align}
where the second identity is true at leading order and obtained after using the equations of motion. For the axial vector boson, they read
\begin{align}
L_{int}^{ A}=g^A_\psi \int \di^3x \, A^\mu\bar\psi \gamma_\mu\gamma_5\psi\,. \label{LagAxVB}
\end{align}
$C_\psi, g_\psi^{A}$  are dimensionless (in natural units) coupling constants and ${\frak f}_{ a}$ is the axion decay constant. Notice that the {\it massive} axial vector boson includes a longitudinal polarization,  whose coupling is similar to that of the  axion field.

\section{Effects of dark matter background on atomic states}\label{sec:AC}

After having described the possible interaction terms between the constituents of the  devices and the DM background, we now move to describe their effect in the evolution of the  relevant atomic states of microwave atomic clocks and co-magnetometers.

The states probed in atomic clocks are hyperfine-split and labeled by their different total angular momentum $F$ and $F-1$, but they have the same angular momentum component $\mz$ along a given axis\footnote{This is determined by a magnetic field with a fixed direction with respect to the ground.},  which we identify with $\hat z$ \cite{vanier1989quantum}. The atoms employed are alkali metals, in particular ${}^{87}$Rb and ${}^{133}$Cs, with the outer electron in the ground state; the $s$-shell. The two states of the atom will be denoted $\left|1\right\rangle$ and $\left|2\right\rangle$ and read, taking Rb for definiteness\footnote{The internal shells do not play a role in our arguments and can be ignored.}:
\begin{subequations}\label{eq:states}
\begin{align}
\left|2\right\rangle\equiv& \left| \mathrm{Rb}^{{F}}_{\lambda}\right\rangle=
\sum_{\lambda_e,\lambda_I} \left| e^{5s}_{\lambda_e }\right\rangle \otimes \left|\mbox{Ncl}^I_{\lambda_I}\right\rangle
\left\langle 1/2,\,\lambda_e,\,I,\, \lambda_I\,| F,\,\lambda\right\rangle,
\\
\left|1\right\rangle\equiv& \left|\mathrm{Rb}^{{   F-1}}_{\lambda}\right\rangle=
\sum_{\lambda_e,\lambda_I} 
\left| e^{5s}_{\lambda_e }\right\rangle \otimes \left|\mbox{Ncl}^I_{\lambda_I}\right\rangle
\left\langle 1/2,\,\lambda_e,\,I,\, \lambda_I\,| F-1,\,\lambda\right\rangle,
\end{align}
\end{subequations}
where $\left\langle j_1,\,\lambda_1,\,j_2,\, \lambda_2\,| J,\,\lambda\right\rangle$ are the Clebsch-Gordan coefficients (non-zero for $\lambda_1+\lambda_2=\lambda$ only) and $I$ is the nuclear spin ($I=3/2$ for the ${}^{87}$Rb  nucleus \cite{Stone:2005rzh}). The case for Cs can be obtained by replacing $5s\to 6s$ and $I=7/2$  \cite{Stone:2005rzh}.

Co-magnetometers use two samples of different atomic species to 
constrain  anomalous dynamics in their spin as described in more detail in sec.~\ref{sec:dev_mg}. 
The set-up in \cite{Allmendinger:2013eya} employs a mixture of ${^3}$He and ${}^{129}$Xe; all electron shells are closed for noble gases so that the  total angular momentum  comes from the nuclear spin, $I=1/2$ in both cases. The spins precess around an homogeneous magnetic field $\vec B$ which induces an energy splitting between the spin states
aligned ($|1\rangle$)  and anti-aligned ($|2\rangle$) with $\vec B$,
\be
|2\rangle =|{\rm {\rm At}}_{-1/2}\rangle, \quad\quad\quad |1\rangle =|{ {\rm At}}_{1/2}\rangle, \label{eq:statesmag}
\ee
where $ {\rm At}$ stands for ${^3}$He or ${}^{129}$Xe.
 The set-up in \cite{Brown:2010dt} works with a  mixture of K ($I_{\rm K}=3/2$) and ${}^3$He and is also sensitive to couplings to the electron spin of the potassium.  
 The K is distributed among different strongly-coupled states of the $F=2$ and $F=1$ multiplets.
The different states in the sample  can be 
written as in \eqref{eq:states}   (the electron is in the $4s$ shell).

In the rest of this section we will describe the time evolution of  these atomic states in the presence of backgrounds for the interactions described in sec.~\ref{sec:TH}. Depending on the mass of the DM candidate it is more convenient to treat the interactions as a collection of scatterings (sec.~\ref{sec:scatt_effect}) or as atomic states evolving in a background field (sec.~\ref{sec:classical}).

\subsection{Cold light particles scattering off atoms}\label{sec:scatt_effect}

For masses $m_\chi$ above few eV,  we can describe DM as a a non-coherent collection of non-relativistic (NR) particles virialized in a halo and with a Maxwellian distribution of velocities  \cite{Gelmini:2015zpa,Lisanti:2016jxe}.  Their interaction with the atomic states is well described as a series of scatterings  of non-relativistic  particles. We deemed useful to summarize this standard material to set conventions and the connection with particle physics (see also app.~\ref{sec:QFT}). We will take DM to be much lighter than the atoms, $m_{\mathrm{At}}\gg m_\chi$, which means  that 
the discussion is also valid for any  non-relativistic particle lighter than the scatterer, e.g. for residual hydrogen atoms, see~\cite{Wolf:2018xlz} for details. In this section we focus on a single DM particle, while the considerations about the flux of DM are deferred to sec.~\ref{sec:devices}, when we discuss the observables. 

We assume the DM particle $\chi$ to have  3-momentum $\bf p_\chi$, whereas the atom's  mean momentum we denote $\bf p_{\rm At}$. 
It is convenient to describe the system with relative and CM coordinates and momenta related to the particle's momenta as:
\begin{align}
{\bf p}_{\At}=&\frac{\mu}{m_\chi}{\bf P}-{\bf p} \,,& &{\bf p}_\chi=\frac{\mu}{m_{\rm At}}{\bf P}+{\bf p}\,,\label{ChngtoCM}
\end{align}
where $\bf P$ is the total and $\bf p$ the relative momentum and $\mu$ is the reduced mass, $\mu\equiv m_\chi m_{\At}/(m_\chi+m_{\At})\simeq m_\chi$. 
The case of light DM implies a  negligible momentum transfer  to the atom ($m_{\rm At}\gg | {\bf q}|\equiv|{\bf p}-{\bf p}'|$) with $\bf p'$ the relative momentum after the scattering. This also means that the kinematics are very different from customary DM searches. The center of mass frame is, to very good approximation, the atomic rest frame, or {\it lab frame}, and the relative momenta that of the DM,
\begin{align}\label{CMapprox}
\bf P&= {\bf p}_{\rm At}+\mathcal O\left(\frac{m_\chi}{m_{\rm At}}\right),  &
{\bf p}&={\bf p}_{\chi} +\mathcal O\left(\frac{m_\chi}{m_{\rm At}}\right),
\end{align}
as clear from eq.~(\ref{ChngtoCM}) in the limit $m_{\At}\gg m_\chi$.
This also means that the atom will not change its direction  noticeably and one can describe the system by the projection onto the atomic states of eq.~\eqref{eq:states} at rest  and by the spatial wave-function of the DM in the lab frame\footnote{The velocity of the atoms in the lab frame is $\lesssim 4\,$m/s in both atomic clocks and magnetometers.}, 
\begin{align}
|\Psi\rangle&= |{\rm At}(t)\rangle\otimes |\chi\rangle, &
\Psi_{\sti}(t,x)&=\langle \sti|  {\rm At}(t)\rangle\langle x|\chi\rangle\equiv e^{-iE_\sti t}e^{-iE_\chi t} c_\sti(t) \chi_\sti(x)\,,\label{eq:wave_in}
\end{align}
where $E_\chi=p_\chi^2/2m_\chi$ and we allow the DM wave-function to depend on the atomic state $\sti=1,2$, as we will see this is the case after the scattering.

 Let us start with the
case of  elastic scattering. 
For an incident DM particle  in a $\bf p_\chi$ momentum eigenstate
scattering off an atomic state $c_1|1\rangle+c_2|2\rangle$,  the out-state far away from the interaction region\footnote{ Meaning at distances much greater than the interaction length, $l_{\rm int}\sim1/m_\Z$; nominally this requires a long enough time $T$ for the DM particle to leave the target's influence, $T\gg\frac{10^{-12}\ \rm eV}{m_\Z}\frac{10^{-3}}{v}\,s $. The case of light mediator ($m_\Z\ll m_\chi v$) is better described as a potential problem, as we do in app.~\ref{app:lightmA}. The  two results coincide except for a a subtlety discussed in sec.~\ref{sec:partDMAC}.\label{foot:range}} reads~\cite{Wolf:2018xlz,goldberger2004collision}: 
\be
\label{eq:wave}
\langle x|\Psi_{\rm out}\rangle
=\sum_{\sti=1,2}c_\sti \,e^{-iE_\sti t}|\sti\rangle\left(e^{i {\bf p}_\chi\cdot {\bf x}} +
\frac{f_\sti(p_\chi\mathbf{\hat x}\,,\,\mathbf p_\chi) e^{i \,p_\chi|{\bf x}|}}{|{\bf x}|}\right) e^{ -iE_\chi t},
\ee
where ${\bf \hat x}\equiv{\bf x}/|{\bf x}|$, $p_\chi\equiv|{\bf p_\chi}|$, we have approximated relative momentum
and position to those of the DM as in eq.~(\ref{CMapprox}) and we omit the overall normalization factor for a plane wave.  
The connection between the scattering matrix generated by the interaction and the amplitude $f_\sti$ is:
\begin{align}
 f_\sti({\bf p}^\prime_\chi,{\bf p}_\chi)={-}\frac{\mu}{2\pi}\mathcal T_\sti({\bf p}'_\chi\,, {\bf p}_\chi)\,. \label{eq:fdefntn}
\end{align}
where $\mathcal T$ is the matrix element for momentum states, cf.  app.~\ref{sec:QFT}. 
 For rotationally  invariant potentials, the only angle on which $f_\sti$  can depend is 
 \be
 \cos\theta={\bf \hat x}\cdot {\bf p}_\chi/p_\chi.
 \ee 
For later convenience  we define $\chi_\sti^{\rm out}(x)$ as
 \begin{align}
\langle \sti |\langle x|\Psi_{\rm out}\rangle 
\equiv e^{-iE_\sti t} e^{-i E_\chi t}
c_\sti\, \chi_\sti^{\rm out}(x).\label{eq:PsiDef}
 \end{align}
The description in terms of wave-packets extends formula~\eqref{eq:wave} in the intuitive manner; 
the unperturbed and spherical waves have probability distributions given by gaussian distributions that follow the original and a radially outgoing trajectory \cite{goldberger2004collision}. Given that the outgoing momentum modulus is the same as the incoming one, there is an overlap of the two distributions   centered in the forward direction. The interference of the overlapping  waves, a genuinely quantum phenomenon,  will change  the probability of detecting the different atomic states. This will be the most relevant effect in the case of particle scattering. Thus, in the following we will be mostly interested in the forward scattering amplitude (we do not show  the dependence on $p_\chi$ to avoid cluttered formulae) 
 \begin{align}
 f_\sti(0)\equiv f_\sti({\bf p_\chi}\,, {\bf p_\chi} ). \label{eq:fwiththeta}
\end{align}
The same interference is not 
possible in the case of inelastic scattering (ionization, level-transitions, etc). This is because  a minimum momentum transfer is required;
for instance, for a $|2\rangle\to |1\rangle$ transition in Rb the energy available is $\approx 6\, \mathrm{GHz}\approx 10^{-5}\, \mathrm{eV}$, which
gives a scattered particle with $m_\chi\sim10\,$eV a velocity shift $\Delta v \sim 10^{7}$cm/s and the two waves decohere almost instantly in lab timescales.

The amplitude $f_\sti$ is complex, with the optical theorem 
\be
{\rm Im}(f(0))=p_\chi \sigma/(4\pi), \label{eq:optical}
\ee 
relating its imaginary part to the cross section, with $\sigma$ the integral of~\cite{goldberger2004collision},
\be
\frac{\di \sigma}{\di\Omega}=|f|^2. \label{eq:cross_f} 
\ee 
For a perturbative interaction, Re($f_\sti$) is proportional to a small coupling. From \eqref{eq:optical} and \eqref{eq:cross_f},  ${\rm Im}(f(0))$ is second order in the expansion and smaller than the real part by $\sim p_\chi f$ (see also \eqref{eq:f_l}).  Given the feeble couplings of DM to ordinary matter, we can thus  neglect Im$(f(0))$ in the remainder of this work.
In the following sub-sections we evaluate the  matrix elements $\mathcal T$ for forward scattering in the different cases described in sec.~\ref{sec:EFT}.

\subsubsection{Effective field theory and  light mediator}\label{sec:contact}

Since the trajectory of the atom is not sizeably deviated by the DM, we can approximate the  atomic matrix elements to have the same in and out states. Furthermore, as mentioned in the previous section, we are interested in the limit of forward scattering, eq.~\eqref{eq:fwiththeta}.
The calculation  at first order in the couplings is straightforward for the models of eq.~(\ref{eq:LagEFT}), while for models with a light mediator,  eq.~\eqref{ModelAVB}, the Lagrangian \eqref{MedLagEFT} reduces to eq.~(\ref{eq:LagEFT}) with the substitution $G_{\psi}\to g^\Z_\psi g^\Z_\chi/m_\Z^2$. 
For each interaction term one finds for the different states $\sti=1,2$: 
\begin{align}
\mathcal T_{\psi,\sti}(0)=G_\psi  \langle \sti | \mJ_\psi(0)|\sti\rangle \times \langle \chi | \mJ_\chi (0)|\chi\rangle\equiv { -} G_\psi\vec J^{\,\psi}_\sti\cdot \vec J_\chi+\mathcal \O(v^2/c^2),\label{eq:MatElEFT}
\end{align}
where we have used the same notation as in eq.~\eqref{eq:fwiththeta} to denote the ${\bf q}\equiv{\bf p}-{\bf p}'=0$ limit and we have defined the  3-vectors $\vec J_{\psi,\chi}$. Notice also that in the expression \eqref{eq:MatElEFT} the currents $\mathcal J(x)$ are evaluated at $x^\mu=0$.  We are interested in couplings such that $\mathcal T_{\psi,1}(0)\neq \mathcal T_{\psi,2}(0)$, since only in these cases
there is a phase shift that one can measure (sec.~\ref{sec:dev_ac}).  

The electron case is the simplest of the ordinary matter matrix elements. In the NR limit, and for the  single electron in \eqref{eq:states}:
\begin{subequations}\label{eq:Jpsie}
\begin{align}
&\left\langle  e \right|  \bar  e(0) \gamma^\mu\gamma_5  e(0)\left|  e \right\rangle=(0,2 \vec  \lambda_ e),\quad 
 \left\langle  e  \right|  \bar  e(0)  \sigma^{\mu\nu}   e(0) \left| e \right\rangle=
 \left(\begin{array}{cc}
 0 &0\\
 0& 2  \epsilon^{ijk}\lambda_ e^k
 \end{array}\right),
\end{align}
\end{subequations}
where $\lambda_e$ is the electron spin, all time-like components vanish to first order in the NR approximation and $\epsilon^{ijk}$ is the antisymmetric tensor in 3 dimensions with $\epsilon^{123}=1$.  Since the matrix element is  evaluated in the $q=0$ limit, the result for an electron bounded in an atom is the same as that for a free one; in general there will be a form factor as detailed in app.~\ref{ap:ele}.
The atomic current coming from the electrons in  \eqref{eq:MatElEFT} is thus\footnote{\label{foot_ep} Specifically one has, $\langle \sti |\bar\psi\gamma^\mu\gamma_5\psi|\sti\rangle=(0,\vec J^{\,\psi}_{\sti})$ while  $\langle \sti | \bar\psi\sigma^{\mu\nu}\psi|\sti\rangle=\mbox{Diag}(0,   \epsilon_{ijk} J^{\psi,k}_{\sti})$. To get to the expression in \eqref{eq:MatElEFT} we move the antisymmetric tensor to the DM current.}, 
\begin{align}
\vec J^{\,e}_{\sti}=(-1)^\sti\frac{\vec\lambda }{F}, \label{eq:Je}
\end{align}
for the  states $\sti$ of 
 eqs.~\eqref{eq:states}  and  where $\vec\lambda$ is the average angular momentum of the atom.

The case in which DM interacts with quarks requires a step-wise connection of the quark interactions with those for nucleons, then for nuclei and finally for the atomic states. The first step was given in eqs.~\eqref{eq:q-NAx}-\eqref{eq:q-NT}. 
For the step from nucleons to nuclei, let us consider the state $\sti$ as a given nucleus Ncl and $\psi\to N(= p, n)$. One can write
\begin{align}
\vec J_{\rm Ncl}^{\,N} = 2 \mathfrak g_{{\rm Ncl}}^N \vec I,
\end{align}
where $\vec I$ is the nuclear spin  and the form factor $\mathfrak g_{{\rm Ncl}}^N$ 
encodes the nuclei-dependent dynamics.
One can evaluate $\mathfrak g_{{\rm Ncl}}^N$ using the nuclear shell model, as done in \cite{Goodman:1984dc,Drukier:1986tm}. 
Here, instead, we take the more accurate numerical values from~\cite{Stadnik:2014xja}, shown in  table~\ref{tab:gNNcl}.
\begin{table}[htp]
\begin{center}
\begin{tabular}{c|c|c|c|c|c}
N\textbackslash Ncl & ${}^3$He & ${}^{39}$K &   ${}^{87}$Rb & ${}^{129}$Xe &   ${}^{133}$Cs \\\hline
 neutron (n) & 	1.000 & -0.064 & 0.248 & 0.730 &-0.206\\
 proton (p)& 	0.000 & -0.536 & 0.752 & 0.270 & -0.572
\end{tabular}\caption{Numerical values for $\mathfrak g_{\rm Ncl}^N$ for the different nucleons and nuclei of interest.\label{tab:gNNcl}}
\end{center}
\end{table}
 
For the alkali elements, the states of eq.~\eqref{eq:states} are eigenstates of total angular momentum, $F$, the sum of the nuclear and electronic contributions. The evaluation of their matrix elements yields, for $\psi=N$
\begin{align}\label{eq:Jq}
\vec J^{N}_{\sti}=
2 \frak{g}_{{\rm Ncl}}^{N}\left(1-\frac{(-1)^\sti}{2F}\right) \vec\lambda.
\end{align}

The evaluation of DM matrix elements, $\langle \chi| \mJ_\chi|\chi\rangle$ corresponding to the operators in 
table~\ref{DMOps} is straightforward.  The final results in terms of the current\footnote{Recall footnote~\ref{foot_ep}.} $\vec J_\chi$ of \eqref{eq:MatElEFT} is 
given in table~\ref{DMMatEl}.   The $\pm$ symbol refers to the particle or antiparticle cases and we distinguish between Majorana and Dirac fermions (specifically we take $\chi\to(\eta_L+\eta_L^c)/\sqrt2$
where $\eta_L$ is left handed fermion to obtain the results for a Majorana fermion).  In the list we can see how the elements depend on either the {\it velocity} or the {\it spin} of the DM (remember we take ${\bf q}=0$) and powers of $m_\chi$ distinguish operators of dimension 5 ($m_\chi^{-1}$) and 6 ($m_\chi^0$). A unit ratio of particles and antiparticles in a medium would cause the velocity dependent currents to cancel on average, whereas this is not the case for spin-dependent interactions. 
\begin{table}[h]
\begin{center}
\begin{tabular}{|cc|ccccc|}
\hline
$\psi=e,q$&&DM &Scalar&D Fermion &M Fermion&Vector Boson\\ \hline
$\mJ_{\psi}:$& $\bar \psi \gamma^\mu \gamma_5\psi $ &$\vec J_\chi:$ & $ (\pm)\vec v_\chi$& ${(\pm) \vec v_\chi}$ &{ 0}& ${ -}(\pm)\vec v_\chi$,\\
&& &  & $2\vec \lambda_\chi$& $2\vec \lambda_\chi$&
\\ \hline
$\psi=e,q$&&DM &Scalar&D Fermion  &M Fermion&Vector Boson\\ \hline
$\mJ_\psi:$ & $\bar \psi \sigma^{\mu\nu} \psi $&
 &---   & ${ -4}\vec \lambda_\chi$  &${ -4}\vec \lambda_\chi$&${ (\pm)}\vec \lambda_\chi/m_\chi$. \\\hline
\end{tabular}
\caption{Dark matter matrix elements, $\vec J_\chi$, for the different currents of table~\ref{DMOps} evaluated at ${\bf q}=0$. `D' and `M' Fermions refer to `Dirac' and `Majorana' respectively. \label{DMMatEl}}
\end{center}
\end{table}

For latter convenience, we summarize the results of this section by writing the difference in the forward scattering amplitudes {\it for a single DM-atom interaction}. The case of atomic clocks reads
\begin{align}
f_{1}(0)-f_{2}(0)
=\frac{{m_\chi }}{\pi} \left( G_N \frak{g}_{{\rm Ncl}}^{N}-G_e  \right) \vec J_\chi\cdot \frac{\vec {\lambda} }{F}, \label{DiffFinForm}
\end{align}
where  $\frak{g}_{{\rm Ncl}}^N$ is given in table~\ref{tab:gNNcl} and the DM currents in table~\ref{DMMatEl}.  $G_N$  in terms of $G_{u,d}$ is shown in  eqs.~\eqref{eq:q-NAx}-\eqref{eq:q-NT}.
Regarding  co-magnetometers,  for both ${}^3$He and  ${}^{129}$Xe the difference for  the states defined along the magnetic axes introduced in eq.~\eqref{eq:statesmag} is
\begin{align}
f_{1}(0)-f_{2}(0)
=\frac{{m_\chi }}{\pi} \left( G_N \frak{g}_{\rm Ncl}^{N} 
 \right) \vec J_\chi\cdot \frac{\vec {B} }{|B|}. \label{DiffFinFormB}
\end{align}
Finally, the K is in a strongly-coupled configuration which interacts with external magnetic fields as the $F=2$ states with 
a slower frequency, see~\cite{Brown:2011exm} for details. To understand how this configuration is 
modified in the presence of a DM background, the relevant quantity is the difference in the scattering amplitude for two consecutive states in the $F=2$ multiplet,  
\begin{align}
\label{eq:diff_f_F2}
\Delta f(0)\equiv  f_{\lambda}(0)- f_{\lambda-1}(0)=\frac{{m_\chi }}{\pi} \left[ \frac{ 3}{4}G_N \frak{g}_{{\rm K}}^{ N} + \frac{G_e}{4} \right]\vec J_\chi\cdot   \frac{\vec {B} }{|B|}.
\end{align}
The general case for alkali metals with arbitrary $F$ can be read from \eqref{eq:f1f2gen}. The last three expressions are  the main results of this section: spin-dependent interactions do affect differently the different angular momentum states in the clock or magnetometers.

\subsubsection{Particle scattering in models of axial boson dark matter} \label{sec:ax_photon}

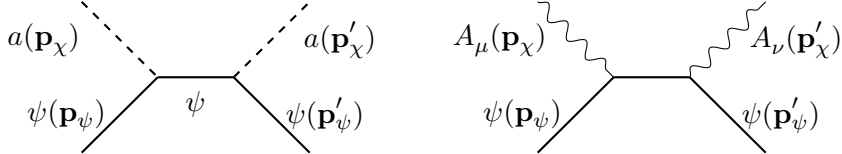
\begin{figure}[h]
\begin{center}
\begin{tikzpicture}
\draw (-2.5,.5) node {$a({\bf p}_\chi)$};
\draw [-, thick] (-2,-1)--(-1,0);
\draw [dashed,thick] (-2,1)--(-1,0);
\draw (-2.2,-.5) node {$ \psi({\bf p}_\psi)$};
\draw[-,thick] (-1,0)--(0,0);
\draw (-.5,-.35) node {$\psi$};
\draw (1.4,.5) node {$a({\bf p}_\chi')$};
\draw [dashed, thick]   (0,0)--(1,1);
\draw [-, thick] (0,0)--(1,-1);
\draw (1.2,-.5) node {$\psi({\bf p}'_\psi)$};
\draw (3.5,.5) node {$A_\mu({\bf p}_\chi)$};
\draw [-, thick] (4,-1)--(5,0);
\draw [style={decorate, decoration={snake}}] (4,1)--(5,0);
\draw (3.8,-.5) node {$ \psi({\bf p}_\psi)$};
\draw[-,thick] (5,0)--(6,0);
\draw (7.4,.5) node {$ A_\nu({\bf p}_\chi^\prime)$};
\draw [style={decorate, decoration={snake}}]   (6,0)--(7,1);
\draw [-, thick] (6,0)--(7,-1);
\draw (7.2,-.5) node {$\psi({\bf p}'_\psi)$};
\end{tikzpicture}
\caption{Diagrams for the axial bosonic DM scattering with a fermion in the atom. \label{fig:dm_pho}}
\end{center}
\end{figure}

In sec.~\ref{sec:axialB} we introduced DM candidates interacting with the SM fields through the renormalizable terms~\eqref{LagAx} and \eqref{LagAxVB}. These operators yield  `Compton' scattering processes between the SM fermions $\psi$ with spin $\vec \lambda_\psi$
and the DM fields through the processes shown in fig.~\ref{fig:dm_pho} (compare with fig.~\ref{Scat3}). 

 The spin-dependent part of the amplitudes in the NR limit and expanding on $m_{{a},A}/m_\psi$ is
\be\label{eq:modelAxAxV}
\begin{split}
\mathcal T^{( a)}({\bf p}_\chi, {\bf p}_\chi^\prime)=&{-i} \frac{C_\psi^2}{2m_{a}^2{\frak f}_{ a}^2}\,{\bf p}_\chi\wedge{\bf p}_\chi' \cdot \vec  \lambda_\psi,\\
\mathcal T^{(A)}({\bf p}_\chi, {\bf p}_\chi^\prime)=&{-2}\frac{i (g^A_\psi)^2}{m_A^2}\,\vec \epsilon\,({\bf p}_\chi)\wedge \vec{\epsilon\,}{}^*({\bf p}_\chi^\prime) \cdot \vec \lambda_\psi,
\end{split}
\ee
where $\vec \epsilon$ is the space-like part of the vector boson polarization vector.  The above result can be inserted in eq.~\eqref{eq:fdefntn} and evaluated at ${\bf p}_\chi={\bf p}_\chi'$. In this limit $\mathcal T^{(a)}$ vanishes and the sensitivity to scattering with axion particles is further suppressed by ${\cal O}(m_{a}/m_\psi)$ factors.

The transverse polarizations for the vector boson have a non-vanishing amplitude. The construction $-i\varepsilon_{ijk} \epsilon^i\epsilon^{*j}$ in \eqref{eq:modelAxAxV} is precisely the spin of the vector particle, $\lambda_A$. The above computation can be taken for $\psi$ an electron or a nucleon and for the clocks it yields:
\begin{align}
f_{1}(0)-f_{2}(0)
=&{\color{green}}\frac{{-1 }}{\pi m_A} \left( (g^A_{N})^2 \frak{g}_{{\rm Ncl}}^{ N}-(g^A_{e})^2 \right)  \frac{\vec \lambda_A\cdot \vec \lambda }{F}. \label{fsAxVB}
\end{align}
This expression is very similar to that of the EFT case \eqref{DiffFinForm} with the substitution $G_\psi=(g^A_{N})^2/m_A^2$.  For the case of magnetometers we have, for the two states of noble gases:\begin{align}
f_{1}(0)-f_{2}(0)
=&{\color{green}}\frac{{ -1 }}{\pi m_A} \left( (g^A_N)^2 \frak{g}_{{\rm Ncl}}^{N}
\right)  \frac{\vec \lambda_A\cdot \vec B }{B}, \label{fsAxVBMag}
\end{align}
 whereas for the states in K with total angular momentum $F=2$:
\begin{align}
\Delta f(0)=\frac{ -1}{\pi m_A} \left[ \frac{ 3}{4}G_N \frak{g}_{{\rm Ncl}}^{N} +\frac{1}{4}G_e \right]\vec \lambda_A\cdot   \frac{\vec {B} }{|B|}.
\label{DeltafK}\end{align}

\subsection{Classical field limit}\label{sec:classical}

 The local energy density of DM is fixed by the properties of the Milky Way DM halo and set to $\rho_{\rm DM}\sim 0.3\ \mathrm{GeV/cm}^3$ in the vicinity of the Solar System. As a result,  the number density  $n_\chi\equiv \rho_\chi/m_\chi$ grows at small DM masses. 
Since the spread in DM momentum should correspond to the values of virial equilibrium, we can estimate the occupation number of momentum states to be
\be
\label{eq:occup}
N_p\sim \left(\frac{15\, \ev}{m_\chi}\right)^4
\frac{\rho_\chi}{0.3\ \mathrm{GeV}/\mathrm{cm^{3}}}.
\ee 
The range below few  eV therefore has  large occupation numbers and the picture of space-separated particles gives way to a continuum description; the DM behaves like a classical field obeying the equations of motion\footnote{This picture is modified by interactions, which introduce a time scale where the classical description is no longer appropriate \cite{Chakrabarty:2017fkd,Hertzberg:2016tal,Dvali:2017ruz}. These references do not agree on the consequence of interactions (self-interactions or gravitational interactions) for the DM configuration in the Milky Way. We consider that the classical description  is  adequate, though our methods can be applied otherwise.}.  In this situation, one no longer describes the interaction with DM as a scattering process but as an atomic system evolving in the DM background. To a first approximation, the latter can be considered as  those of a free field. The field is therefore a superposition of plane waves
\be
e^{-i\omega t +i\vec k\cdot \vec x+i\phi_{\vec k}}, \label{eq:pw}
\ee 
with $\omega=(m_\chi^2+k^2)^{1/2}$ and a momentum dependent phase $\phi_{\vec k}$.  The behavior as a `cold' and clustering substance at early times in these scenarios
is a consequence of the initial conditions. More concretely,  a coherent massive  field displaced from its minimum  behaves as cold dark matter for times much longer than the typical oscillating time (set by the inverse of the mass) \cite{Preskill:1982cy,Abbott:1982af,Nelson:2011sf,Li:2013nal}. 
These configurations can be generated 
by a phase transition in the primordial universe or during the last $e$-folds of inflation, see e.g.  \cite{Graham:2015ifn,Nelson:2011sf,Arias:2012az,Alexander:2016aln,Graham:2015rva}.

The distribution of the plane waves \eqref{eq:pw} is set by  cosmic evolution and virialization in the DM halo. They satisfy
a Maxwell-Boltzmann distribution in the galactic rest frame with {\it a priori} random phases\footnote{We ignore other inhomogeneities that may arise due condensation phenomena at different scales, e.g. \cite{Sikivie:2009qn,Guth:2014hsa,Schive:2014hza,Schwabe:2016rze,Levkov:2018kau,Widdicombe:2018oeo}. They do not seem relevant for the DM distribution seen by  direct detection experiments. } \cite{Arvanitaki:2017nhi,Foster:2017hbq}. For instance, for the real vector field in the laboratory frame this translates into
\be
A_\m(x,t)    =\int \di^3v\,  {\cal F}({\bf v},{\bf v}_{\rm lab}) \hat n_\mu({\bf v}) e^{-i k_\m x^\m}+c.c.  \label{eq:A_dist}
\ee
where $\hat n_\mu({\bf v})$ is a transverse ($\hat n_\mu k^\mu=0$) arbitrary  vector  including a random phase.  ${\cal F}$ represents the normalized distribution, approximated by
\be\label{eq:virial_distr}
{\cal F}({\bf v},{\bf v}_{\rm lab})={\cal N}\left(\frac{1}{\pi v_0^2}\right)^{3/2}\exp\left[-\frac{({\bf v}-{\bf v}_{\rm lab})^2}{v_0^2}\right],
\ee
where $v_{\rm lab}\approx v_\odot \sim 10^{-3}$, with $v_\odot$ being the   velocity of the Sun in the  DM rest frame.  Finally, $v_0\sim 10^{-3}$ is the width of the distribution of virial equilibrium on the Milky Way~\cite{Arvanitaki:2017nhi}. 
  Given that the distribution is dominated by  velocities $v\sim 10^{-3}$,  
any given configuration at a particular time is coherent for  $10^6$ oscillations 
\cite{Graham:2015ifn}. This defines the mass below which the distribution behaves as a coherent medium for a time interval $T$,
\be
m_{\rm coh}\equiv 10^{-9}\,{\rm eV} \left(\frac{{s}}{T}\right),\label{eq:mcoh}
\ee
which follows from imposing that the phase of the different typical momenta differs by less than  $1$ in time $T$.
Finally, the normalization of \eqref{eq:A_dist} is set by comparing the energy density of the configuration 
with $\rho_{DM}$.

We will consider a monochromatic wave, which gives the correct description of the effect for
the cases we will study. 
In particular we will take the following expressions for (real or complex) scalars and vector bosons: 
\begin{subequations}\label{eq:testClFields}
\begin{align}
a=&\frac{\sqrt{{\rho_\chi}}}{\sqrt{2}m_\chi}e^{-ik_\mu  x^\mu }+c.c.,  & \chi=&\frac{\sqrt{{\rho_\chi}}}{\sqrt{2}m_\chi}e^{{-ik_\mu  x^\mu } },\\
A^\mu=&\frac{\sqrt{{\rho_\chi}}}{\sqrt{2}m_\chi}\epsilon^\mu e^{-ik_\mu  x^\mu }+c.c. , & \chi^\mu=&\frac{\sqrt{{\rho_\chi}}}{\sqrt{2}m_\chi}\epsilon^\mu e^{{-ik_\mu  x^\mu } },
\end{align}
\end{subequations}
 where for the complex field we have assumed it is made up of particles, if anti-particles are also present one would add a $e^{ik_\mu x^\mu}$ term. The effect in the atomic states  is that of an external time-dependent background. The corresponding Hamiltonian is obtained after substituting the previous field configurations in the interactions of sec.~\ref{sec:TH} and evaluating them at the atom's position.
The two types of interactions considered in this work, contact (sec.~\ref{sec:EFT}) and 3-body (sec.~\ref{sec:axialB}), scale differently with the amplitude of the oscillation and therefore $\rho_\chi$. Let us now make explicit the Hamiltonian they generate for a fermion $\psi$.
\begin{itemize}
\item {\it Effective field theory.} The operators in table~\ref{DMOps}  have two powers of the field and a net non-averaging-out effect {\it for each frequency}. We  collectively write the Hamiltonian of interaction as 
\begin{align}
H_{\rm int}={-}\frac{2\,G_\psi\rho_\chi}{m_\chi} \vec  \lambda_\psi \cdot \vec J_\chi^{\rm\, cl.}
\label{eq:HclEFT}
\end{align}
where $\vec J_{\chi}^{\rm \,cl.}$ is $\mJ_\chi$ evaluated for $\chi$ as given in eqs.~\eqref{eq:testClFields} and rescaled by a factor $\sqrt{m_\chi/\rho_\chi}$ which has been taken out front. With  this definition $\vec J^{\rm cl}_\chi$ coincides with $\vec J_\chi$ of the particle description, compare table~\ref{Jclass} and table~\ref{DMMatEl}.  The current proportional to the {\it velocity} can be considered constant during the time of the experiment for basically all the masses. Indeed, for masses $m_\chi \lesssim m_{\rm coh}$,
a single measurement will see a particular velocity with probability distribution given by eq.~\eqref{eq:virial_distr}. In the opposite regime, there is an averaging effect from different frequencies and the surviving current is  the laboratory's velocity with respect to the DM frame. The case for the {\it spin} current is different:  the survival of the effect in the regime $m_\chi>m_{\rm coh}$ requires a mechanism to produce polarized DM. The effect will tend to average out otherwise.  Once again we recall that the analysis is valid for light mediator as well with the substitution $G_\psi=g^\Z_\chi g^\Z_\psi/m_\Z^2$, 
as follows from \eqref{MedLagEFT}. 
\begin{table}[h]
\begin{center}
\begin{tabular}{|cc|cccc|}
\hline
$\psi=e,q$&&DM &Scalar&Fermion &Vector Boson\\ \hline
$\mJ_{\psi}:$& $\bar \psi \gamma^\mu \gamma_5\psi $ &$\vec J_\chi^{\rm\,cl.}:$ & $\pm \vec v_\chi$& -- & $-(\pm)\vec v_\chi$,\\
&& &  & -- & 
\\ \hline
$\psi=e,q$&&DM &Scalar&Fermion &Vector Boson\\ \hline
$\mJ_\psi:$ & $\bar \psi \sigma^{\mu\nu} \psi $ & $\vec J_\chi^{\rm\,cl.}:$& -- 
& -- & $ \pm\vec \lambda_\chi/m_\chi$. \\\hline
\end{tabular}
\caption{DM currents for the operators of tab~\ref{DMOps}  in the classical field case as defined in \eqref{eq:HclEFT}, cf. tab~\ref{DMMatEl} for the particle regime. The $\pm $ signs correspond to a field made of particles or antiparticles. The ansatz of eq.~\eqref{eq:testClFields} selects the upper sign.
\label{Jclass}}
\end{center}
\end{table}
\item { \it DM as an axial boson.}
The Hamiltonians corresponding 
to the interactions \eqref{LagAx} and \eqref{LagAxVB},  once evaluated in the 
previous solutions, are 
\begin{align}
\hspace{-.1cm} H_{a}=-\frac{C_\psi\sqrt{2\rho_\chi}}{{\frak f}_{ a}} \vec  \lambda_\psi \cdot \vec v  \cos( m_{ a} t +\phi_0),  \ \,\,
H_{A}= 2g^A_\psi \frac{\sqrt{{2}\rho_\chi}}{m_A}  \vec \lambda_\psi  \cdot  {\rm Re}[\vec \epsilon \,e^{-im_A t+i\phi_0}],\label{eq:Haxion-VB}
\end{align}
where  we have introduced an arbitrary constant phase $\phi_0$.
These cases scale linearly with the amplitude of oscillation.
\end{itemize}
The main difference between the two cases is the oscillating contribution for axial bosons that {\it always} averages out the effect at leading order for  times
 $t\gg 2\pi m^{-1}_{a,A}$, much shorter than the decoherence time.  This is due, both to the interaction being 4-point vs the 3-point interaction of the axion {\it and} the fact that the interactions do not vanish with ${\bf q}\to0$. This last remark is less evident so let us illustrate it with  a real scalar field $\chi$ and the interaction $\bar\psi \gamma_\mu \gamma_5\psi \partial^\mu \chi^2$ which we discarded in table~\ref{DMOps} since indeed it vanishes for ${\bf q}=0$. If one inputs the monochromatic solution $\chi=\chi_0\cos(k_\mu x^\mu)$ in this operator the outcome is an interaction that averages-out despite being 4-body.
 
 In the cases in which the DM effect is constant during the measurement time,
the contribution can be easily understood as a energy splitting of atomic states analogous to the Zeeman effect. For instance, for the case \eqref{eq:HclEFT} and the two states of \eqref{eq:states}
\begin{align}\label{DeltaEClass}
\delta E_{2}-\delta E_1=
 \frac{2 \rho_\chi   }{  m_\chi  } \left( G_N \frak{g}_{{\rm Ncl}}^{N}-G_e  \right)  \frac{\vec J_\chi^{\rm\, cl.}\cdot\vec {\lambda} }{F}.
\end{align} 
The time dependent cases have averaging effects  and may display resonant behaviour; we discuss the former in the next section while we leave the latter for future study.

\section{Atomic clocks and magnetometers in the presence of light dark matter}\label{sec:devices}

We now describe
how the effects discussed  in the last section can be used to constrain the possible interactions of DM with matter from experiments. We present the atomic clocks in sec.~\ref{sec:dev_ac}  and magnetometers in 
sec.~\ref{sec:dev_mg}. More details for the case of atomic clocks are given in the accompanying paper~\cite{Wolf:2018xlz}.  The  bounds we will derive do not require an improvement of current technology, but 
a reinterpretation of current data or modifications of existing devices.  The comparison with existing constraints  is deferred to sec.~\ref{sec:comp}.

\subsection{Atomic clocks in the presence of light dark matter}
\label{sec:dev_ac}



The scattering of background-gas particles is already recognized as a source of systematic uncertainty for atomic clocks \cite{vanier1989quantum,2012arXiv1204.3621G,2013PhRvL.110r0802G,2015RvMP...87..637L}. These effects modify the clock operation, be it by 
frequency shifts or losses of fringe amplitude, as shown in detail in  Ref.~\cite{Wolf:2018xlz}.
\begin{figure}[h]
\begin{center}
\begin{tikzpicture}
\draw[fill=orangeh,draw=gray] (-170pt,28pt) circle (10pt);
\draw [orangeh] (-170pt,28pt) ellipse (15pt and 3pt);
\draw [orangeh] (-170pt,28pt) ellipse (3pt and 15pt);
\draw[fill=black,draw=gray] (-65pt,46pt) circle (4pt);
\draw [->] (-60pt,45pt)--(-30pt,35pt);
\draw (-45pt,30pt) node {$\bf p_\chi$};
\draw (-160pt,0)--(160pt,0);
\draw (-140pt,85pt)--(-140pt,-2pt);
\draw [style={decorate, decoration={snake,amplitude=10pt}}] (-140pt,75pt)--(-110pt,75pt);
\draw (-125pt,93pt) node  {$\pi/2$ pulse};
\draw (-110pt,85pt)--(-110pt,-2pt);
\draw [<->, thick] (-110pt,0pt)--(97pt,0pt);
\draw  (-150pt,65pt)--(145pt,65pt);
\draw  (-140pt,85pt)--(-110pt,85pt);
\draw[fill=black,draw=gray] (35pt,25pt) circle (4pt);
\draw [->] (40pt,25pt)--(70pt,30pt);
\draw (55pt,15pt) node {$\bf p_\chi'$};
\draw[fill=orangeh,opacity=0.8,draw=gray] (0pt,28pt) circle (10pt);
\draw [orangeh,opacity=0.8] (0pt,28pt) ellipse (15pt and 3pt);
\draw [orangeh,opacity=0.8] (0pt,28pt) ellipse (3pt and 15pt);
\draw (-13pt,47pt) node {$\frac{1}{\sqrt 2}$};
\draw[fill=blueh,opacity=0.5,draw=black] (5pt,35pt) circle (10pt);
\draw [blueh,opacity=0.2] (5pt,35pt) ellipse (15pt and 3pt);
\draw [blueh,opacity=0.2] (5pt,35pt) ellipse (3pt and 15pt);
\draw (-18pt,17pt) node {$\frac{-i}{\sqrt 2}$};
\draw [<->, thick] (-140pt,0pt)--(-110pt,0pt) node [anchor=north east] {$t_1\,\,\,$};
\draw (0pt,-8pt) node {$T$};
\draw (97pt,85pt)--(97pt,-2pt);
\draw [style={decorate, decoration={snake,amplitude=10pt}}] (97pt,75pt)--(127pt,75pt);
\draw (112pt,93pt) node  {$\pi/2$ pulse};
\draw (97pt,85pt)--(127pt,85pt);
\draw (127pt,85pt)--(127pt,-2pt);
\draw [<->, thick] (97pt,0pt)--(127pt,0pt) node [anchor=north east] {$t_1\,\,\,$};
\draw (147pt,47pt) node {$\frac{1-e^{i\Delta \omega T}}{2}$};
\draw[fill=orangeh,opacity=0.5,draw=gray] (175pt,28pt) circle (10pt);
\draw [orangeh,opacity=0.5] (175pt,28pt) ellipse (15pt and 3pt);
\draw [orangeh,opacity=0.5] (175pt,28pt) ellipse (3pt and 15pt);
\draw (149pt,10pt) node {$\frac{1+e^{-i\Delta \omega T}}{2i}$};
\draw[fill=blueh,opacity=0.8,draw=black] (180pt,35pt) circle (10pt);
\draw [blueh,opacity=0.5] (180pt,35pt) ellipse (15pt and 3pt);
\draw [blueh,opacity=0.5] (180pt,35pt) ellipse (3pt and 15pt);
\end{tikzpicture}
\caption{Scheme for the Ramsey sequence. The horizontal axis represents time. In orange (blue) we  depict  the ground (excited) state $|1\rangle$ ($|2\rangle$). During the Ramsey time $T$ the atoms can interact with DM particles of momentum $\bf p_\chi$.  See main text for details.\label{fig:Ramseq}}
\end{center}
\end{figure}
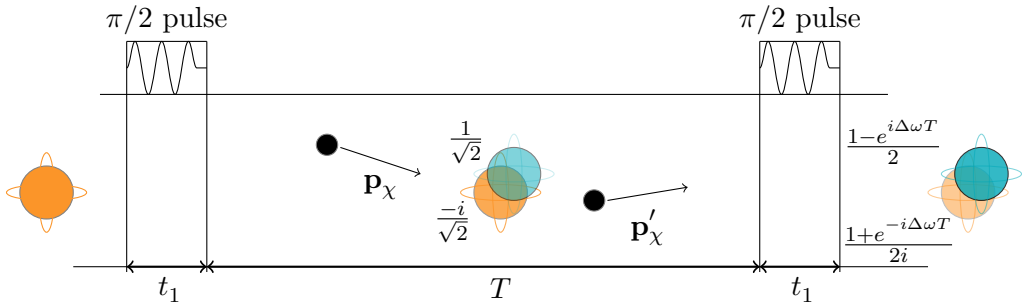

In this section we make explicit how the  spin-dependent interactions with DM on the two hyperfine-split states \eqref{eq:states} affect the atomic clock's operation. 
In particular we focus on a Ramsey sequence, schematically displayed in fig.~\ref{fig:Ramseq}. 
In this sequence, atoms in the ground state $|1\rangle$ are subjected to a light pulse of frequency $\omega$ for a time $t_1$ such that they come out in a superposition state. The latter is left free for a long time $T$ (the fiducial time is $T\approx 0.5 \, \mathrm{s}$) except for the possible interaction with the background  and other noise sources. Finally another pulse of the same specifications as the first one is applied and the final state is measured.

In the absence of new interactions,  the standard choices for the Ramsey sequence yield the probabilities of detection for each state at the end of the process \cite{weinberg2013lectures}
\begin{align}
P_1=&\sin[\Delta\omega T/2]^2, & P_2=&\cos[\Delta\omega T/2]^2, \label{eq:P1P2}
\end{align}
where\footnote{The reader acquainted with neutrino physics might find the following analogy useful: the light pulses can be taken to be ``production" and ``detection" with the association of the outcome states (superposition of energy states) to the interaction basis. During the longer  time $T$ the system   oscillates freely.
The probabilities in eqs.~\eqref{eq:P1P2} can be interpreted as the outcome of oscillations where nonetheless we can `tune' the energy difference via $\omega$.}
\be
\Delta \omega\equiv \omega-(E_2-E_1). \label{eq:Dw}
\ee
The light frequency  $\omega$ can be locked to the energy split by adjusting it to the value  $\omega_{\rm max}$  that maximises $P_2$. In the presence of a background, be it particles or a field, the evolution of the system is modified as made explicit in the respective  subsections below.

\subsubsection{Particle dark matter}\label{sec:partDMAC}
Since the free-fall time between pulses $T$ is much larger than the duration of the pulses $t_1$ we look at DM particle scattering during the interval $T$.
 Up to irrelevant phases, the wave-function of the DM-atom system after the second pulse is 
\begin{align}
\Psi_1(t,x)=&\frac12\left(\chi_1^{\rm out}(x)-e^{i\Delta \omega T}\chi_2^{\rm out}(x)\right),\\
\Psi_2(t,x)=&-\frac i2\left(\chi_2^{\rm out}(x)+e^{-i\Delta \omega T}\chi_1^{\rm out}(x)\right),
\end{align}
where the out states are given in eqs.~(\ref{eq:wave}) and (\ref{eq:PsiDef}).

As previously remarked, the leading effect will come from forward scattering where there is no momentum transfer and the trajectory of the atoms is unchanged.  The detection probabilities in each of the states at the end of the sequence when subjected to a flux of DM particles read: 
\begin{align}
P_2=& \int \di^3x|\Psi_2(t,x)|^2=
\cos[\Delta \omega\, T/2]^2 
+\frac{\pi n_\chi v \,T}{  p_\chi }\mathrm{Re}[\bar f_1(0)-\bar f_2(0)] \sin[\Delta \omega T], \label{eq:P2}\\
P_1=& \int \di^3x|\Psi_1(t,x)|^2=
\sin[\Delta \omega\, T/2]^2 
-\frac{\pi n_\chi v \,T}{   p_\chi }\mathrm{Re}[\bar f_1(0)-\bar f_2(0)] \sin[\Delta \omega T]
\label{eq:P1},
\end{align} 
where $n_\chi$ is the DM particle density $n_\chi=\rho_{\chi}/m_\chi$ and  $n_\chi v \,T$ is the DM flux per unit area around the atom.  
Notice that the pre-factors in the DM contribution are velocity independent and that the probabilities $P_1$, $P_2$ add up to $1$ at this order. The averaged amplitude $\bar f$ is
\begin{align}
\label{eq:av}
\bar f_\sti(0)=&\frac{1}{N_{sc}}\sum_a^{N_{sc}} f^a_\sti(0)\,, & N_{sc}=& \frac{ n_\chi v\,T  }{\kappa}\simeq 10^{12}  \left(\frac{T}{s}\right) \left(\frac{\rm eV}{m_\chi}\right)^3\left(\frac{\rho_\chi}{0.3\,{\rm GeV}/{\rm cm^{3}}} \right) \,,
\end{align}
where  $N_{sc}$ counts the number of scatterers that pass by the atom and $\kappa$ is the DM probability flux per unit area at the atom's position\footnote{For a Gaussian wave-function with spread $d$, $\Psi\sim e^{-(p-p_0)^2d^2}$, one has $\kappa=1/(2\pi d^2)$ at the center of the distribution. As described in more detail in \cite{Wolf:2018xlz}, $N_{sc}$ is modified for $m_\chi \gtrsim 10^4$~eV, since at these masses the wavepacket's  size is smaller or of the order of the size of the atom at the relevant temperatures ($\mu$K) for clocks, $d\approx 10^{-8}$~m. Our bounds are never competitive at these `high' masses, and we only consider the limit described above.  Magnetometers operate at much higher temperatures and hence are more localized.}.
To estimate  $N_{sc}$, we considered a dispersion in momentum  of the order of $m_\chi v_0$.  We note that $N_{sc}\sim 1$ marks a threshold in DM mass, 
\be
\label{eq:msc}
m_{\rm sc}\equiv 10 \left(\frac{T}{s}\right)^{1/3}\left(\frac{\rho_\chi}{0.3\,{\rm GeV}/{\rm cm^{3}}}\right)^{1/3}\, \rm keV.
\ee 
Above this mass value, DM particles are too sparse for a single atom to encounter more than one of them in time $T$.  In this regime $N_{sc}$ should be taken as a probability: if there are $N_{at}$ in the experiment $N_{sc}N_{at}$ of them will be traversed by  DM wave-packets, which in turn means that for $N_{sc}<1$ the signal deteriorates as $N_{sc}$. For smaller masses there will be a large number of scatterers per-atom and the final effect depends on whether the interactions add up or average out in  (\ref{eq:P2}) and (\ref{eq:P1}).  In the case in which the interaction depends on the velocity of DM there is a coherent contribution from the different scatterers representing the
velocity of the detector with respect to the DM rest frame, $\bar v\sim v_\odot$. For DM-spin-dependent case, $\vec J_\chi\propto \vec \lambda_\chi$, the net effect tends  to average out unless there is a net polarization of DM. If DM is unpolarized and the spin of each scatterer is taken to be a random variable, the effect diminishes as $1/\sqrt{N_{sc}}$ for each atom, the DM being a  source of `noise'.  Furthermore,  even in the case in which $N_{sc}\sim1$, the device is composed  of $N_{at}$ atoms (a standard value is  $N_{at}\sim 10^6$). A key question for the averaging is whether each atom sees different DM particles or if some of them are the same. To determine this we estimate the number of atoms that a DM particle sees as it flies through the sample. This is the volume of the experiment that the DM packet sweeps times the atom's density. For a  contact interaction and a sample of size $L$, this number is $L/(mv)^2\times N_{at}/L^3$. For atomic clocks, $L\sim$ cm, $N_{at}\sim 10^6$, and above $\sim~{\cal O}(10)$~eV every atom sees different DM and so the total suppression is $1/\sqrt{N_{at} N_{sc}}$. This will not be the case for co-magnetometers due to the high density of atoms.
On the other hand if DM is polarized the effect is unsuppressed, yet the dynamical process to obtain such polarization is unclear\footnote{One can speculate about the  possibility of DM interacting with the galactic magnetic fields and acquiring a net polarization. Determining if this is allowed by current constraints is beyond the purpose of this work. }. Finally, a  coherent signal for times longer than $T$  may be important for different observables. For instance, if we want to use daily modulation of the signal. 

The case of a light mediator, $m_\chi v\gg m_\Z$, generates different thresholds. As discussed in app.~\ref{app:lightmA}, the atom sees not only the DM that passes {\it over} it (impact parameter $< m_\chi v$), but also that within a $1/m_\Z$ radius. The effect of this extra shell is the same as that in eqs.~(\ref{eq:P1}) and (\ref{eq:P2}), but the average is now over $N\sim n_\chi l_{\rm int}^3= n_\chi/m_\Z^3$ DM particles. The equivalent of eq.~\eqref{eq:msc} is thus
\be
m_{\rm med}\equiv(10^{-2}{\rm eV}/m_\Z)^3\, \rm eV, \label{eq:mmed}
\ee  
and one should also be careful when computing the amount of atoms seen by a DM particle.

The modification of the detection probability present  in   \eqref{eq:P2} changes the pulse frequency that maximizes $P_2$. For the EFT four-point interaction case and the light mediator:
\be
{\rm Contact\ Interaction\,or\,Light\ Mediator},\qquad\Delta \omega_{\rm max}= \frac{2\rho_\chi   }{  m_\chi  } \left( G_N \frak{g}_{{\rm Ncl}}^{N}-G_e \right)\vec{\bar J}_\chi\cdot \frac{\vec {\lambda} }{F}\label{eq:domega_flux1},
\ee
where $\Delta \omega_{\rm max} \equiv \omega_{\rm max}-(E_2-E_1)$, and we have used \eqref{DiffFinForm}. From this expression it follows that the effect can be interpreted as a contribution to the energy difference of the two states, just like eq.~\eqref{DeltaEClass} in the classical case; what is more, the two expressions are the same, since $\vec J_\chi^{\rm cl.}=\vec J_\chi$ as given in table~\ref{Jclass}. 
This suggests a smooth transition from the particle to the field description -- even though our expressions are not strictly valid for intermediate occupation numbers. 

For the case of axial vector, in the limit  of particle description $ m_A> 10\ {\rm eV}$, from \eqref{fsAxVB}
\be \label{eq:domegaAVB}
{\rm Axial\ Vector},\qquad\Delta \omega_{\rm max} =  - \frac{ 2 \rho_\chi  }{ m_A^3  }   \left( (g^A_{N})^2 \frak{g}_{{\rm Ncl}}^{N}-(g^A_{e})^2 \right)  \frac{\vec{ \bar  \lambda}_A\cdot \vec \lambda }{F}. 
\ee
Here $\bar\lambda_A$ is the averaged spin over scatterers and atoms in the sample, which yields the same suppression as in the unpolarized case of the spin-spin interaction of the EFT.

 Finally, as we emphasized before, the effects of \eqref{eq:domega_flux1} and \eqref{eq:domegaAVB} are present even if the momentum transfer is zero. This is an important difference as compared to various traditional  DM  searches based on scattering of DM, which allows one to explore light dark matter scenarios.

It is worth pausing and pondering the nature of the effect we described, specially in contrast with conventional DM direct detection searches.
We focus in the case  in which the mediator is much heavier than DM so that the EFT applies; that is
$m_\chi^2 G\sim (m_\chi/m_\Z)^2\ll 1$.
In conventional WIMP DM searches the number of events $N$ is the product ({total} flux)$\times$(cross section($\sigma$))$\simeq$ flux$\times m_\chi^2/m_\Z^4$, so that $N$ can be taken to be the number of particles
that pass within a radius $r\sim\ m_\chi\,/m_\Z^2$. The effect in atomic clocks on the other hand modifies the probability as the product (flux) $\times$(scattering amplitude($\mathcal T$))/(velocity) $\simeq$ $n_\chi T/ m_\Z^2$ as made explicit in eqs.~\eqref{eq:fdefntn} and \eqref{eq:P2}. This effect is proportional to the {\it potential} generated by the DM particles within a radius $r\sim 1/m_\Z$. This is indeed analogous to neutrino forward scattering and hence a `$G_F$' effect as opposed to the `$G_F^2$' effect in WIMP searches. This remark is aimed at providing some intuition about the effect at high masses. Notice, however,  that the searches here described are not competitive for conventional WIMP parameters (see below).
\subsubsection{Dark matter in the classical field limit}

In the classical field limit, the contact interactions of the EFT in table~\ref{DMOps} have a non-vanishing average value which  contributes to the hyperfine splitting, eq.~\eqref{DeltaEClass}. Hence, the formulae \eqref{eq:P1P2} are simply modified by adding the quantity \eqref{eq:Dw} to the energy difference:
\be
{\rm Contact\ Interaction\,or\,Light\ Mediator},\qquad\Delta \omega_{\rm max}= \frac{2 \rho_\chi   }{  m_\chi  }  \left( G_N \frak{g}_{{\rm Ncl}}^{N}-G_e \right)\vec J_\chi^{\,\rm cl.}\cdot \frac{\vec {\lambda} }{F}\label{eq:domega_flux2}.
\ee
This result coincides with that of the particle regime, \eqref{eq:domega_flux1}, with the substitution $\vec J_\chi\mapsto \vec J_\chi^{\,\rm cl.}$, and $\vec J_\chi^{\,\rm cl.}$ can be thought of as the macroscopic average of the microscopic property $\vec J_\chi$.
Also in this case the current $\vec J_\chi^{\rm\, cl.}$ may average out if it is not coherent during the Ramsey sequence. For the vector case $\chi_\mu$ this 
means that $m_\chi\lesssim m_{\rm coh}$, cf. \eqref{eq:mcoh}. As explained, in sec.~\ref{sec:classical}, for the velocity current we do not have this extra caveat since the velocity of the detector with respect to the reference frame of the galaxy will emerge in the average of the different flows for masses above $m_{\rm coh}$.

In contrast, in the axial boson case the interaction Hamiltonian generated by the DM oscillates around 0. In this situation,  the evolution of the atom's wave-function  generates the probabilities (at first order in the interactions):
\begin{align}
P_2=\cos[\Delta \omega\, T/2]^2{  -}\frac{C_\psi\sqrt{\rho_\chi}}{2\sqrt{2}{\frak f}_{ a}m_{ a}}\left( \vec J^{\,\psi}_1-\vec J^{\,\psi}_2\right) \cdot{\rm Im}\left[\vec v\, e^{-i\phi_0}(e^{-im_{ a} T}-1) \right] \sin[\Delta \omega T],\\
P_1=\sin[\Delta \omega\, T/2]^2{  +}\frac{C_\psi\sqrt{\rho_\chi}}{2\sqrt 2 {\frak f}_{a}m_{ a}}\left( \vec J^{\,\psi}_1-\vec J^{\,\psi}_2\right)\cdot {\rm Im}\left[\vec v \, e^{-i\phi_0}(e^{-im_{ a} T}-1)\right]\sin[\Delta \omega T], 
\end{align}
where $\Delta \omega$ is defined in \eqref{eq:Dw}.
The axial vector boson case is
obtained substituting $-C_\psi\vec v/(2{\frak f}_{ a}) \to g^A_\psi\vec\epsilon /m_A$. 
Assuming that $\Delta \omega\,T\ll 1$, one can write the shift of frequency as
\begin{align}
&{\rm Axion}\ \eqref{LagAx},\ \ \Delta \omega_{\rm max} =- \frac{ \left( C_N \frak{g}_{{\rm Ncl}}^{N}-C_e \right)\sqrt{2\rho_\chi}   }{ {\frak f}_{ a}  }  \frac1T\int_0^T\di t \, {\rm Re}[ \frac{\vec{\lambda}\cdot \vec v}{F}e^{-im_{ a} t-i\phi_0}], \label{eq:domega_flux}\\
&{\rm Axial\ Vector}\, \eqref{LagAxVB},\ \, \Delta \omega_{\rm max} = \frac{2 \left( g^A_N \frak{g}_{{\rm Ncl}}^{N}-g^A_e \right)\sqrt{2\rho_\chi}   }{ m_A  }  \frac1T\int_0^T \di t \, {\rm Re}[\frac{\vec{\lambda}\cdot\vec \epsilon}{F}e^{-im_{A} t-i\phi_0}] . \label{eq:domega_flux3}
\end{align}
From these expressions it is clear that for  $m_{a,A} \gtrsim 10^{-15}(s/T)${\,eV}  there is a loss of sensitivity inversely proportional to the mass\footnote{One could in principle look for resonant effects at all mass scales by tuning $\Delta \omega$ to the mass scale. We do not discuss this possibility here.}.
Similarly, recall that non-coherent effects appear after $10^6$ oscillations. This is important if we use daily modulation of the signal to detect the background. 
From \eqref{eq:mcoh},  coherence of one day  requires masses below $\sim 10^{-14}\,$eV.

\subsubsection{Sensitivity of atomic-clocks to dark matter interactions with matter}\label{sec:AC_DM}

After having identified the frequency $\omega_{\rm max}$ that maximises $P_2$ as our handle on the interactions between the atomic 
clock and the background, we now examine the size of the interactions that can be measured by current devices.

The main effect is absent in unpolarised atoms,  as can be seen in eqs.~\eqref{eq:Je} and \eqref{eq:Jq}. 
One thus needs to  run the clock on spin polarized states, as done routinely to calibrate for magnetic effects \cite{vanier1989quantum,2012arXiv1204.3621G}. The resulting magnetic sensitivity reduces the performance, but in the dual fountain clock FO2 \cite{2012arXiv1204.3621G}, running simultaneously on $^{133}$Cs and $^{87}$Rb, this limitation can be overcome by a particular combination of the observables whilst retaining sensitivity to the DM interaction, as we now explain.
When a polarised sample is selected by the use of  magnetic fields one generates  the energy difference \cite{vanier1989quantum} 
\begin{align}
\Delta E(B)\equiv E_2(B)-E_1 (B)= \mu_B\left({\mathsf g}_e-{\mathsf g}_I \right) \frac{\vec B\cdot \vec \lambda}{F}, 
\end{align}
where $\mu_B$ is the Bohr's magneton and  ${\mathsf g}_e$ and ${\mathsf g}_I$ are the electron and nuclear   Land\'e factors. The previous expression resembles the effect of DM, in particular both depend on total angular momentum in the same way, cf. eq.~\eqref{DiffFinForm}. 
To resolve this degeneracy, one can compare the effect in two different atoms in the same $\lambda$ state, e.g. Rb ($F=2$) and Cs ($F=4$). The ratio of magnetic energy shifts is dominated by the electron contribution, and given by $\Delta E^{\rm Rb}(B)/ \Delta E^{\rm Cs}(B)\simeq2$ to order $\sim 5\times10^{-4}$~\cite{vanier1989quantum}.  The same ratio in the case of DM reads:  $2\left( G_N \frak{g}_{{\rm Rb}}^{N}-G_e  \right)/\left( G_N \frak{g}_{{\rm Cs}}^{N}-G_e  \right)$. In the coupling to quarks we have $\frak{g}_{{\rm Rb}}^{N}/\frak{g}_{{\rm Cs}}^{N}-1\sim {\cal O}(1)$, cf. table~\ref{tab:gNNcl}, and the effect can be distinguished from the magnetic shift without further suppression (barring the tuned case where the couplings produce the same effect as in the SM). The electron case  is partially degenerate with the magnetic shift which deteriorates the sensitivity by a factor $\sim 5\times 10^{-4}$.

Furthermore, since the computation of $E_2-E_1$ entering in \eqref{eq:Dw}  can not be performed to sufficient accuracy for the atoms of interest, one can not use absolute information about $\Delta \omega_{\rm max}$ to hunt for interactions during the Ramsey process. In other words, if the effect of the DM is constant, it can not be {distinguished from atomic effects}. 
There are different ways to overcome this difficulty. First, if the DM currents produce a {\it coherent effect} lasting for several hours one can look for changes in the  scalar product of the spin of the probe with the DM current. These  are present since the polarization of the sample is selected by a magnetic field pointing in a fixed direction rotating with the Earth.  These variations can be measured by comparing with clocks working with unpolarized states,  insensitive to the  interactions we considered.  Notice also that the relative variations  are ${\cal O}(1)$  and that this search strategy is that of Lorentz violation searches, see app.~\ref{app:LV}. This can   be  used to {\it confirm} the DM origin of a signal  by placing the detector at different latitudes  or changing the direction of the magnetic field\footnote{Other effects
related to annual modulation
or, for strong enough interactions, shielding of DM by the 
the Earth 
 or extra-shielding may be used as extra confirmation of the astrophysical origin of the effect.}.  Secondly, the non-coherent part of the currents represents  an irreducible source of noise for the clock. Thus, instead of looking for the effect of the DM in the precise value of $\Delta \omega_{\rm max}$, one can use the level of the noise to bound its interaction with the sample. The disadvantage of this approach is that the effect is reduced by the standard suppression factor related to noise sources.
 
The precision on the phase achieved in Rb/Cs clocks is $\Delta \omega^{\rm exp}_{\rm max}  \lesssim  10^{-5}\,$rad/s \cite{2012arXiv1204.3621G,2014Metro..51..108G}.   This requires combining data taken over $\sim 10^6\,$s, while each measurement lasting $0.5\, $s has a precision of $\Delta \omega^{\rm exp}_{\rm max}  \lesssim  10^{-3}\,$rad/s. For the contact interactions and models with light mediator of sec.~\ref{sec:EFT}, we can plug the best clock precision into \eqref{eq:domega_flux1}.  After combining the Cs and Rb results to suppress the magnetic sensitivity, one gets:
\begin{align}
G_\psi \bar J_\chi&<3\times 10^{3}\frac{1}{\Upsilon_\psi} \frac{F_{\rm Rb}}{\lambda} \left(\frac{\Delta \omega^{\rm exp}_{\rm max}}{ 2\times 10^{-5}\rm rad/s}\right)\left(\frac{0.3\, \mathrm{GeV/cm^3}}{\rho_{DM}}\right)\left(\frac{m_\chi}{\mathrm{eV}}\right)\mathrm{GeV}^{-2}. \label{eq:Gg}
\end{align}
The factor $\Upsilon_\psi$ depends on the coupling that we are considering. For nucleons, it represents the difference $\Upsilon_N\approx \mathfrak g^N_{\rm Rb}-\mathfrak g^N_{\rm Cs}\sim 0.5( n), 1.3(p)\,$ cf. table~\ref{tab:gNNcl}. For the electron, it represents the suppression from the degeneracy with magnetic effects, $\Upsilon_e\approx 5\times 10^{-4}$.
The previous expression  is also valid for the $\Z_\mu$-mediated interactions with the substitution  $G_\psi=g^\Z_\psi g^\Z_\chi/m_\Z^2$, recall \eqref{MedLagEFT}. 
It also holds for the particle and field regimes \eqref{eq:domega_flux2} of the EFT case once $\bar J_\chi$ is regarded as a generalized average  (or classical current). As we discussed before, this current may  include an extra source of suppression depending on the nature of the interactions and the DM mass.
These bounds are compared with existing ones in sec.~\ref{sec:part_bounds}, figs.~\ref{Fig:elec_1} and \ref{Fig:nn_cl1}.

The previous analysis applies for the particle regime of the models of DM as an axial boson.  However, even the most promising case of the axial vector 
does not produce competitive bounds, as can be shown by the substitution $G_\psi\vec J_\psi\to g_\psi^2/m_A^2 \vec \lambda_A$ (eq.~\eqref{eq:domegaAVB}). The conclusion is different in the ultra-light mass regime. For the cases of  configurations coherent for at least several  days (we take $10$~days as a very conservative limit), $m_{a,A}\lesssim 10^{-15}$eV, the expressions (\ref{eq:domega_flux}) and (\ref{eq:domega_flux3}) yield
\begin{align}
&\frac{{\frak f}_{ a}}{C_\psi} \gtrsim  { 1.6\times 10^5}\Upsilon_\psi\frac{\lambda}{F_{\rm Rb}}\left(\frac{2\times10^{-5}{\rm rad\ Hz}}{\Delta \omega^{\rm exp}_{\rm max}}\right)\sqrt{\frac{\rho_{DM}}{0.3\, \mathrm{GeV/cm^3}}}\left(\frac{v}{10^{-3}}\right){\rm GeV},\label{eq:ax_a}\\
&g^A_\psi \lesssim  { 3}\times 10^{-38}\frac{1}{ \Upsilon_\psi } \frac{F_{\rm Rb}}{\lambda} \left(\frac{\Delta \omega^{\rm exp}_{\rm max}}{2\times 10^{-5}{\rm rad\ Hz}}\right)\sqrt{\frac{0.3\, \mathrm{GeV/cm^3}}{\rho_{DM}}}\left(\frac{m_A}{\mathrm{10^{-20}eV}}\right), \label{eq:ax_b} 
\end{align}
for the axion and axial-vector couplings in sec.~\ref{sec:axialB}. These bounds are compared with present bounds in sec.~\ref{sec:compaxfld}, figs.~\ref{fig:axion_b} and \ref{fig:axiV_b}. 

 Finally, it is common practice to compare the sensitivity of experiments looking for DM in the plane `DM mass-cross section'. 
Atomic clocks are mostly sensitive to  forward scattering amplitudes whose connection to total cross-section is model dependent. To establish it, we start from the
standard relation between the matrix element and the differential cross-section, eq.~\eqref{eq:cross_f}. For the cases with a `light' mediator we can approximate
\begin{align}
\label{eq:diffsigma}
\frac{\di \sigma}{\di\cos\theta}\simeq & \frac{2\pi f(0)^2}{\left({\bf q}^2/m_\Z^2+1\right)^2},
\end{align}
where we have neglected the $\bf q$ dependence in the numerator, valid for an estimate in the cases here discussed.  The limit of contact interaction corresponds to $m_\Z\gg \bf q$. The bound we established applied for {\it differences} of amplitudes. If the leading contribution to the cross-section does not cancel in these differences, the expression \eqref{eq:Gg} translates into
\begin{align}
\frac{\di \sigma_N}{\di\cos\theta}
&\leq \frac{6 \times10^{-39}}{\left({\bf q}^2/m_\Z^2+1\right)^2} \left[\left(\frac{\Delta \omega^{\rm exp}_{\rm max}}{2\times 10^{-5}{\rm rad\, Hz}}\right)\left(\frac{0.3\, \mathrm{GeV/cm^3}}{\rho_{DM}}\right)\label{eq:boundDs}
\left(\frac{m_\chi}{\mathrm{eV}}\right)^2\right]^2 {\rm cm}^2.
\end{align}
In the previous expression there will be a factor ${\mathcal O}(\mathfrak g_{\rm Rb}^N/\mathfrak g_{\rm Cs}^N-1)^2\sim 1$ that we ignore. The sensitivity in the case of electrons is worsened by a factor $10^7$.
One  sees that if the mediator is lighter than
the momentum transfer of the process $\bf q$, the sensitivity to cross section improves. This fact, together with the quick sensitivity improvement as one moves to lower DM masses,
will be relevant when contrasting atomic clocks with other bounds and searches in sec.~\ref{sec:comp}. 

The cross-sections \eqref{eq:boundDs} may be large enough to generate scattering or absorption by the Earth or the atmosphere. The total cross-sections for which this starts being important are  \cite{Erickcek:2007jv,Riedel:2016acj,Kavanagh:2016pyr,Emken:2017qmp} 
\be
\sigma_{\oplus}\sim 10^{-35}\, \mathrm{cm}^2,\  \quad
\sigma_{\mathrm{atm}} \sim10^{-28} \, \mathrm{cm}^2.
\ee
Depending on the model, DM can be absorbed or thermalized at higher cross sections, which modifies the properties of the distribution at the  detector. 
For our experimental set-ups, only the possibility of distortions by the atmosphere may constitute a real challenge, and we will always work with $\sigma<\sigma_{\rm atm}$. This limit may be overcome by using atomic clocks in space, which is already planned for other tests of fundamental physics \cite{LAURENT2015540,2017arXiv170903256L}. 

\subsection{Magnetometers in the presence of light dark matter}\label{sec:dev_mg}

The use of magnetometers to constrain axionic DM of very small masses was already suggested in \cite{Graham:2017ivz}. Here we extend this analysis in two ways: first, we provide the nuclear factors that connect the bounds to fundamental couplings.  Secondly, we consider the different DM candidates of sec.~\ref{sec:TH}, including the regime of particle scattering. As mentioned in sec.~\ref{sec:AC}, we will focus on the set-ups \cite{Allmendinger:2013eya,Brown:2010dt}.  An important remark is that these devices operate  with a number of atoms  many orders of magnitude larger than for atomic clocks. 

\subsubsection{Particle dark matter}\label{sec:partDMMG}

For the set-ups that we will consider, it is important to understand how the spin states behave when exposed to  DM collisions. The spin states are interacting with a magnetic field and hence evolving according to  the Hamiltonian:
\begin{align}
H_{\rm int}=- \gamma \vec B \cdot \vec {\lambda}\,, \label{eq:HintB}
\end{align}
with $\vec {\lambda }$ the total angular momentum operator and $\gamma$ the gyromagnetic ratio of a given atom. 

Let us consider  the interaction within a time interval $t$ with the DM particles of number density $n_\chi$. In the approximation of elastic scattering, the time evolution after a single passage of a DM particle off an  atomic state (which we describe in the basis of energy eigenstates~\eqref{eq:statesmag}) is given by eq.~\eqref{eq:wave}. The time evolution of the average for a given {\it atomic observable} $\hat O$ is found by tracing over the DM states. Assuming 
a flux of DM as done  in \eqref{eq:av}, to first order in the amplitudes $f_\sti$,
\begin{align}
\hspace{-.17cm}\int \di^3x|\chi(x)|^2 \langle{\rm At}(t)|  \hat O|{\rm At}(t)\rangle=
\langle {\rm At}_0(t)|\hat O| {\rm At}_0(t)\rangle\hspace{-.05cm}+\hspace{-.05cm} \frac{i2\pi n_\chi v \, t }{p_\chi}\langle{\rm At}_0(t)|\hspace{-.1cm}\left[\hat O\,,\, \hat  f\,\right]\hspace{-.1cm}|{\rm At}_0(t)\rangle,\label{eq:obsO}
 \end{align}
where we used $\hat O^\dagger =\hat O$ and $|{\rm At}_0(t)\rangle$ is the standard evolution without scattering. The operator $\hat f$ is defined as 
\be
\hat f|\sti\rangle=\bar f_\sti(0) |\sti\rangle,
\ee
where $\bar f_\sti(0)$ was introduced in \eqref{eq:P2}.
Only if $\bar f_\sti(0)$ depends on the spin state there will be an effect at this order.  Since $\hat f$ and $H_{\rm int}$ are already diagonal in the spin basis,  the new effect is a modification of the precession of the atom spin. To illustrate this, take the spin-$1/2$ states of \eqref{eq:statesmag} and a small time interval $\delta t$ such that \eqref{eq:HintB} and the DM effect are treated as a perturbation. The operator $\vec { \lambda}$ evolves as 
\begin{align}
\delta\langle \vec \lambda\rangle=\delta t\left(\left( E_2-E_1\right)-\frac{2\pi n_\chi v}{p_\chi}\left(\bar f(0)_2-\bar f(0)_1\right)\right)\langle
\vec\lambda\rangle \wedge \vec u_B \label{DeltaS}\,,
\end{align}
where   $\vec u_B=\vec B/|B|$ and $E_2-E_1=\gamma |B|$ is the Larmor frequency.   For higher spins states we note that $f_2-f_1$ is the difference in decay amplitudes for states differing by one unit in their spin along the magnetic field direction and the equivalent of eq.~\eqref{DeltaS} contains $\Delta f$ as in eqs.~\eqref{eq:diff_f_F2} and \eqref{DeltafK}.

Therefore, the effect of the DM scattering is a modification of the magnetic field to an effective field $\beta$ which modifies the precession frequency
 $\omega$ 
\begin{align}
\label{eq:betaeff}
\beta &\equiv  B+\frac{2\pi n_\chi }{m_\chi \gamma }\left(\bar f(0)_1-\bar f(0)_2\right)\,,& \omega \equiv \,&\gamma \beta=\gamma\left( B+\frac{2\pi n_\chi }{m_\chi \gamma }\left(\bar f(0)_1-\bar f(0)_2\right)\right)\,. 
\end{align}
The contribution of DM effects to the frequency  reads as in the atomic clock case, eqs.~\eqref{eq:P2}, \eqref{eq:P1} and \eqref{eq:domega_flux1}. Indeed, one can think of both experiments as sensitive to DM via the measure of the atomic state's energy split.
The expression for the difference of amplitudes is given in eq.~\eqref{DiffFinFormB} for the EFT in terms of the DM current $J_\chi$ which can be either its velocity or spin and in eq.~\eqref{fsAxVBMag} for the axial vector boson case. The expression for the average $\bar f$ is given in eq.~\eqref{eq:av}. 

The He-Xe co-magnetometer described in~\cite{Allmendinger:2013eya} measures the deviations from magnetic couplings by comparing the Larmor frequencies of the two noble gases. In our case, using eq.~\eqref{eq:betaeff} and 
eq.~\eqref{DiffFinFormB} one obtains:
\begin{align}
\omega_{\rm He}-\frac{\gamma_{\rm He}}{\gamma_{\rm Xe}}\omega_{\rm Xe}=\frac{{ 2}\rho_\chi}{m_\chi}G_N\left( {\mathfrak g}_{\rm He}^{N}-\frac{\gamma_{\rm He}}{\gamma_{\rm Xe}}{\mathfrak g}_{\rm Xe}^{N}\right)\vec{\bar J}_{\chi}\cdot \vec{u}_B\,, \label{eq:ptcl_HeXe}
\end{align}
where $\bar J_\chi$ is given in table~\ref{DMMatEl} for the EFT. In the axial vector  mediator  $G_\psi\to g^\Z_\psi g^\Z_\chi/m_\Z^2$.

 The self-compensating  K-He magnetometer of \cite{Brown:2010dt}  is sensitive to the different  shift from the magnetic field for electrons in K and the He nucleus. 
For K,   the effective magnetic field reads:
\begin{align}
\beta_e= B+{ 2}\frac{\rho_\chi}{m_\chi} \frac{ 3 G_N \frak{g}_{{\rm K}}^{ N}  +G_e}{{ 4}\gamma_e}(\vec{\bar  J}_\chi \cdot \vec u_B)\,, \label{eq:beta_e}
\end{align}
where $\gamma_e$ is the electron gyromagnetic ratio. In contrast, for He the nuclear spin gives the main contribution and the effective field it feels is:
\begin{align}
\label{eq:beta_He}
\beta_{\rm He}= B+2\frac{\rho_\chi}{m_\chi}\frac{G_N \frak{g}_{{\rm He}}^{ N}}{\gamma_{\rm He}} (\vec {\bar  J}_\chi \cdot \vec u_B)\,.
\end{align}
In the difference $\beta_e-\beta_{\rm He}$ the magnetic field cancels and one has a clean probe of DM effects.  Given that $\gamma_{\rm He}/\gamma_e\sim m_e/m_N$, the sensitivity to DM-electron interactions is considerably worse than the nucleon's one, as in the atomic clock case. 

The average effective magnetic field  $\beta$ depends on the type of DM current as detailed in the atomic clock case; a key difference is that the number of atoms that a DM particle sees is  very large $N_{at}/(Lm_\chi v)^2\gg1$ given $N_{at}\sim 10^{22}$ (for He in K-He, $L\approx$\ cm is the size of the sample). The total number of un-correlated\footnote{In the case of the K-He co-magnetometer atomic collisions might correlate some of the events discussed here. We ignore such possible correlations in our order of magnitude estimates.} events or crossings is then not $N_{at}N_{sc}$ but rather this quantity divided by the number of atoms that one DM particle sees, $N_{sc}(Lm_\chi v)^2$, which is independent of $N_{at}$. The `noise' suppression is then $1/(\sqrt{N_{sc}}m_{\chi}vL)$.

\subsubsection{Dark matter in the classical field limit} \label{sec:DMcl_mag}

For very low DM masses the field behaviour of DM induces an extra interaction term in the Hamiltonian. The simplest instance to deal with is the case in which the perturbation is time independent, where \eqref{eq:HintB} is complemented by  \eqref{eq:HclEFT}. This means in practice an `anomalous' magnetic field around which the spin will precess and a modified Larmor frequency. Take He for instance,
the energy states of the total Hamiltonian have a splitting 
\begin{align}
E_2-E_1=\omega_{\rm He}&=\left|\gamma_{\rm He} \vec B +{2}\mathfrak g^{N}_{\rm He} n_\chi G_N \vec J_\chi^{\rm\,cl}  
\right| \simeq \gamma_{\rm He} B+2n_\chi \mathfrak g^{N}_{\rm He} G_N  \vec J_\chi^{\rm\, cl} \cdot \vec u_B\,, 
\end{align}
and we recover \eqref{eq:beta_He} with $\vec {\bar  J}_\chi \to \vec J_\chi^{\rm\,cl}$.
We notice that the effect is proportional to the DM current projection onto the magnetic field. The relevant quantities for  He-Xe and K-He co-magnetometers read:
\begin{align}\label{deltaomegaXe}
&\omega_{\rm He}-\frac{\gamma_{\rm He}}{\gamma_{\rm Xe}}\omega_{\rm Xe}=\frac{{ 2}\rho_\chi}{m_\chi}G_N\left( {\mathfrak g}_{\rm He}^{N}-\frac{\gamma_{\rm He}}{\gamma_{\rm Xe}}{\mathfrak g}_{\rm Xe}^{N}\right)\vec J_{\chi}^{\rm\,cl}\cdot \vec u_B\,,
\end{align}
and
\begin{align}
&\beta_e-\beta_{\rm He}
={ 2}\left(\frac{ 3 G_N \frak{g}_{{\rm K}}^{ N}  +G_e }{ 4\gamma_e} -\frac{G_N \frak{g}_{{\rm He}}^{ N}}{\gamma_{\rm He}}\right)\frac{\rho_\chi}{m_\chi}(\vec J_\chi^{\rm\,cl} \cdot \vec u_B). \label{eq:b_diff}
\end{align}

The axion and axial vector boson cases at low masses generate  the time dependent Hamiltonian \eqref{eq:Haxion-VB}. In the limit  $ m_{a,A}\ll 2\pi/t$, 
the effect can be taken constant and    is given by the substitution $G_\psi \rho_\chi \vec J_\chi /m_\chi\to C_\psi\sqrt{\rho_\chi}\,\vec v\,/\sqrt{2}f_a$ (axion), $G_\psi\rho_\chi \vec J_\chi /m_\chi\to g_\psi\sqrt{2\rho_\chi}\,\vec \epsilon\,/m_A$ (axial vector boson) in eqs.~(\ref{deltaomegaXe},\ref{eq:b_diff}).   For He-Xe, $7$ measurement runs, each lasting  $\sim 24$~hours, are performed~\cite{Allmendinger:2013eya}. The total time of observation is  $t\sim 10^6\,$s. For the K-He co-magnetometer of~\cite{Brown:2010dt},  each data point is determined in $\sim\,$s, the runs last for several days, and they used data spanning $t \sim143$ days for their constraints.
For larger masses, the DM field oscillates during the experiment's data taking  and the search strategy differs from the conventional searches for a constant background \cite{Graham:2017ivz}. We describe this case in more detail in app.~\ref{sec:Ht}.

\subsubsection{Sensitivity of magnetometers to dark matter interactions with matter}\label{sec:AC_Mag}

The two cases described in the previous sections generate the same effect as a  frame anomalously coupled to spin; an apparent violation of Lorentz symmetry, in our case generated by the DM `medium'.
The different systematics for constraining such frame with daily modulation effects in co-magnetometers
have been considered in  searches for Lorentz violation \cite{Gemmel:2010ft,Allmendinger:2013eya,Brown:2010dt}. Their results are easily 
translated to bounds in DM-SM couplings for a certain  mass range, see also \cite{Graham:2017ivz}.  We make this comparison more 
explicit in app.~\ref{app:LV}.

The results in~\cite{Allmendinger:2013eya} imply $\Delta\omega/2\pi={ 4} \times 10^{-10}$Hz for the frequency difference in eq.~\eqref{deltaomegaXe}. This
 translates into a bound on the coupling of DM. One gets:
\begin{align}
G_\psi \bar J_\chi< 0.35 \frac{1}{\Upsilon^m_\psi}\left(\frac{\Delta \omega}{2\pi \times 0.4\,{\rm nHz}}\right)\left(\frac{0.3\,{\rm GeV/cm}^3}{\rho_\chi}\right)\left(\frac{m_\chi}{\rm eV}\right){\rm GeV}^{-2},\label{eq:GgMag}
\end{align}
where the constant $\Upsilon^m_\psi$ is an order one factor (similar to $\Upsilon_\psi$ for atomic clocks in \eqref{eq:Gg}); e.g. for He-Xe, sensitive to nucleon couplings $\psi=N$, one has $({\mathfrak g}_{\rm He}^{N}-\mathfrak g_{\rm Xe}^{N}\gamma_{\rm He}/\gamma_{\rm Xe})$. Let us emphasize that this estimate is valid for both the particle and field regimes.
When compared with the bound from eq.~\eqref{eq:Gg}, we notice that the magnetometers are orders of magnitude more sensitive than atomic clocks when we study the same averaged current $\bar J_\chi$. 
On the other hand for axions and axial bosons: 
\begin{align}
&\frac{{\frak f}_{ a}}{C_\psi} \gtrsim 3\times 10^{9}\Upsilon_\psi^m\left(\frac{\Delta \omega^{\rm exp}_{\rm max}}{{\rm 2\pi \times 0.4\,  nHz}}\right)\sqrt{\frac{0.3\, \mathrm{GeV/cm^3}}{\rho_{DM}}}\left(\frac{v}{10^{-3}}\right)\, {\rm GeV},\label{eq:ax_maga}\\
&g^A_\psi <  {2}\times 10^{-42}\frac{1}{\Upsilon_\psi^m}\left(\frac{\Delta \omega^{\rm exp}_{\rm max}}{{\rm 2\pi\times 0.4\, nHz}}\right)\sqrt{\frac{0.3\, \mathrm{GeV/cm^3}}{\rho_{DM}}}\left(\frac{m_A}{\mathrm{10^{-20}eV}}\right), \label{eq:ax_mag}
\end{align}
to be compared with \eqref{eq:ax_a} and \eqref{eq:ax_b}.

 The K-He co-magnetometer of \cite{Brown:2010dt} has a sensitivity a factor 4 less precise (cf. app.~\ref{app:LV}) but is also sensitive to the couplings to the electron spin.  As 
shown  explicitly in \eqref{eq:b_diff}, the ratio $\gamma_{\rm He}/\gamma_e$ implies that the bounds on the couplings to electrons are 
also suppressed by an ${\cal O}(10^{-3})$ factor. When extending the  bounds  in \cite{Brown:2010dt,Allmendinger:2013eya} for constant  background during the campaign to other situations we will focus on the K-He since its sensitivity to measure anomalous magnetic fields has been established as $2\ {\rm fT/\sqrt{Hz}}$ \cite{Brown:2010dt}, while that of He-Xe at short times is harder to determine.

\section{Comparison with present bounds}\label{sec:comp}

We now compare our estimates with previous bounds, both for electron and nucleon interactions. For nucleons we chose to display the neutron case in the figures of this section; the proton is obtained by the order one rescalings  $\Upsilon^{(m)}_p/\Upsilon_n^{(m)}$, with $\Upsilon^{(m)}$ as in in eqs.~(\ref{eq:Gg}-\ref{eq:ax_b}) ((\ref{eq:GgMag}-\ref{eq:ax_mag}) for magnetometers), given by differences of the nuclear form factors in tab.~\ref{tab:gNNcl}. Since the sensitivity is better at low masses we focus on  sub-MeV DM scenarios.  Finally, as shown in tables~\ref{DMMatEl} and \ref{Jclass}, the velocity dependent interactions require  an asymmetry in the number of particles-antiparticles of DM.

\subsection{Axial fields at small masses}\label{sec:compaxfld}

\begin{figure}[h]
  \centering
    \includegraphics[width=0.48\textwidth]{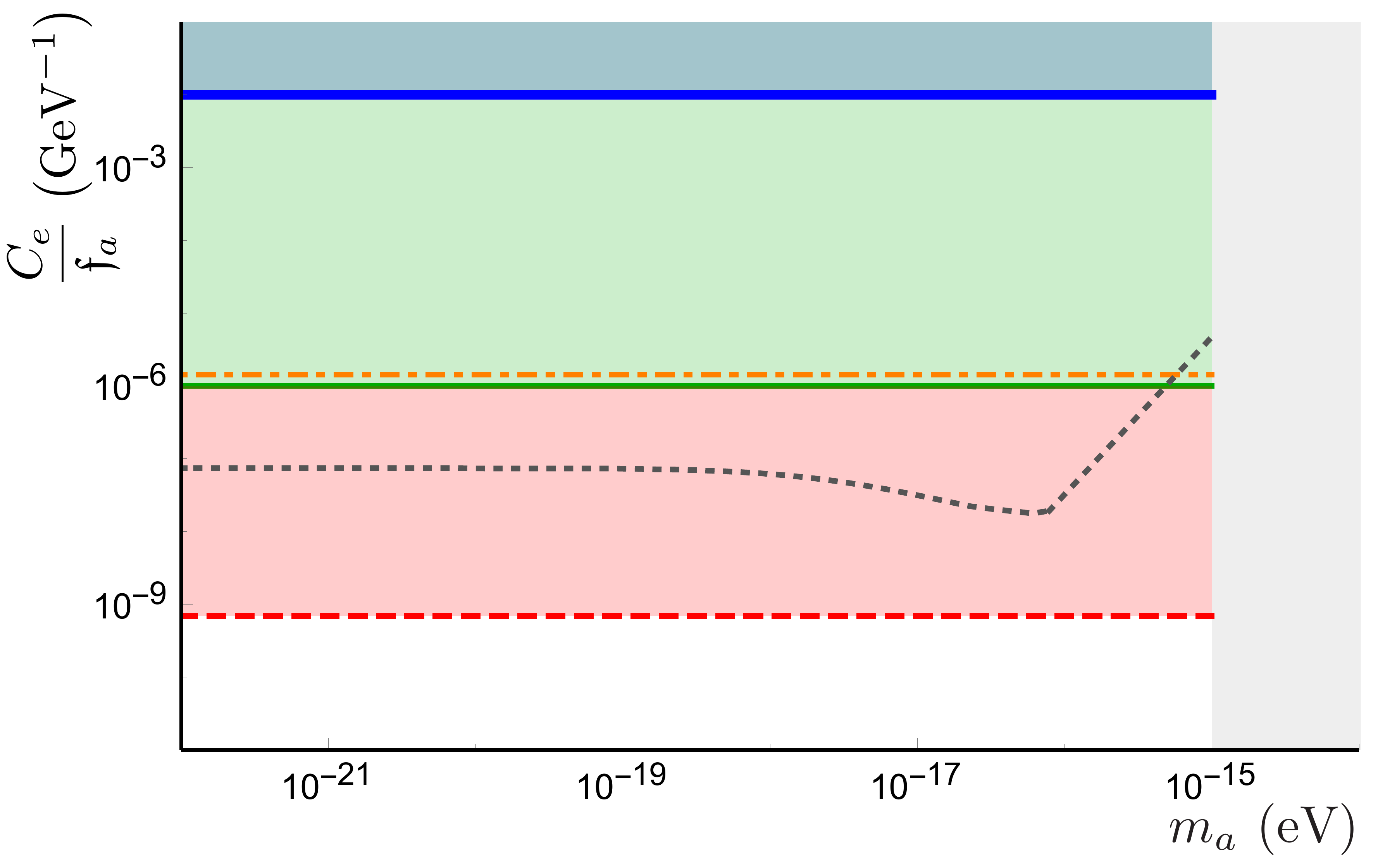}\hfill
  \includegraphics[width=0.48\textwidth]{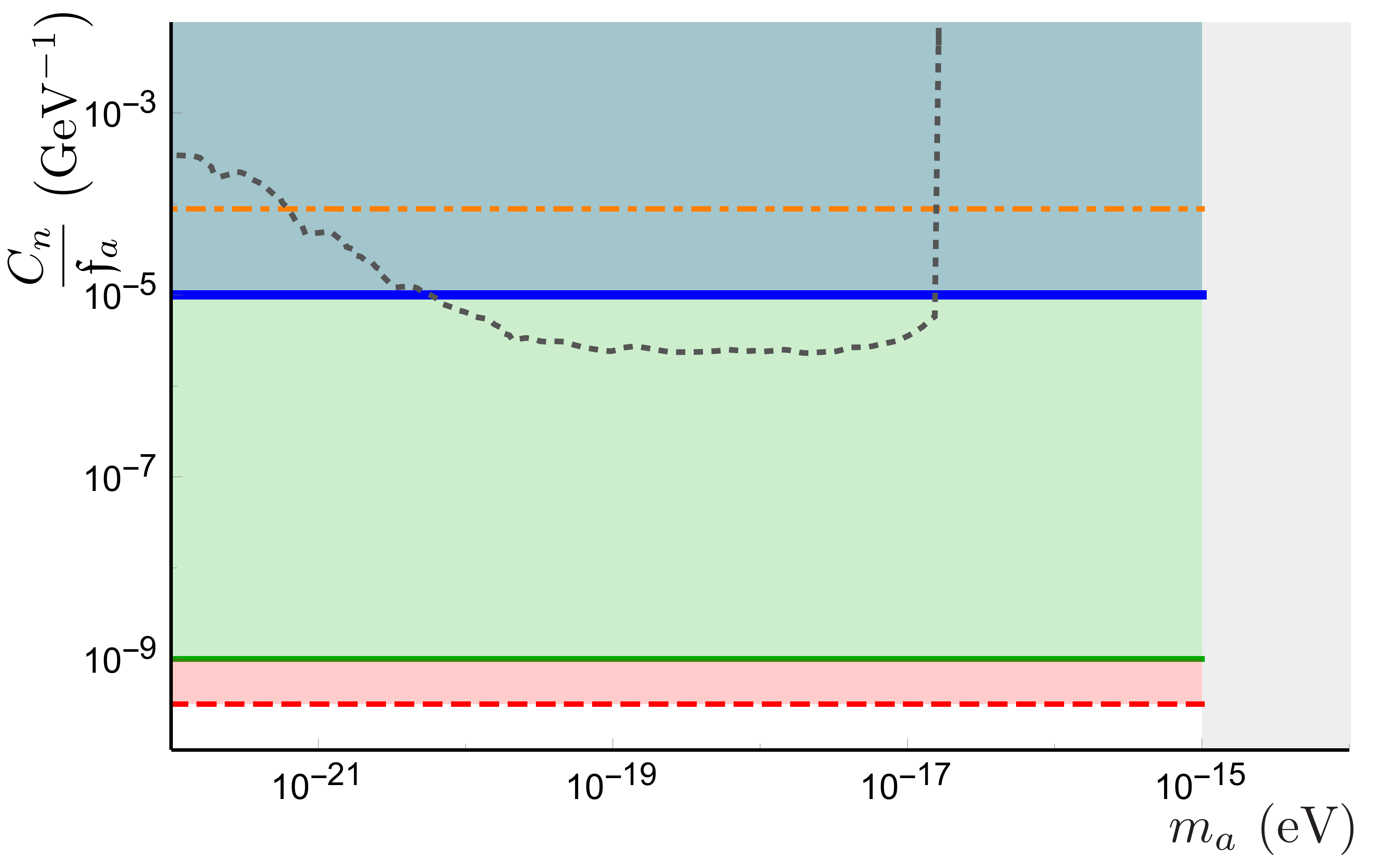}
  \caption{Expected bounds for the coupling of axion DM fields to electrons (left) and nucleons (right). The 
   thick-solid (blue) lines represent atomic clocks, while the thin-solid (green) line are constraints from  magnetometers.  The dotted-dashed (orange) lines represent the bounds from fifth-force experiments  \cite{Terrano:2015sna} (e) and \cite{Vasilakis:2008yn} (n). 
 The red dashed lines represent the bounds coming from energy loss in stars \cite{Raffelt:2012sp}. Finally, the dotted (black) lines are expected \cite{Graham:2017ivz} (e) and current \cite{Abel:2017rtm} (n) laboratory bounds}.
\label{fig:axion_b}
\end{figure}

\begin{figure}[h]
  \centering
      \includegraphics[width=0.48\textwidth]{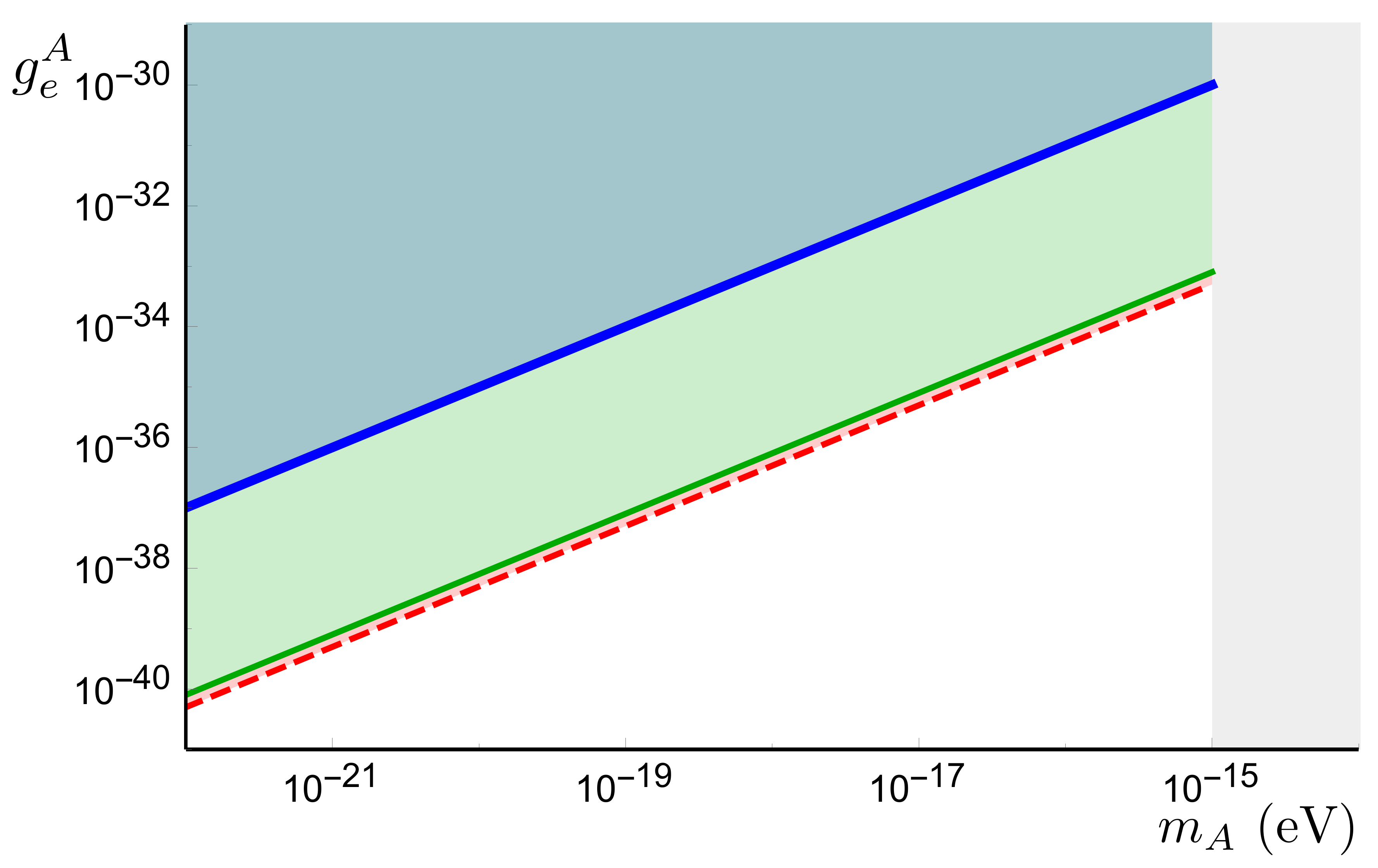}\hfill
  \includegraphics[width=0.49\textwidth]{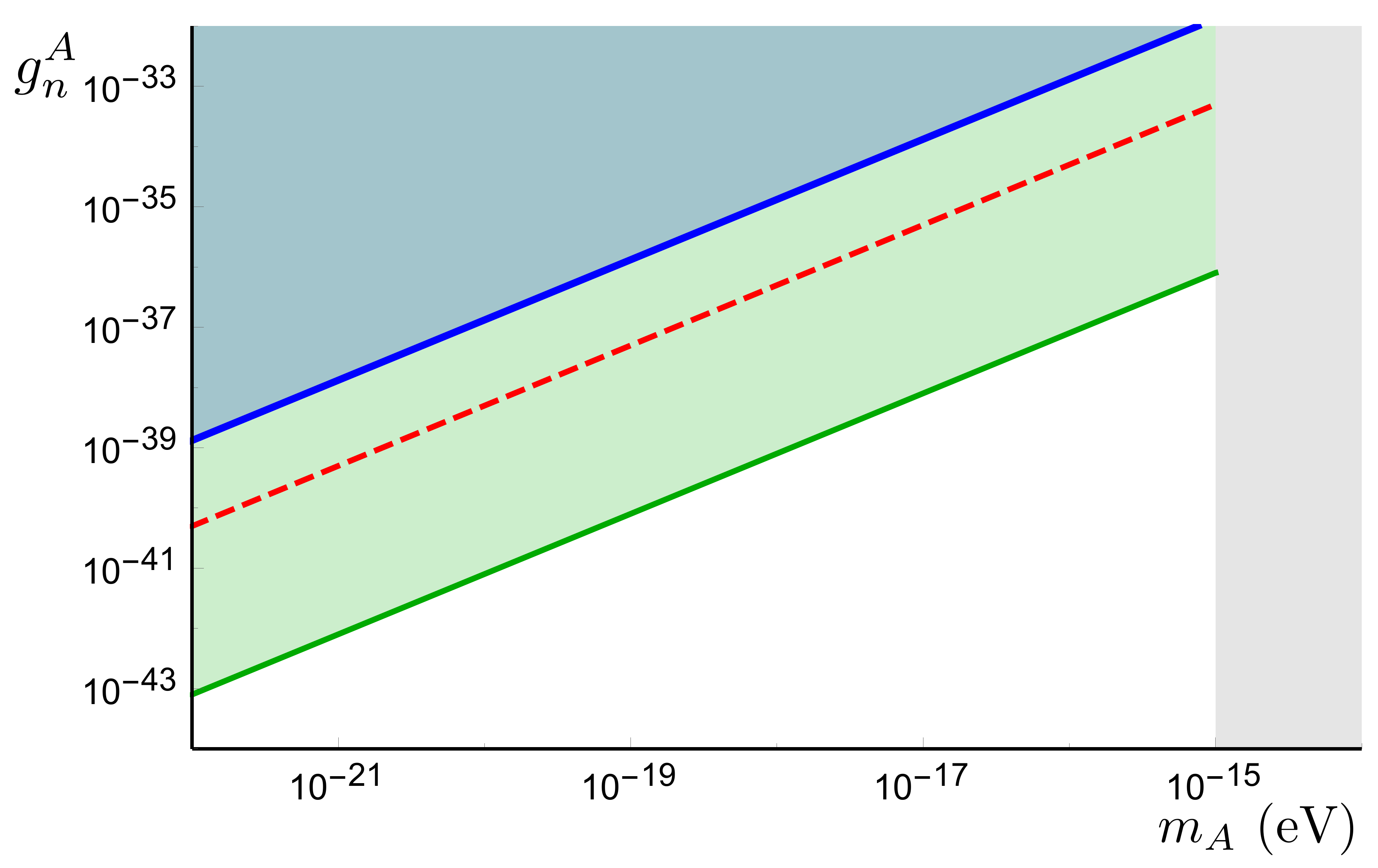}
  \caption{Expected bounds on the coupling of axial field DM to electrons (LHS) and neutrons (RHS) from atomic clocks, thick-solid (blue) line, and magnetometers, thin-solid (green) line. The red dashed line represents the bound coming from stellar energy loss \cite{Raffelt:2012sp}.  }\label{fig:axiV_b}
\end{figure}

Let us start with the bounds of expressions \eqref{eq:ax_a}, \eqref{eq:ax_b}, \eqref{eq:ax_maga}   and  \eqref{eq:ax_mag}  for axion and axial bosons at very low masses. 
A recent summary of constraints on these models can be found in  \cite{Safronova:2017xyt}.
Our results are summarized in figs.~\ref{fig:axion_b} and~\ref{fig:axiV_b}.  We focus on the cases coherent for one day (to detect the modulation) and that do not oscillate during the time of {\it each} measurement. For the atomic clocks $T\sim 0.5\,$s, the same order of magnitude as for K-He~\cite{Brown:2010dt}. We thus restrict to $m_{a,A} < 10^{-15}\,$eV. Coherence during $10^6$ seconds starts to deteriorate above $\sim10^{-16}$ eV \cite{Graham:2017ivz}.  However, the sensitivity decreases mildly in the range of masses here considered, and we neglect this effect. 
Finally, the results in~\cite{Allmendinger:2013eya,Brown:2010dt} assume that the effect remains constant for several hours, $m_{a,A} < 10^{-20}\,$eV. In the case of  \cite{Brown:2010dt}, the measurement takes place at scales of seconds and hence the bounds remain essentially the same for masses up to $10^{-15}\,$eV. This is not so for \cite{Allmendinger:2013eya} (see app.~\ref{sec:Ht} and ref.~\cite{Graham:2017ivz}). Since the bounds of the two experiments differ by a factor 4, we focus on the K-He case and the larger span of masses.

For the axion case, the limits here presented for {\it current} sensitivity are weaker than other astrophysical bounds, but  are in some cases 
stronger than other laboratory experiments. This is true for both  the couplings to electrons and neutrons. In the latter an order of magnitude improvement of sensitivity in magnetometers would change the picture.  

The expressions \eqref{eq:ax_b}  and \eqref{eq:ax_mag} bound the coupling to an  axial vector boson.  In this case
we find they exceed all other bounds for nucleons, whereas for electrons they are at the same level as the strongest constraints as shown in fig.~\ref{fig:axiV_b}. In the plots we only compare with the bounds from star cooling by emission of the axial boson $A_\mu$, since the rest of constraints are much weaker. This process
is dominated by the longitudinal mode at low masses given that  its coupling is enhanced as $(E/m_A)$. This means that one can  translate the bound on the axion decay constant $f_a/C_\psi\geq 10^{9}\,$GeV by using $2f_a/C_\psi\sim~m_A/g_\psi^A$.

\subsection{Models with an axial  spin-1 mediator}\label{sec:part_bounds}

The models including a mediator are subject to extra constraints from the DM-DM and SM-SM interactions. 
Here we focus on the cases with an axial vector boson $\Z_\mu$ mediator, eq.~\eqref{ModelAVB}. We will explore the scalar and fermion models with $\mathcal J_\chi^\mu=i\chi^\dagger \overleftrightarrow\partial^\mu \chi$ and $\mathcal J_\chi^\mu=\bar\chi \gamma^\mu\gamma_5\chi$ respectively. We  drop the superindex $\Z$ in $g_\psi^\Z$ to avoid cluttered expressions in the rest of this section. These two models serve as representative of the spin-velocity and spin-spin interaction of ordinary and dark matter, cf. table~\ref{DMMatEl}.  In the limit of heavy mediator $m_\Z\gg m_\chi$ one recovers an EFT with operators given  in the first two entries of table~\ref{DMOps} for the scalar and fermionic case respectively.
 For the DM spin we will only consider the case of unpolarized DM. The bounds assuming a fractional polarization are easily retrieved from our previous formulae  and the expressions in this section.  
 
The DM self-interaction is controlled by ($g_\chi, m_\Z, m_\chi$). The viscosity cross section of DM-DM processes, with the forward divergences removed~\cite{Knapen:2017xzo} and in the non-relativistic Born approximation, reads in the two models :
\begin{align}\label{SIDMScalar}
\sigma_{\chi\chi}^V&\equiv \int \di\Omega \sin^2\theta\frac{ \di\sigma}{ \di\Omega}=\frac{(g_\chi)^{ 4}}{\pi m_\chi^2v^4}\left(\frac{y^4+2y^2+2}{y^2(2+y^2)}\log\left(1+y^2\right)-1\right),&&{\rm scalar}\ \chi,\\\label{SIDMFermion}
\sigma_{\chi\chi}^V&= \frac{2(g_\chi)^{4}}{\pi m_\chi^2v^4}\left(\frac{y^4+5y^2+5}{y^2(2+y^2)}\log\left(1+y^2\right)-\frac{60-y^4}{24}\right),&&{\rm fermion}\ \chi,
\end{align}
where $y\equiv m_\chi v/m_\Z$.   We note that for the fermionic case the longitudinal mode in $\Z_\mu$ necessarily contributes (chiral symmetry is broken by a massive fermion $\partial_\mu(\bar\chi\gamma^\mu\gamma^5\chi)\neq 0$) but not in the scalar case\footnote{The breaking of the symmetry in the SM sector will, at some loop order, leak into the DM and break the symmetry there, but since this is model dependent  we do not pursue this possibility.}.
If $\chi$ constitutes all the DM,  the previous cross-section is bounded by  \cite{Tulin:2017ara}:
\be
\sigma_{\chi\chi}/m_\chi \lesssim 1\, \mathrm{cm}^2/{\rm gr}\,. \label{eq:SIDM}
\ee
In the limits of heavy or light mediator of the {\it scalar case}, this translates into the bounds
\be
(g_{\chi})^2< \left\{ \begin{array}{c} 3\times 10^{ -18}\left(\frac{m_\chi}{\rm eV}\right)^{3/2}\frac{1}{\sqrt{\log(y)}},\quad \ m_\Z\ll  m_\chi v, \\  0.2\sqrt{\frac{\rm keV}{m_\chi}}\left(\frac{m_\Z}{\rm MeV}\right)^2,\qquad\,\,\,\,\, \ m_\Z\gg m_\chi v,\end{array}\right.\label{eq:gDM}
\ee
whereas for the fermionic case a bound like the one for the $m_\Z\gg m_\chi v$ case above  applies in both $y\ll 1$ and $y\gg1$ regimes due to the contribution from the longitudinal $\Z_\mu$ mode.
This bound disappears quickly if the field $\chi$ does not make up for {\it all} the DM. Following ~\cite{Knapen:2017xzo,Pollack:2014rja}, we consider this constraint as  irrelevant for $\rho_{\chi}\lesssim 0.05\, \rho_{\rm DM}$.  
 When no other bound exists on $g_{\chi}$, we { will saturate it by $g_\chi=1$, which is still a safe choice regarding the applicability of our perturbative calculation.} 
Other considerations we made so far should be also revisited if $\rho_\chi < \rho_{\rm DM}$. A {\it modified} Tremaine-Gunn constrain may also apply for fermionic DM. For instance,  assuming that the component $\chi$ is virialized in the DM halos one can estimate  $m_\chi \gtrsim 100\, ( \rho_\chi/ \rho_{\rm DM})^{1/4}$eV. 
Also the scale at which the occupation number is bigger than unity is modified as shown in \eqref{eq:occup}. 
Finally, the number of particles interacting with the clock is also different, recall \eqref{eq:av}.  This affects the value of the mass for which the averaging of the spin may be relevant or where the number of interactions drop below one. These considerations are taken into account in the following when we consider cases with $\rho_\chi<\rho_{\rm DM}$.

The bounds on the couplings of the mediator to SM fields rely on many observations, with the leading constraint varying with $m_\Z$. For mediators with sub-GeV masses, there are very strong constraints from astrophysics, in particular star and SN cooling; for masses above the GeV, the dominant bounds come from accelerator experiments where SM particles decay to (or collide to produce) the mediator, setting $g_{\psi}^2/m_\Z^2\lesssim G_F$ \cite{Dror:2017ehi,Dror:2017nsg}.

 With both sets of constraints, one can derive bounds on the DM-SM couplings ($g_\chi, g_\psi$) as a function of the masses $m_\chi$ and $m_\Z$. In contrast, in the EFT limit of the model the couplings and masses appear in the combination $g_{\chi,\psi}/m_\Z$ and one can constrain the effective DM-SM coupling $g_\chi g_\psi/m_\Z^2$ as a function of $m_\chi$ alone.

\subsubsection*{DM-electron interactions}

\begin{figure}
\begin{subfigure}{.5\textwidth}
\centering
   \includegraphics[width=\textwidth]{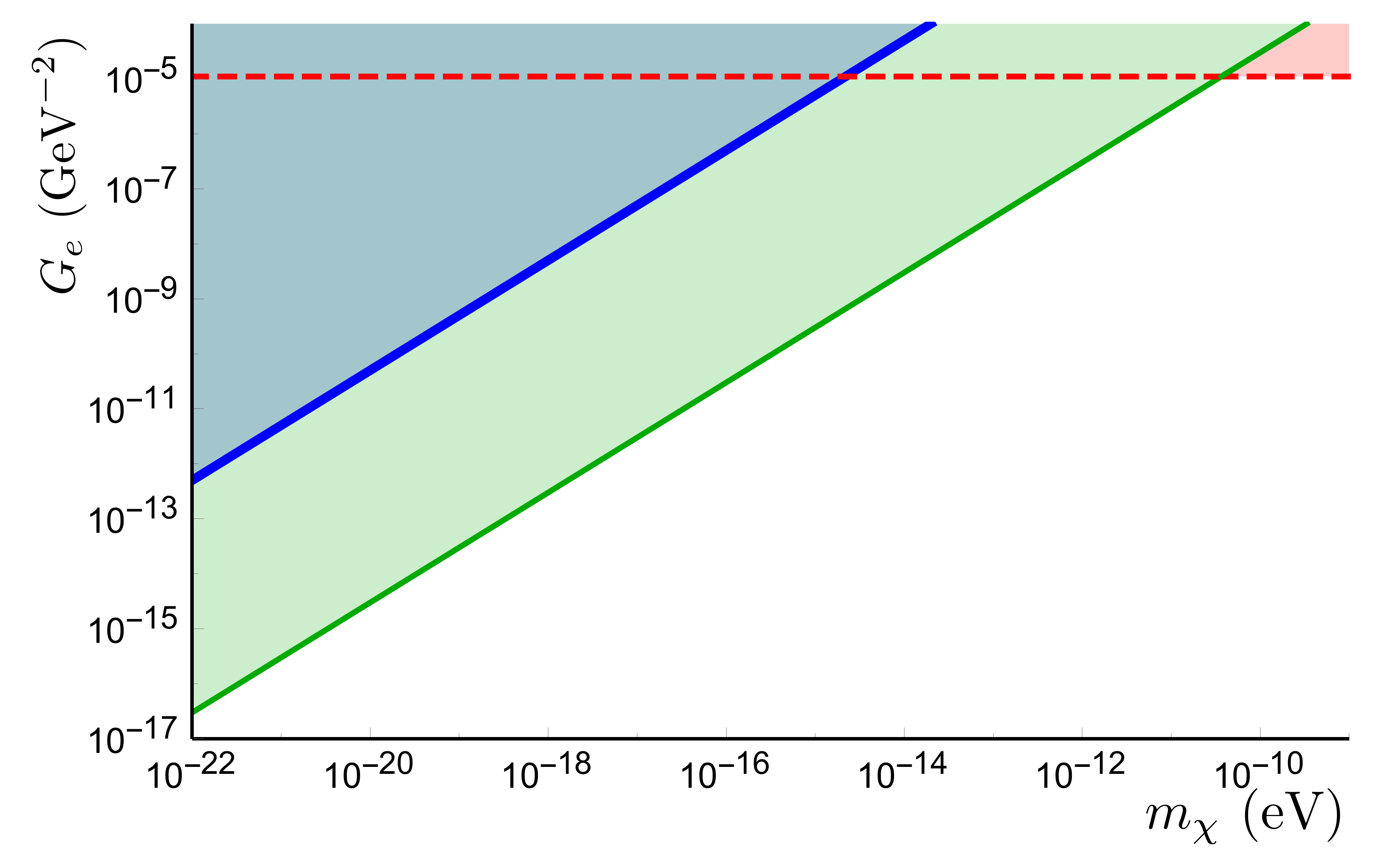} 
\end{subfigure}
\begin{subfigure}{.5\textwidth}
\centering
   \includegraphics[width=\textwidth]{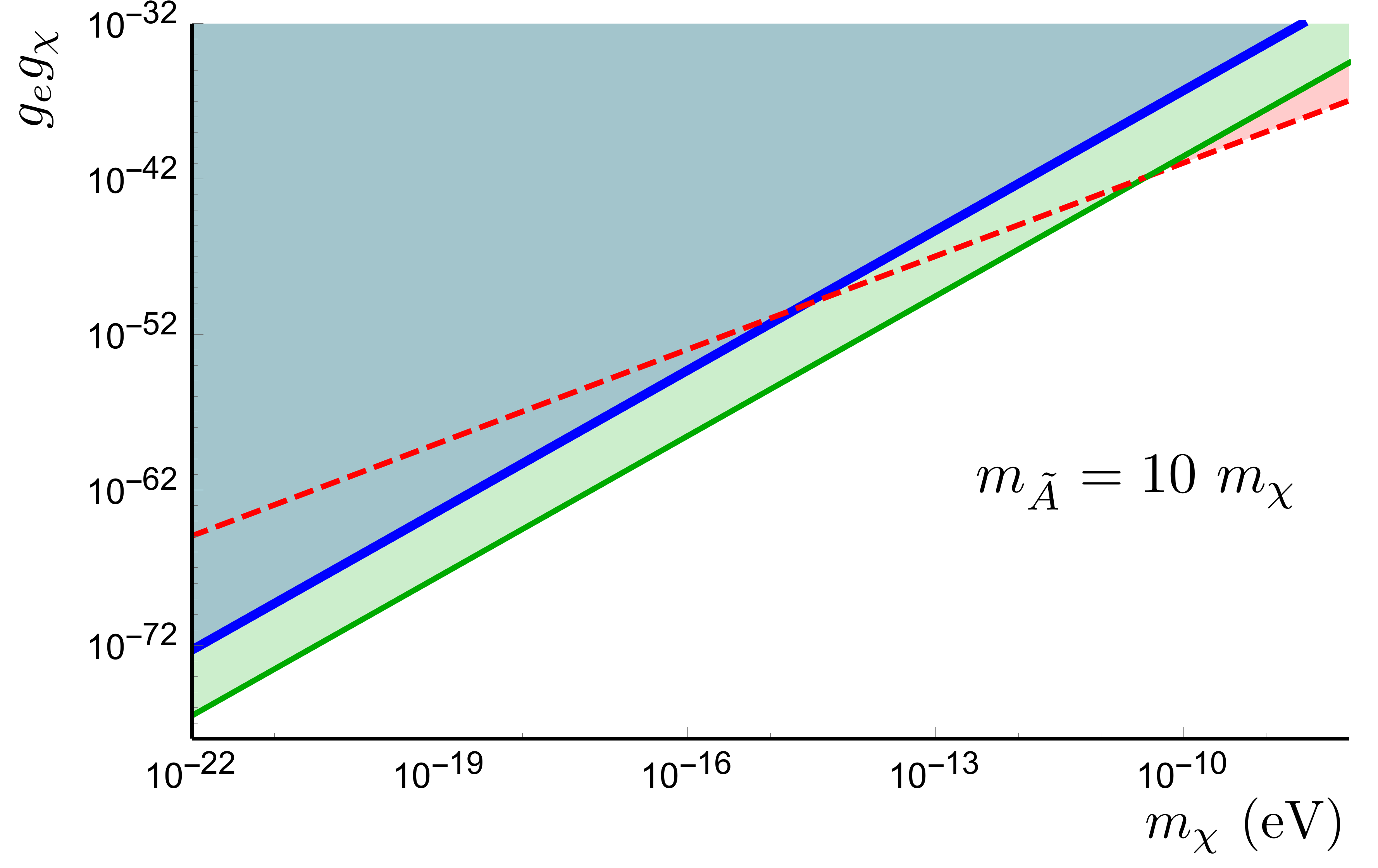} 
\end{subfigure}
   \caption {Left:  Expected constraints on DM-electron coupling $G_{e}\equiv  g_e g_\chi/m_\Z^2$ 
 for the scalar DM case with a heavy mediator ($m_\Z\geq 0.1$~GeV) from atomic clocks -- thick-solid (blue) line--, magnetometers -- thin-solid (green) line--, and DM pair emission in stars --dashed (red) line. Right: Expected 
bounds on the product of DM and electron couplings of the mediator $\Z$ with mass $m_\Z=10\,m_\chi$; same line color coding, with the star cooling bound coming from $\Z$ emission.
 } \label{Fig:elec_1}
   \end{figure}

The cross-section of DM with electrons at small DM masses is constrained from many different sources, see e.g.  \cite{Davidson:2000hf,Raffelt:1996wa,Raffelt:1999tx,Essig:2015cda,Alexander:2016aln,Knapen:2017xzo,Safronova:2017xyt}. We have already remarked that for $m_\Z\leq0.1\,$ GeV star and SN cooling set the strongest constraints on $g_e$. In the opposite regime, production at LEP though  $e^+e^- \to \gamma \Z_\mu$ sets $g_e<10^{-4}$; however, for this range of $m_\Z$ the bound on  \eqref{eq:gDM} is very weak. A tighter constrained arises from extending the star cooling constraints by considering the emission of $\chi\chi^\dagger$ pairs. In particular cooling through $e\gamma\to e\chi\chi^\dagger$ gives an energy loss rate per unit mass:
\begin{align}
\epsilon_{\chi\chi}=\frac{240 \alpha}{\pi^4} Y_e\frac{G_e^2 T^8_*}{m_Nm_e^2}\simeq {\rm erg\, g}^{-1} {\rm s}^{-1}\left( \frac{T_*}{10^8{\rm K}}\right)^8\frac{G_e^2}{G_F^2},
\end{align}
with $G_e\equiv g_e g_\chi/m_\Z^2$, $T_*$ the star temperature and $Y_e$ the ratio of electrons to nucleons. The previous quantity should satisfy $\epsilon_{\chi\chi}\lesssim 10\,  {\rm erg\, g}^{-1} {\rm s}^{-1}$ \cite{Raffelt:1996wa}, which we conservatively implement as $G_e< G_F$.
When compared with \eqref{eq:Gg} and \eqref{eq:GgMag}, and recalling the suppression of ${\cal O}(10^{-3})$ of the electronic case,  these expressions set  competitive bounds  for heavy mediators. We show the cases $m_\Z=0.1\,$GeV and $m_\Z=10\,m_\chi$, where these bounds are relevant till  $m_\chi\leq 10^{-10}$eV, as
fig.~\ref{Fig:elec_1} displays. These small masses are only compatible with bosonic DM, at least for substantial DM fractions. 

\subsubsection*{DM-nucleon interactions}

\begin{figure}[h]
\begin{subfigure}{.5\textwidth}
\centering
   \includegraphics[width=\textwidth]{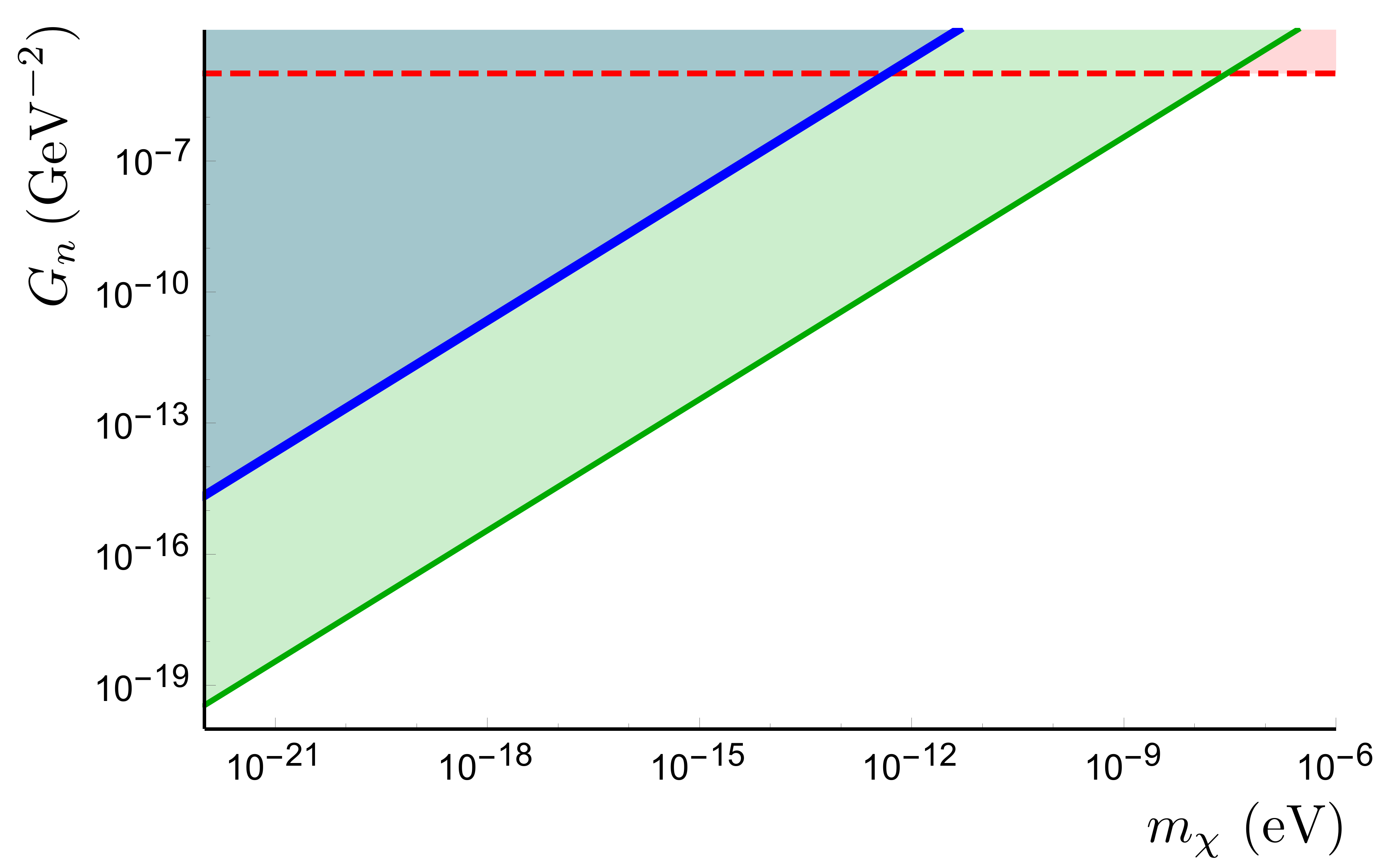} 
\end{subfigure}
\begin{subfigure}{.5\textwidth}
\centering
   \includegraphics[width=\textwidth]{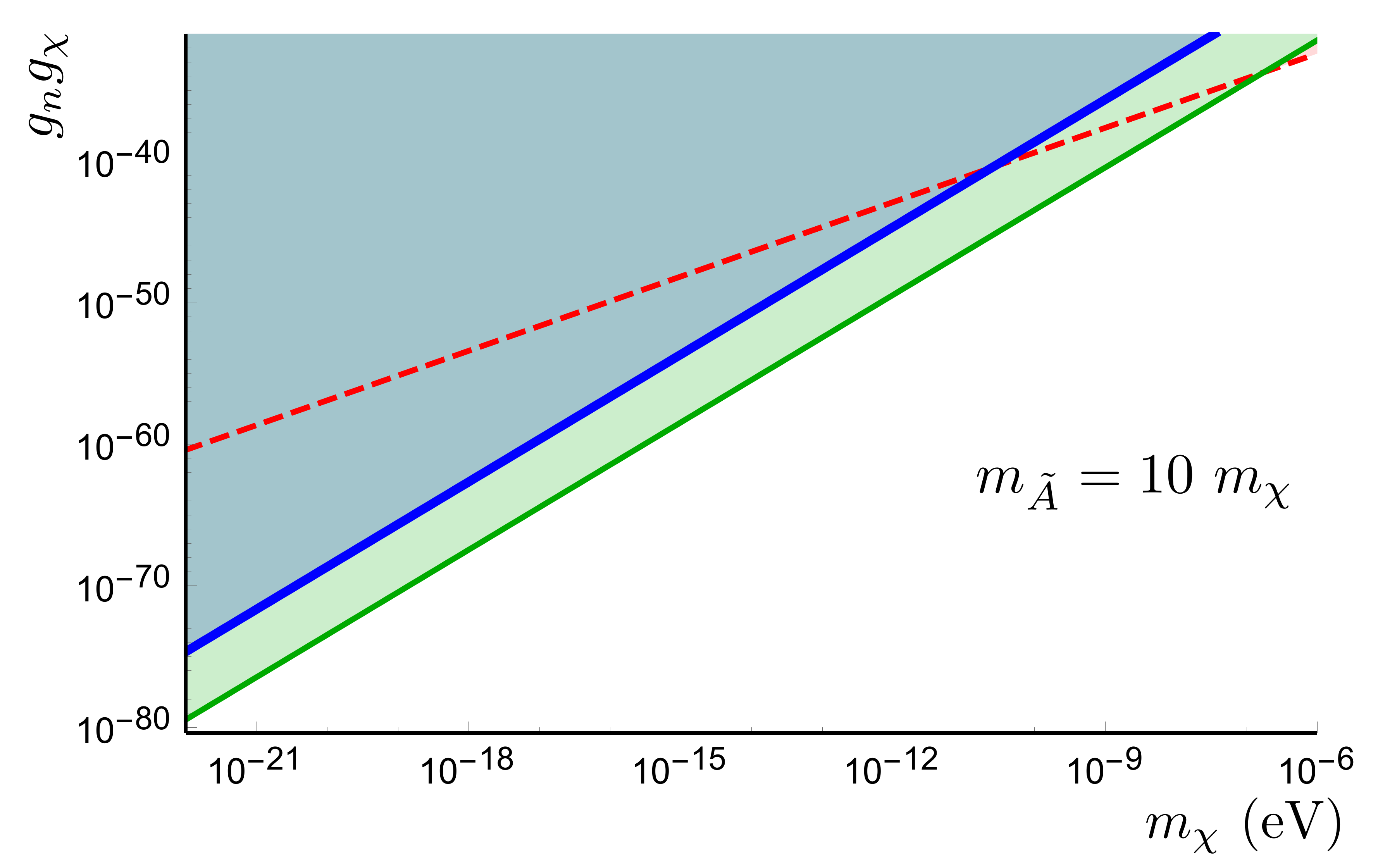} 
\end{subfigure}
   \caption {
  Left:  Expected constraints on DM-neutron coupling $G_{n}\equiv  g_n g_\chi/m_\Z^2$ 
 for the scalar DM case with a heavy mediator ($m_\Z\geq 0.1$~GeV) from atomic clocks --thick-solid (blue) line --, magnetometers -- thin-solid (green) line--, and DM pair emission in stars -- dashed (red) line. Right: Expected
bounds on the product of DM and neutron couplings of the mediator $\Z$ with mass $m_\Z=10\,m_\chi$; same line color coding, with  star cooling bound coming from $\Z$ emission.}\label{Fig:nn_cl1}
   \end{figure}

The nucleon case has the same type of constraints as electrons for $m_\Z\leq 0.1\,$GeV, whereas above this value constraints from flavour violating decays of Kaon and B mesons \cite{Dror:2017nsg} into the mediator + SM take over. For masses $m_\Z>\,$GeV the constraint \eqref{eq:gDM} is not relevant, and $G_N\equiv g_N g_\chi/m_\Z^2$ is better bound  via invisible decays mediated by $\Z_\mu$. Given that we only assume couplings to $u$ and $d$, the pion invisible decay gives the strongest constraint:
\begin{align}
\Gamma_{\pi\to\chi\chi}=\frac{f_\pi^2 m_\pi^3  (g_u-g_d)^2g_\chi^2}{\pi (m_\Z^2)^2}\leq 10^{-15}\rm\, GeV.
\end{align}
In the case of heavy mediator ($m_\Z>m_\chi v$ ), the sensitivity of magnetometers and atomic clocks to $G_N$ is competitive for $m_\chi\leq 10^{-5}$ eV as shown in fig.~\ref{Fig:nn_cl1}.

\begin{figure}[h]
\begin{subfigure}{.5\textwidth}
\centering
   \includegraphics[width=\textwidth]{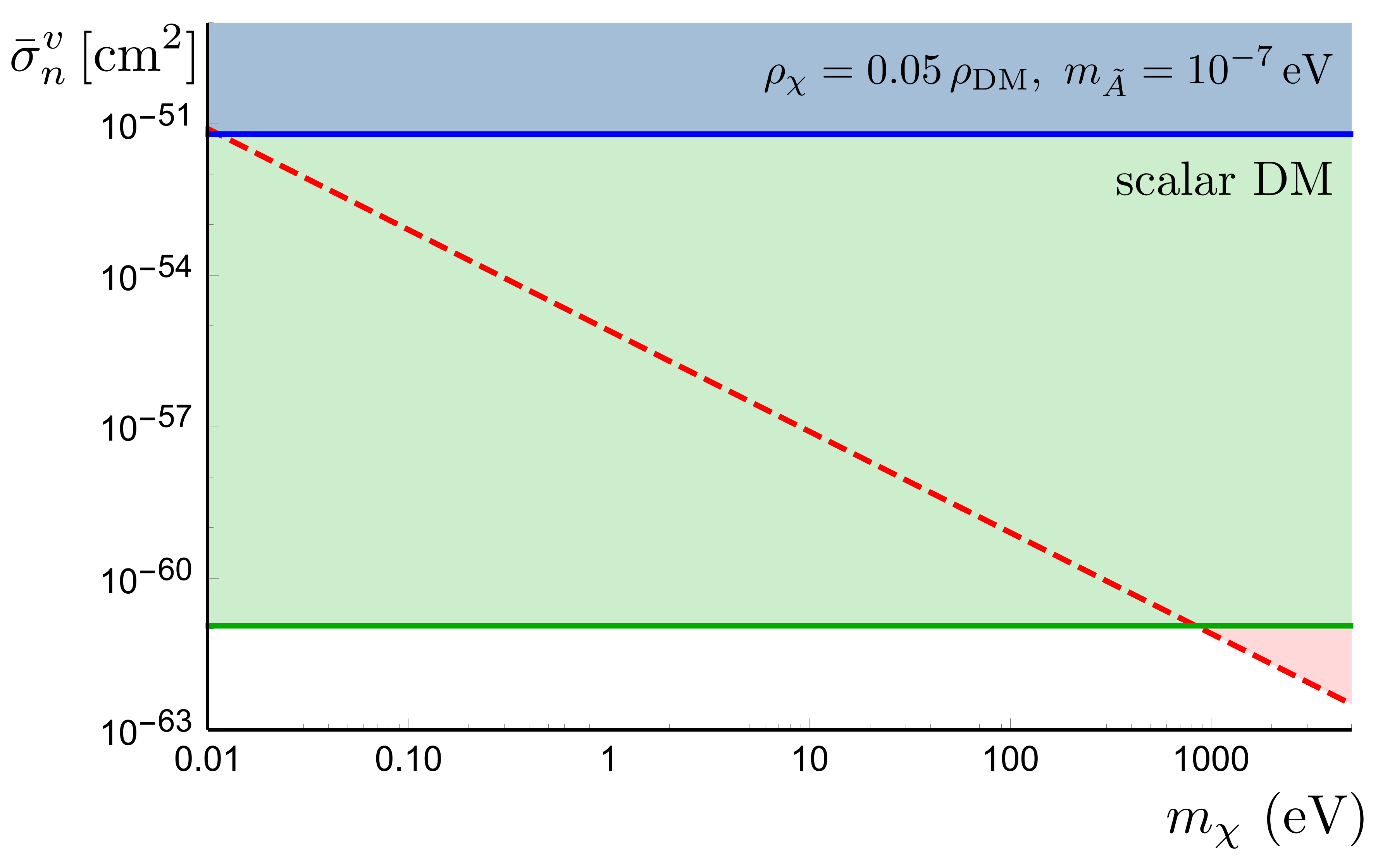} 
\end{subfigure}
\begin{subfigure}{.5\textwidth}
\centering
   \includegraphics[width=\textwidth]{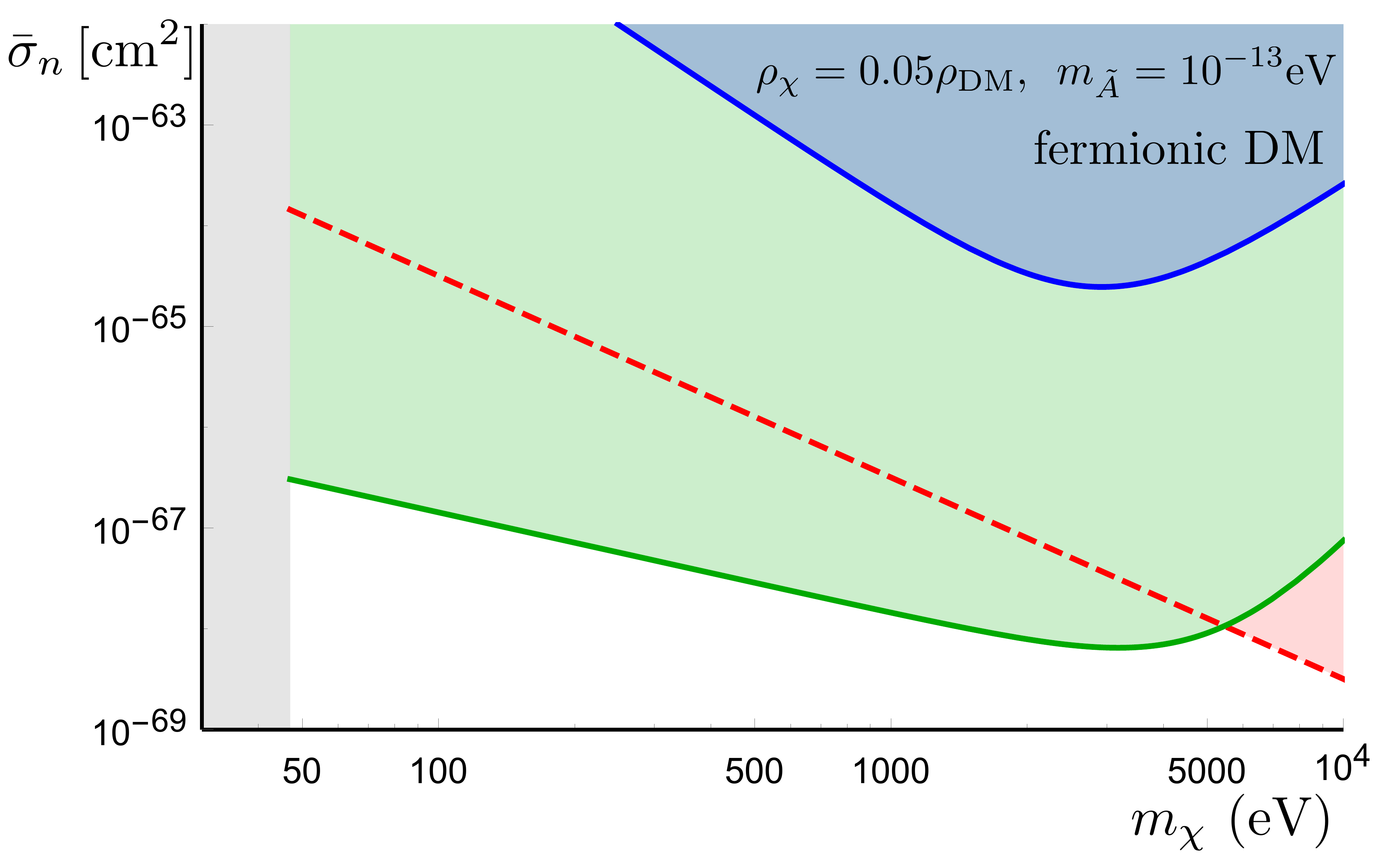} 
\end{subfigure}
\caption{Expected constraints on DM-neutron cross-sections for DM masses corresponding to the particle regime. Both panels show cases $\rho_\chi=0.05\,\rho_{\rm DM}$ and  $g_\chi=1$. Left panel: scalar case with mediator mass $m_\Z=10^{-7}\,$eV. Same colour code as in fig.~\ref{Fig:nn_cl1}. Right panel: fermionic DM case with $m_\Z=10^{-13}$eV. The gray area is the Tremaine-Gunn bound for this $\rho_\chi$.\label{FigPartComp}}
\end{figure}

Our results also imply relevant constraints for higher DM masses in the case of a light mediator, $m_\Z\ll m_\chi v$.  The 
comparative improvement is due to the propagator of the mediator, $1/({\bf q}^2+m_\Z^2)$, being enhanced in the forward limit (${\bf q}\to0$)
(that co-magnetometers and atomic clocks are sensitive to) with respect to the case of momentum transfer which typically has ${\bf q}\sim m_\chi v$.
 Remarkably this is true for both velocity and spin dependent couplings. If one further assumes $\rho_\chi< \rho_{\rm DM}$ so that the bound on $g_\chi$ is relaxed, higher DM masses can be reached with a smaller hierarchy in $m_\Z/m_\chi$. For instance, in fig.~\ref{FigPartComp} we show the velocity-dependent\footnote{ These bounds are derived assuming an asymmetry in  particle-antiparticle for DM which results in a net result proportional to the average velocity. If this asymmetry is absent, one can apply similar ideas as those for the spin-dependent (non-coherent) situation described below.}  case with
 $\rho_\chi=0.05\,\rho_{\rm DM}$ and $m_\Z\sim 10^{-7}\,$eV compared to the strongest constraint, again SN/star cooling via production of the longitudinal mode of $\Z_\mu$.  Recall from the paragraph above \eqref{eq:mmed} that for these light mediator masses the atom `senses' DM within a radius $1/m_\Z$ and the average is over the velocity of $n_\chi/m_\Z^3$ DM particles. To compare with other bounds for light mediators we plotted 
 \begin{align}
\bar \sigma_N \equiv \frac{(g_N g_\chi)^2}{4\pi}\frac{\mu_{\chi-N}^2}{\left(m_\Z^2+v^2m_\chi^2\right)^2}, \quad\quad\quad \bar\sigma_N^v\equiv \bar \sigma_N v^2,
\end{align}
for the spin-dependent case and  velocity-dependent case respectively.

The case of fermionic DM (with spin dependent coupling) can also be constrained from our methods in the limit of light mediator.  If there is a net polarization of the DM particles one can recycle the velocity-dependent results via the substitution $v\to \langle \lambda_\chi \rangle$ and hence rescale the bounds on cross section by a factor  $(\langle \lambda_\chi \rangle/v)^2$ (which is to say the plot on the left of fig.~\ref{FigPartComp} corresponds to a per-mile polarization). The effect of the unpolarized part of the signal is suppressed by the averaging over the number of scatterings and atoms as described in secs.~\ref{sec:partDMAC} and \ref{sec:partDMMG}. In particular, the effect is suppressed by $N_{sc}$ on both the two possible regimes: if $N_{sc}>1$ there will be a statistical average of $1/\sqrt{N_{sc}}$ while if $N_{sc}<1$ the effect is linearly suppressed with $N_{sc}$, since only this fraction of the atoms in the sample are affected. On the other hand depending on whether every atom sees different or common DM particles to the rest of atoms an extra suppression of $1/\sqrt{N_{at}}$ or $1/(Lm_\chi v)$ is present as is the case for atomic clocks and magnetometers respectively.
 This effect is understood as a noise contribution to each measurement and  is independent of  daily modulation.
 As an example where our results may be interesting, we show the bounds in the right panel of fig.~\ref{FigPartComp} for  the case with $\rho_\chi=0.05\,\rho_{\rm DM}$ and $m_\Z\sim 10^{-13}\,$eV.  We remain agnostic about the origin of such a hierarchy of masses in the dark sector. One sees that co-magnetometers still do better than atomic clocks with the peak sensitivity given by $N_{sc}\sim1$ itself dictated by the time of exposure of the measurements.
In the case of light mediator there is an effect from DM particles within $1/m_\Z$ on top of the interactions with those that pass within $1/m_\chi v$. However, for the un-polarized case and parameters of the plot  this effect averages out very efficiently. 

\section{Cosmic neutrinos}\label{sec:neutrinos}

Given their sensitivity to backgrounds of low-mass particles, it is natural to explore to which extent atomic clocks and magnetometers are affected by {\it astrophysical neutrinos}. 
Some early ideas in this direction can be found in  \cite{1998BrJPh..28...72E}. 
The background of neutrinos on Earth has different sources and fluxes depending on the energy scale: at low energies the dominant source is the cosmic neutrino background while at higher energies neutrinos coming from the Sun  dominate \cite{Becker:2007sv}. Neutrinos  interact  {\it via} weak interactions, which include the 4-fermion axial vector interactions we described in table~\ref{DMOps} with $G_\psi\approx G_F$. All our formulae are  valid in the non-relativistic limit, which is satisfied by at least a major component of cosmic neutrinos~\cite{Cocco:2007za}. Their translation into the relativistic case is straightforward. Independently of other considerations that may reduce the total effect (e.g. the polarization of the background, degeneracies, etc.) recall from eqs.~\eqref{eq:P1}, \eqref{eq:ptcl_HeXe} 
that the phase of the system under study is modified with the flux per unit area and time $\flux$ times the scattering amplitude $f_\nu\sim \sqrt{\sigma_\nu}$ as
\begin{align}
\Delta \varphi\simeq\frac{\flux}{p_\nu}\sqrt{\sigma_\nu}T\simeq\flux\, \frac{G_F E_\nu}{p_\nu}T, \label{eq:esti_neu}
\end{align}
where we have used $\sigma_\nu\sim G_F^2 E_\nu^2$. This estimate displays two promising features: $i)$ the effect is enhanced by a factor $1/v$ in the non-relativistic case, $ii)$ the effect is coherent and linear in $G_F$.
  This process is an application of the Stodolsky effect~\cite{Stodolsky:1974aq,Langacker:1982ih}; other effects are proportional to the $G_F^2$ and hence further suppressed. The expression \eqref{eq:esti_neu} shows that the set-ups we described  could probe the most abundant source of neutrinos, regardless of their energy. The flux for cosmic neutrinos is  $10^{12}\,{\rm cm}^{-2}s^{-1}$ \cite{Ringwald:2004np}, while for solar neutrinos  the number is $10^{11}{\rm cm}^{-2}s^{-1}$~\cite{Becker:2007sv}. After considering the typical velocity of these backgrounds, we obtain the phase shifts:
\begin{align}
\Delta\varphi &\sim 10^{-18}\left(\frac{T}{s}\right) {\rm\,\, [cosmic\,\, neutrinos]}; &\Delta\varphi &\sim 10^{-22}\left(\frac{T}{s}\right)  {\rm\,\, [solar\,\, neutrinos]}. &
\end{align}
Co-magnetometers are sensitive to $\Delta\varphi\sim 10^{-9}\, T/s$, while atomic clocks can reach a sensitivity of  $\Delta\varphi\sim 10^{-5}\, T/s$. Thus, independently of other difficulties we discussed, the effects of astrophysical neutrinos are many orders of magnitude away from the accuracy of current devices. 
These numbers are not so pessimistic in the context of other attempts to   detect coherent effects of the cosmic neutrino background, see e.g.~\cite{Langacker:1982ih,Domcke:2017aqj}. 
On a positive note, there is ample space for exploration of DM signals. 

\section{Summary  and future prospects}\label{sec:conclu} 

As current DM searches return null results, the conventional ideas for DM are being revised, triggering extensions of paradigmatic models and new possibilities altogether. The theoretical landscape is naturally broadened towards  candidates with masses well below the GeV. For the heavier end of this spectrum, thermal production in the primordial Universe is still viable whereas lighter DM can arise via freeze-in, a  misalignment mechanism or other out-of-equilibrium processes. In practice, one can populate the whole mass spectrum down to the lightest (bosonic) viable candidate of mass $m_\chi\sim 10^{-21}\,$eV.  Testing these  {well-motivated} DM models  requires experimental techniques beyond the  use of recoil of matter produced by DM collisions. Indeed, the deposited energy for a target of mass $m_T$ scales as $\sim (m_\chi v)^2/m_T$ and hence lighter candidates are soon below experimental thresholds.  DM does however leave an imprint on matter in certain zero momentum transfer processes, which are natural places to look for light candidates. This work considered two concrete examples:  atomic clocks and magnetometers. 

Atomic clocks measure the transition frequency between two hyperfine split states to great accuracy via interference effects. 
DM coupled to the spin of matter affects this interference for polarized samples even if there is no momentum transfer. Atomic magnetometers monitor the precession of atomic angular momenta under a magnetic field. This precession is affected by DM when it couples to spin since it contributes to the Zeeman splitting.  These effects are daily modulated when the DM current is coherent on the scale of days (the angle between the DM and the spin of the sample changes as the Earth rotates). They may also 
generate an irreducible source of noise in the experiment if they are not coherent.

\begin{figure}[h]
\begin{center}
\includegraphics[width=0.8\textwidth]{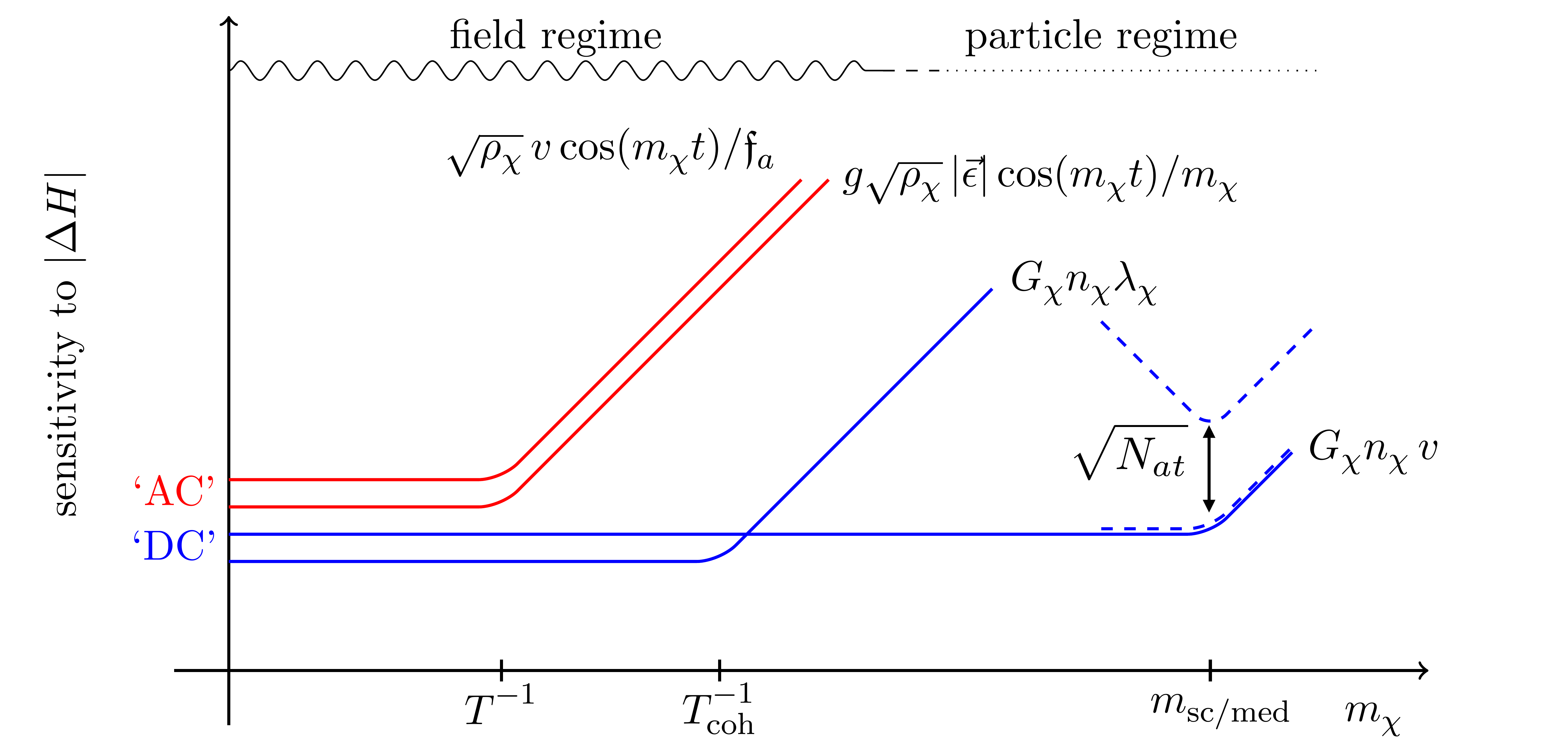}
\caption{Qualitative depiction of the sensitivity to the contribution to the energy difference (frequency) in the different regimes and types of DM coupling. Solid (dashed) lines correspond to bosonic (fermionic) DM candidates, the scaling is merely symbolic and meant to represent a loss of sensitivity at the different thresholds $T_i$ \label{Fig:Concl}, see sec.~\ref{sec:comp} for plots.}
\end{center}
\end{figure}

Our results are qualitatively summarized in fig.~\ref{Fig:Concl}.   In the horizontal axes we show the span in DM mass to which atomic clocks and magnetometers are sensitive.  It includes different regimes. The `particle regime' refers to the masses for which the DM-SM interaction is described as a scattering process. The `field regime' refers to the masses for which a description of SM particles interacting with a DM `field' is appropriate. 
The loose frontier between these two regimes is a few eV and it scales with $\rho_\chi$ if the species $\chi$ does not make up all of the DM, cf. eq.~\eqref{eq:occup}. Even if our calculations were performed deep in these two regimes, we found that the formulas can be extrapolated within regimes straight-forwardly.

To explain the rest of relevant mass thresholds, we introduce the observable  shown in  the vertical axes  of  fig.~\ref{Fig:Concl}.  This  is the  effective energy split  induced by  DM-SM interactions,
\begin{align}
\Delta H= C \vec J_\chi\cdot\vec \lambda,
\end{align}
where $\vec \lambda$ represents the spin of the matter state,  $\vec J_\chi$ is a DM current and $C$ sets the interaction strength and is in general time dependent. Different DM models  generate different possibilities depending on the {\it type of coupling} and {\it state of DM}: 
\begin{itemize} 
\item[\it i)] {\it Type of coupling.}  On   general grounds, DM can couple to $\vec \lambda$ either through their relative velocity or its spin  (in the limit of negligible momentum transfer):
\begin{align}
\vec J_\chi=\left\{\begin{array}{c}
\vec v\,, \qquad {\rm\,\,\, DM\,\, wind}\,,\\
\vec \lambda_\chi\,,\qquad{\rm DM \,\,noise} \,.
\end{array}\right.
\end{align} 
Although the velocity-dependent interaction has a $v\sim 10^{-3}$ suppression, it has a non-zero average which is not the case for a spin-dependent interaction unless there is a net polarization. One can entertain the thought of the two cases as DM wind which we would seek with a `vane' or DM noise which would turn up in the experiment. The question of averaging leads to the abundance of DM  and in turn to the state of DM, the next point.
\item [\it ii)] {\it State of DM}. It  affects the time dependence of the coefficient $C$ and the  average $\langle\vec J_\chi\rangle$:
\begin{itemize} 
\item[\it a)] Field regime: in the ultralight mass region,  DM acts as a coherent field oscillating with frequency $m_\chi$ with a coherence lasting $10^6$ oscillations. If DM couples linearly to matter as in the axial boson cases, $C$ itself oscillates with the same period. In the case of quadratic coupling, the interaction Hamiltonian can have a DC component as we find in the EFT case. The sensitivity therefore depends on the typical time of measure in the experiment $T$; for masses above $T^{-1}$ sensitivity to the AC case worsens whereas for even higher masses, when coherence does not last throughout a measurement, $m_\chi> (Tv^2)^{-1}\equiv T_{\rm coh}^{-1}$, sensitivity to the DC spin coupling case worsens. This dependence of $C$ on DM properties is made explicit in fig.~\ref{Fig:Concl}.
\item[\it b)] Particle regime: in this case, it is useful to think of the atom as traversing the DM medium and experiencing matter effects as neutrinos do when they travel through ordinary matter. The relevant quantity is then the effective potential, proportional to $n_\chi \langle J_\chi\rangle$.   This average is different for the cases of  heavy  or light mediator $\tilde A$, the boundary of these cases being $m_\Z\sim m_\chi v$. In the former case, this is related to the number of particles that go through the atom $N_{sc}$, while in the second case one should consider the particles passing within a radius $1/m_\Z$.
The spin or velocity coupling cases hence differ quite drastically: for the latter there is  a net effect, the DM wind, whereas for the former there is an averaging-out and loss of sensitivity unless there is a net polarization. In the totally unpolarized case, the DM effect can be understood as an irreducible source of noise, $\propto1/\sqrt{N_{sc}}$.  
The upper threshold in both spin and velocity dependent cases is the mass scale when a single atom barely sees DM particles, $N_{sc}\sim1$ or $n_\chi/m_\Z^3\sim 1$ which define $m_{\rm sc}$ and $m_{{\rm med}}$ (see sec.~\ref{sec:partDMAC}) in fig.~\ref{Fig:Concl}.
A final consideration for the evaluation of net current  $\langle J_\chi\rangle$ is that experiments have a number of atoms (typically $10^6$ for atomic clocks and a macroscopic sample for magnetometers) which leads to further averaging effects as discussed in  in secs.~\ref{sec:partDMAC} and \ref{sec:partDMMG}.
 An extra consideration in this case is how many atoms does a single DM particle see as it goes through the sample.
\end{itemize}
\end{itemize}

The effects summarized in the previous points and in fig.~\ref{Fig:Concl} generate competitive bounds to several DM models that we illustrated  in figs.~\ref{fig:axion_b}, \ref{fig:axiV_b}, \ref{Fig:elec_1}, \ref{Fig:nn_cl1} and \ref{FigPartComp}.  In all these plots green (thin solid line) represents the bounds from magnetometers while blue (thick blue line) are those from atomic clocks. Present bounds are shown with a different color scheme. The lesson is clear: both experiments are very sensitive at low masses, and probe regions of parameters previously unconstrained. Magnetometers are more powerful in the determination of total phase shifts, and hence tend to yield better constraints. Still, both technologies are quite independent and it makes sense to explore them simultaneously. 

There are many future developments possible. It would be interesting to perform the experiments with the atomic clocks in the configuration here suggested ($m_F\neq 0$) to provide realistic constraints. On the same footing, it would be interesting to reanalyze the data from co-magnetometers including the DM effects we discussed. It is also worth considering  whether other experiments sensitive to differences in atomic phases can yield bounds on couplings to other SM currents. As an example, the phase shift generated in two populations of different momentum could generate bounds on vector couplings.  Given the sensitivity shown above, one could also consider  to use clocks or magnetometers as detectors in certain particle physics set-ups,  as in BSM models where the particle background need not be DM. Finally, we have been rather naive about cosmological consequences of the models we  considered. It would be desirable to understand  which models are preferred when put in a more complete cosmological context, see e.g. \cite{Green:2017ybv} for a study in this direction.

\section*{Acknowledgements}

We have benefited from discussions with many colleagues.  
We are particularly grateful to Talytha Barbosa, Kfir Blum, Marco Drewes,  Jacopo Ghiglieri, Jacobo L\'opez-Pav\'on, Matthew McCullough, Paolo Panci, Maxim Pospelov, Javier Redondo,  Ben Safdi,  Kai Schmidt-Hoberg, Mikhail Shaposhnikov,   Satoshi Shirai, Sergey Sibiryakov, Sean Tulin, Alfredo Urbano, Tien Tien Yu, Jes\'us Zavala and  Jure Zupan. 
We also acknowledge the use of the web-digitizer  \url{https://automeris.io/WebPlotDigitizer/}.

\appendix

\section{Quantum Field Theory and Scattering conventions}\label{sec:QFT}

Our convention for the metric is $(+,-,-,-)$. We write the fields (bosonic or fermionic)   as 
\begin{align}
&\Phi_a=\int\frac{\di^3p}{2E_{\bf p} (2\pi)^3} \left(\alpha_a a_{\bf p} e^{{ -}ip^\mu x_\mu}+\beta_a b_{\bf p} ^\dagger  e^{{}ip^\mu x_\mu}\right),
\end{align}
where $\alpha,\beta$ carry the index associated to the Lorentz representation of the field, i.e. they are $1$ for scalars, spinors $u,v$ for fermions and polarization vectors $\epsilon_\mu$ for vector bosons. We  split the scattering matrix for momentum eigenstates in the non-relativistic limit as
\begin{align}
S({\bf p}_\chi\,{\bf p}_{\At} \to& {\bf p}_\chi^\prime\,{\bf p}_{\At}^\prime )=\nonumber
\\
 &\left\langle{\bf p}'\,{\bf P'} | {\bf p}\,{\bf P} \right\rangle-i(2\pi)^4\delta({\bf p}^2/2\mu -{\bf p}^{\prime 2}/2\mu)\delta^{(3)}({\bf P}^\prime-{\bf P}) \mathcal T ({\bf p}^\prime,{\bf P}',{\bf p},{\bf P})\,, \label{eq:Smatrix}
\end{align}
where one-particle momentum states are defined as 
\be
|{\bf p}\rangle=\frac{1}{\sqrt{2E_{\bf p}}}a^\dagger_{\bf p} |0\rangle, \label{eqstate}
\ee
 with $a_{\bf p}^\dagger$ a creation operator. Creation and annihilation operators satisfy  
\be
\left[a_{\bf p} \,,a_{\bf k} ^\dagger\right]_{\pm}=2E_{\bf p}  (2\pi)^3\delta^{(3)}(p-k),
\ee 
 where $\pm$ represents commutator or anti-commutator depending on the spin of the particle. 
The normalization of momentum states is 
 \be
 \left\langle {\bf p}|{\bf k}\right\rangle = (2\pi)^3\delta^{(3)}({\bf p}-{\bf k}), \label{eq:norm_p}
 \ee 
 and the measure in momentum space is $\di^3 p/(2\pi)^3$. These conventions differ from custom in particle physics, and in particular $\mathcal T$ is related to the Lorentz invariant matrix element $\mathcal M$ by factors of $\sqrt{2E}_{\bf p}$ for each external particle. We define $f$  as, 
\begin{align}
 f({\bf p}^\prime,{\bf p})={-}\frac{\mu}{2\pi}\mathcal T({\bf p}'\,, {\bf p})\,, 
\end{align}
where we have suppressed the dependence on $\bf P$ of the matrix element $\mathcal T$. Its connection to the scattering process can be read from \eqref{eq:wave}. In the Born approximation,
\be
\left\langle {\bf P}^\prime,{\bf p}^\prime| H_{\rm int}|{\bf P}\,,{\bf p}\right\rangle=(2\pi)^3\delta^{(3)}({\bf P}^\prime-{\bf P})\mathcal T ({\bf p}',{\bf P}', {\bf p},{\bf P})\,, \label{eq:tdefntn}
\ee
where $H_{\rm int}$ is the interacting Hamiltonian built out of the Lagrangian in eqs.~\eqref{eq:LagEFT}.

For the computation of $\mathcal T$ we require the evaluation of currents in the states of \eqref{eqstate} as given in eq.~\eqref{eq:MatElEFT} both for the DM and SM pieces. With the conventions here provided and in the limit $q\to 0$ the evaluation is straightforward and given in tabs.~\ref{DMMatEl} and eq.~\eqref{eq:Jpsie} for DM and a free fermion. The states of interest are atoms. The expressions for the scattering amplitude for each of the two hyperfine split states in them are:
\be
\label{eq:f1f2gen}
\begin{split}
f_1(0)
&=\frac{m_\chi}{\pi}\vec J_\chi\cdot \vec \lambda \left(  \frac{2F+1}{2F} g_{\tiny \mbox{Ncl}}^NG^a_N -\frac{1}{2F}G_e   \right) ,\\
f_2(0)&=
\frac{m_\chi}{\pi}\vec J_\chi \cdot \vec \lambda \left(  \frac{2F-1}{2F} g_{\tiny \mbox{Ncl}}^NG^a_N +\frac{1}{2F}G_e   \right) ,
\end{split}
\ee
 where $\vec\lambda$ is the average spin of the atom (the same for the two states which can be taken to be $(0,0,\lambda)$).

Another standard result of scattering theory is the relation between the scattering cross section and the amplitude of  forward scattering. In particular, the optical theorem relates the imaginary part of the scattering element to the total cross-section  \cite{goldberger2004collision}. 
For our set-up, the observables of interest \eqref{eq:P1}  and \eqref{DeltaS} depend on the {\it real} part of the difference in forward amplitudes\footnote{In the main text, in particular from \eqref{eq:fwiththeta} onwards, we did not include an explicit dependence on $p_\chi$ in this quantity for presentation purposes.} 
\be
\mathrm{Re}[\bar f_1(0,p_\chi)-\bar f_2(0,p_\chi)],
\ee
whose connection to the cross-section can be established as follows. 
Let us first introduce a partial-wave expansion for the amplitudes \cite{goldberger2004collision}, 
\be
f(0,k)=\sum_{l=0}^\infty (2l+1)f_l(k),
\ee
where the subindex $l$ refers to the decomposition of the amplitude in Legendre polynomials. 
From unitarity
\be
f_l(k)=\frac{e^{i\delta_l}\sin \delta_l}{k}, \label{eq:f_l}
\ee
with $\delta_l$ real. Thus, at order $\O(\delta_l)$, $f_l \sim \delta_l/k$,  which is real.

\section{Contribution of the electronic wave-function}\label{ap:ele}

To evaluate the contribution of the unpaired electron to the  matrix element in  an alkali atom, one needs to consider  the corresponding wave-function. For instance,  taking the $5 s$ wave-function of ${}^{87}$Rb
and approximating the atom to be at rest before and after the scattering:
\begin{align}
\mathcal T_e &=\int \frac{{\rm d}^3{\bf k}}{(2\pi^3)}\frac{{\rm d}^3{\bf k}'}{(2\pi)^3}\hat \psi_{5s}^*({\bf k}')\hat \psi_{5s}({\bf k})2G_e\vec S_e\cdot \vec J_\chi  \delta^{(3)}\left({\bf k}+{\bf p}-{\bf k}^\prime-{\bf p}^\prime\right){(2\pi)^3}\nonumber\\
&= \int {\rm d}^3 x \,\psi_{5s}^*(x)\psi_{5s}(x)e^{i{\bf x}({\bf p}-{\bf p}^\prime)} 2G_e\vec S_e \cdot \vec J_\chi.
 \label{ElecScatt2}
\end{align}
Since the momentum carried by the electron\footnote{As discussed, we neglect ionization.} in the final state $\bar {\bf p}_e$ is a fraction $\sim m_e/m_{\rm Ncl}$ of the total momentum we can neglect it and evaluate the space integral 
\be
\int \di^3x\, \psi^*_{ns}(x) \psi_{ns}(x) e^{i\bf (p-p')x}=4\int_0^\infty  \di z\,  z^2 e^{-2z} \frac{\sin(\hat p z)}{\hat p z}, \label{eq:int_q}
\ee 
where $\hat p\equiv a_0 |{\bf p}-{\bf p} '|$, with $a_0= 2.68 \times 10^{-4}/\mathrm{eV}$ the Bohr radius. This factor is very close to unity for DM masses below GeV. In addition in our set-ups we are sensitive to forward scattering, a limit in which there is no momentum transfer and the previous integral reduces to $\int\di^3 x |\psi(x)|^2=1$.

\section{Light mediator and effective potential}\label{app:lightmA}
The case of light mediator, i.e. interaction length greater than de Broglie wavelength, $1/m_\Z\gg 1/m_\chi v$, presents some qualitative differences which nonetheless do not translate into quantitative differences in our estimates.
In particular in this case DM particles within a distance $1/m_\Z$ source a potential for matter. Given the interaction of \eqref{ModelAVB} in the NR limit one has:
\begin{align}
V=- \vec J_\psi \cdot\int \di^3x \,n_\chi (x)\, \frac{g_\chi g_\psi }{4\pi |x|}e^{-|x|m_\Z}\vec J_{\chi}(x)=-\frac{2g_\chi g_N n_\chi }{m_\Z^2}\langle \vec J_{\chi}\rangle \cdot \vec \lambda_\psi\, ,
\end{align}
where the average is over $n_\chi /m_\Z^3$ particles. The result reads like the effect of scattering~\eqref{eq:domega_flux1} and the field regime contribution to the Hamiltonian~\eqref{eq:HclEFT} for light mediator, i.e. $G_\psi\to g_\psi g_\chi/m_\Z^2$, and hence in both cases the contact interaction limit extends to light mediator.

\section{Coherently oscillating DM and magnetometers}\label{sec:Ht}

The treatment of oscillating magnetic-like fields is also standard in quantum mechanics. Let us
treat the interaction \eqref{eq:HclEFT} as a perturbation to the standard Larmor oscillations generated by \eqref{eq:HintB}, $\vec \lambda(t)=\vec \lambda_0(t)+\delta \vec \lambda$ where
\begin{align}
\vec \lambda_0(t)=\cos(\gamma Bt) \vec \lambda(0) +\sin(\gamma Bt)  \vec \lambda(0)\wedge\vec u_B+(1-\cos(\gamma Bt))\vec u_B (\vec u_B \cdot \vec \lambda(0)).
\end{align}
For the axion case in~\eqref{eq:Haxion-VB} we find,
\begin{align}\label{deltaFt}
\delta \vec \lambda=\frac{C_\psi\sqrt{2\rho_\chi}}{\mathfrak{f}_a}&\Bigg[ \left(\frac{\sin((m_a+\gamma B)t)}{m_a+\gamma B}+\frac{\sin((m_a-\gamma B)t)}{m_a-\gamma B}\right)\vec\lambda(0) \wedge\vec  v\\ \nonumber
+&\left(\frac{\cos((m_a+\gamma B)t)}{m_a+\gamma B}-\frac{\cos((m_a-\gamma B)t)}{m_a-\gamma B}\right)(\vec v (u_B\cdot \lambda(0))-\vec u_B(v\cdot \lambda(0))  )\\ \nonumber
+&\left(\frac{2\sin(m_at)}{m_a}-\frac{\sin((m_a+\gamma B)t)}{m_a+\gamma B}-\frac{\sin((m_a-\gamma B)t)}{m_a-\gamma B}\right)\vec \lambda(0) \wedge \vec u_B (u_B\cdot v)\Bigg]\,,
\end{align}
with $\vec \lambda(0)$ the total spin at $t=0$. The axial vector boson case is
obtained by substituting $-C_\psi\vec v/(2{\frak f}_{ a}) \to g^A_\psi\vec\epsilon /m_A$. One can see that if the axion mass is well below Zeeman's energy splitting the  first term of the last parenthesis in~\eqref{deltaFt} dominates and there is a suppression with the mass $m_a^{-1}$. This effect will be relevant for masses above the inverse of the typical measurement time of the experiment.  As discussed in sec.~\ref{sec:DMcl_mag}, the He-Xe co-magnetometer is based on runs of several hours and a total time of  $10^6$s to achieve its sensitivity.  This means that a suppression at $m_a \sim 10^{-20}$eV is expected unless one reanalyses the data  (the spin is monitored at much shorter times by SQUIDS), see also \cite{Graham:2017ivz}. For K-He magnetometers,  the polarization of the electrons is measured in time scales that may be pushed to $m$s, while data from $~143$ days is used for the final sensitivity.  However, the signal is not suppressed in each measurement, and one can get sensitivities of  $2\ {\rm fT/\sqrt{Hz}}$ \cite{Brown:2010dt}. This means that masses as high as  $m_a\sim10^{-12}$eV can be constrained. 

\section{Dark matter currents and preferred frame effects}\label{app:LV}

In some of the models we described the effect of DM in the clocks or magnetometers is equivalent to that of a constant background vector field anomalously coupled to spin.
The presence of such a frame has been constrained in different studies motivated by models  that violate Lorentz invariance, e.g.~\cite{Kostelecky:1999mr}. In particular, the co-magnetometers of~\cite{Brown:2010dt} and~\cite{Allmendinger:2013eya,Gemmel:2010ft} were used to test the coupling $\bar\psi\gamma^{\mu}\gamma_5\psi b_\mu $ with $b_\mu$ being a constant 4-vector. In our case, the DM currents are  the source of Lorentz violation. In the NR limit, the explicit connection is shown in table~\ref{tab:DMLV}.
\begin{table}[h]
\begin{center}
\begin{tabular}{|c|c|c|c|}
\hline
 &EFT& Axion& Axial Vector\\ \hline
$\vec b$&$\frac{\rho_\chi}{m_\chi} G_\chi \langle \vec J_\chi \rangle$& $ \frac{C_\psi \sqrt{2\rho_\chi}}{\mathfrak f_a}\langle \vec v \cos(m_a t)\rangle$ &
$\frac{ 2g_\psi \sqrt{2\rho_\chi}}{m_A}\langle {\rm Re}(\vec \epsilon \cos(m_a t))\rangle$
\\ \hline
\end{tabular} 
\caption{DM currents and preferred frames}\label{tab:DMLV}
\end{center}
\end{table}

The estimates of sec.~\ref{sec:AC_Mag} are a translation of the bounds $b< 8.4\times 10^{-34}\,$GeV of~\cite{Allmendinger:2013eya} and $b <3.7\times10^{-33}\,$GeV  from~\cite{Brown:2010dt}. We assume (for convenience) that the oscillating function in the axial cases of table~\ref{tab:DMLV} is equal to one. This assumption is justified, since even in the lightest cases one could run campaigns at different times of the year to explore the different phases of oscillation. For the cases where the current oscillates, the connection to the bounds is more subtle, see sec.~\ref{sec:Ht} and \cite{Graham:2017ivz}.

\bibliography{biblioLDMAC}

\providecommand{\href}[2]{#2}\begingroup\raggedright\begin{thebibliography}{100}

\bibitem{2010pdmo.book..121G}
G.~{Gelmini} and P.~{Gondolo}, {\it {DM production mechanisms}},  in {\em
  Particle Dark Matter : Observations, Models and Searches} (G.~{Bertone},
  ed.), p.~121, Cambridge University Press, 2010.
\newblock \href{http://xxx.lanl.gov/abs/1009.3690}{{\tt arXiv:1009.3690}}.

\bibitem{Bertone:2004pz}
G.~Bertone, D.~Hooper, and J.~Silk, {\it {Particle dark matter: Evidence,
  candidates and constraints}},  {\em Phys. Rept.} {\bf 405} (2005) 279--390,
  [\href{http://xxx.lanl.gov/abs/hep-ph/0404175}{{\tt hep-ph/0404175}}].

\bibitem{Goodman:1984dc}
M.~W. Goodman and E.~Witten, {\it {Detectability of Certain Dark Matter
  Candidates}},  {\em Phys. Rev.} {\bf D31} (1985) 3059.

\bibitem{Drukier:1986tm}
A.~K. Drukier, K.~Freese, and D.~N. Spergel, {\it {Detecting Cold Dark Matter
  Candidates}},  {\em Phys. Rev.} {\bf D33} (1986) 3495--3508.

\bibitem{ji2017astroparticle}
X.~Ji, {\it Astroparticle physics: Dark matter remains elusive},  {\em Nature}
  {\bf 542} (02, 2017) 172--173.

\bibitem{Aprile:2017iyp}
{\bf XENON} Collaboration, E.~Aprile et~al., {\it {First Dark Matter Search
  Results from the XENON1T Experiment}},  {\em Phys. Rev. Lett.} {\bf 119}
  (2017), no.~18 181301, [\href{http://xxx.lanl.gov/abs/1705.06655}{{\tt
  arXiv:1705.06655}}].

\bibitem{Gelmini:2015zpa}
G.~B. Gelmini, {\it {TASI 2014 Lectures: The Hunt for Dark Matter}},  in {\em
  Theoretical Advanced Study Institute in Elementary Particle Physics: Journeys
  Through the Precision Frontier: Amplitudes for Colliders (TASI 2014) Boulder,
  Colorado, June 2-27, 2014}, 2015.
\newblock \href{http://xxx.lanl.gov/abs/1502.01320}{{\tt arXiv:1502.01320}}.

\bibitem{Lisanti:2016jxe}
M.~Lisanti, {\it {Lectures on Dark Matter Physics}},  in {\em Theoretical
  Advanced Study Institute in Elementary Particle Physics: New Frontiers in
  Fields and Strings (TASI 2015) Boulder, CO, USA, June 1-26, 2015}, 2016.
\newblock \href{http://xxx.lanl.gov/abs/1603.03797}{{\tt arXiv:1603.03797}}.

\bibitem{Essig:2013lka}
R.~Essig et~al., {\it {Working Group Report: New Light Weakly Coupled
  Particles}},  in {\em Proceedings, Community Summer Study 2013: Snowmass on
  the Mississippi (CSS2013): Minneapolis, MN, USA, July 29-August 6, 2013},
  2013.
\newblock \href{http://xxx.lanl.gov/abs/1311.0029}{{\tt arXiv:1311.0029}}.

\bibitem{Alexander:2016aln}
J.~Alexander et~al., {\it Dark sectors 2016 workshop: Community report},  2016.
\newblock \href{http://xxx.lanl.gov/abs/1608.08632}{{\tt arXiv:1608.08632}}.

\bibitem{Knapen:2017xzo}
S.~Knapen, T.~Lin, and K.~M. Zurek, {\it {Light Dark Matter: Models and
  Constraints}},  {\em Phys. Rev.} {\bf D96} (2017), no.~11 115021,
  [\href{http://xxx.lanl.gov/abs/1709.07882}{{\tt arXiv:1709.07882}}].

\bibitem{Baur:2015jsy}
J.~Baur, N.~Palanque-Delabrouille, C.~Yche, C.~Magneville, and M.~Viel, {\it
  {Lyman-alpha Forests cool Warm Dark Matter}},  {\em JCAP} {\bf 1608} (2016),
  no.~08 012, [\href{http://xxx.lanl.gov/abs/1512.01981}{{\tt
  arXiv:1512.01981}}].

\bibitem{Randall:2016bqw}
L.~Randall, J.~Scholtz, and J.~Unwin, {\it {Cores in Dwarf Galaxies from Fermi
  Repulsion}},  {\em Mon. Not. Roy. Astron. Soc.} {\bf 467} (2017), no.~2
  1515--1525, [\href{http://xxx.lanl.gov/abs/1611.04590}{{\tt
  arXiv:1611.04590}}].

\bibitem{Tremaine:1979we}
S.~Tremaine and J.~E. Gunn, {\it {Dynamical Role of Light Neutral Leptons in
  Cosmology}},  {\em Phys. Rev. Lett.} {\bf 42} (1979) 407--410.

\bibitem{Boyarsky:2008ju}
A.~Boyarsky, O.~Ruchayskiy, and D.~Iakubovskyi, {\it {A Lower bound on the mass
  of Dark Matter particles}},  {\em JCAP} {\bf 0903} (2009) 005,
  [\href{http://xxx.lanl.gov/abs/0808.3902}{{\tt arXiv:0808.3902}}].

\bibitem{DiPaolo:2017geq}
C.~Di~Paolo, F.~Nesti, and F.~L. Villante, {\it {Phase space mass bound for
  fermionic dark matter from dwarf spheroidal galaxies}},  {\em Mon. Not. Roy.
  Astron. Soc.} {\bf 475} (2018), no.~4 5385--5397,
  [\href{http://xxx.lanl.gov/abs/1704.06644}{{\tt arXiv:1704.06644}}].

\bibitem{Kobayashi:2017jcf}
T.~Kobayashi, R.~Murgia, A.~De~Simone, V.~Ir{\v s}i{\v c}, and M.~Viel, {\it
  {Lyman-$\alpha$ constraints on ultralight scalar dark matter: Implications
  for the early and late universe}},  {\em Phys. Rev.} {\bf D96} (2017), no.~12
  123514, [\href{http://xxx.lanl.gov/abs/1708.00015}{{\tt arXiv:1708.00015}}].

\bibitem{Bar:2018acw}
N.~Bar, D.~Blas, K.~Blum, and S.~Sibiryakov, {\it {Galactic Rotation Curves vs.
  Ultra-Light Dark Matter: Implications of the Soliton -- Host Halo Relation}},
   \href{http://xxx.lanl.gov/abs/1805.00122}{{\tt arXiv:1805.00122}}.

\bibitem{Marsh:2015xka}
D.~J.~E. Marsh, {\it {Axion Cosmology}},  {\em Phys. Rept.} {\bf 643} (2016)
  1--79, [\href{http://xxx.lanl.gov/abs/1510.07633}{{\tt arXiv:1510.07633}}].

\bibitem{Hui:2016ltb}
L.~Hui, J.~P. Ostriker, S.~Tremaine, and E.~Witten, {\it {Ultralight scalars as
  cosmological dark matter}},  {\em Phys. Rev.} {\bf D95} (2017), no.~4 043541,
  [\href{http://xxx.lanl.gov/abs/1610.08297}{{\tt arXiv:1610.08297}}].

\bibitem{Kouvaris:2016afs}
C.~Kouvaris and J.~Pradler, {\it {Probing sub-GeV Dark Matter with conventional
  detectors}},  {\em Phys. Rev. Lett.} {\bf 118} (2017), no.~3 031803,
  [\href{http://xxx.lanl.gov/abs/1607.01789}{{\tt arXiv:1607.01789}}].

\bibitem{Hochberg:2016ajh}
Y.~Hochberg, T.~Lin, and K.~M. Zurek, {\it {Detecting Ultralight Bosonic Dark
  Matter via Absorption in Superconductors}},  {\em Phys. Rev.} {\bf D94}
  (2016), no.~1 015019, [\href{http://xxx.lanl.gov/abs/1604.06800}{{\tt
  arXiv:1604.06800}}].

\bibitem{Essig:2017kqs}
R.~Essig, T.~Volansky, and T.-T. Yu, {\it {New Constraints and Prospects for
  sub-GeV Dark Matter Scattering off Electrons in Xenon}},  {\em Phys. Rev.}
  {\bf D96} (2017), no.~4 043017,
  [\href{http://xxx.lanl.gov/abs/1703.00910}{{\tt arXiv:1703.00910}}].

\bibitem{Essig:2011nj}
R.~Essig, J.~Mardon, and T.~Volansky, {\it {Direct Detection of Sub-GeV Dark
  Matter}},  {\em Phys. Rev.} {\bf D85} (2012) 076007,
  [\href{http://xxx.lanl.gov/abs/1108.5383}{{\tt arXiv:1108.5383}}].

\bibitem{Graham:2012su}
P.~W. Graham, D.~E. Kaplan, S.~Rajendran, and M.~T. Walters, {\it
  {Semiconductor Probes of Light Dark Matter}},  {\em Phys. Dark Univ.} {\bf 1}
  (2012) 32--49, [\href{http://xxx.lanl.gov/abs/1203.2531}{{\tt
  arXiv:1203.2531}}].

\bibitem{Essig:2015cda}
R.~Essig, M.~Fernandez-Serra, J.~Mardon, A.~Soto, T.~Volansky, and T.-T. Yu,
  {\it {Direct Detection of sub-GeV Dark Matter with Semiconductor Targets}},
  {\em JHEP} {\bf 05} (2016) 046,
  [\href{http://xxx.lanl.gov/abs/1509.01598}{{\tt arXiv:1509.01598}}].

\bibitem{Lee:2015qva}
S.~K. Lee, M.~Lisanti, S.~Mishra-Sharma, and B.~R. Safdi, {\it {Modulation
  Effects in Dark Matter-Electron Scattering Experiments}},  {\em Phys. Rev.}
  {\bf D92} (2015), no.~8 083517,
  [\href{http://xxx.lanl.gov/abs/1508.07361}{{\tt arXiv:1508.07361}}].

\bibitem{Essig:2012yx}
R.~Essig, A.~Manalaysay, J.~Mardon, P.~Sorensen, and T.~Volansky, {\it {First
  Direct Detection Limits on sub-GeV Dark Matter from XENON10}},  {\em Phys.
  Rev. Lett.} {\bf 109} (2012) 021301,
  [\href{http://xxx.lanl.gov/abs/1206.2644}{{\tt arXiv:1206.2644}}].

\bibitem{Knapen:2016cue}
S.~Knapen, T.~Lin, and K.~M. Zurek, {\it {Light Dark Matter in Superfluid
  Helium: Detection with Multi-excitation Production}},  {\em Phys. Rev.} {\bf
  D95} (2017), no.~5 056019, [\href{http://xxx.lanl.gov/abs/1611.06228}{{\tt
  arXiv:1611.06228}}].

\bibitem{Hochberg:2015fth}
Y.~Hochberg, M.~Pyle, Y.~Zhao, and K.~M. Zurek, {\it {Detecting Superlight Dark
  Matter with Fermi-Degenerate Materials}},  {\em JHEP} {\bf 08} (2016) 057,
  [\href{http://xxx.lanl.gov/abs/1512.04533}{{\tt arXiv:1512.04533}}].

\bibitem{Hochberg:2016sqx}
Y.~Hochberg, T.~Lin, and K.~M. Zurek, {\it {Absorption of light dark matter in
  semiconductors}},  {\em Phys. Rev.} {\bf D95} (2017), no.~2 023013,
  [\href{http://xxx.lanl.gov/abs/1608.01994}{{\tt arXiv:1608.01994}}].

\bibitem{Schutz:2016tid}
K.~Schutz and K.~M. Zurek, {\it {Detectability of Light Dark Matter with
  Superfluid Helium}},  {\em Phys. Rev. Lett.} {\bf 117} (2016), no.~12 121302,
  [\href{http://xxx.lanl.gov/abs/1604.08206}{{\tt arXiv:1604.08206}}].

\bibitem{Essig:2016crl}
R.~Essig, J.~Mardon, O.~Slone, and T.~Volansky, {\it {Detection of sub-GeV Dark
  Matter and Solar Neutrinos via Chemical-Bond Breaking}},  {\em Phys. Rev.}
  {\bf D95} (2017), no.~5 056011,
  [\href{http://xxx.lanl.gov/abs/1608.02940}{{\tt arXiv:1608.02940}}].

\bibitem{Brax:2017xho}
P.~Brax, S.~Fichet, and G.~Pignol, {\it {Bounding Quantum Dark Forces}},  {\em
  Phys. Rev.} {\bf D97} (2018), no.~11 115034,
  [\href{http://xxx.lanl.gov/abs/1710.00850}{{\tt arXiv:1710.00850}}].

\bibitem{Fichet:2017bng}
S.~Fichet, {\it {Quantum Forces from Dark Matter and Where to Find Them}},
  {\em Phys. Rev. Lett.} {\bf 120} (2018), no.~13 131801,
  [\href{http://xxx.lanl.gov/abs/1705.10331}{{\tt arXiv:1705.10331}}].

\bibitem{Budnik:2017sbu}
R.~Budnik, O.~Chesnovsky, O.~Slone, and T.~Volansky, {\it {Direct Detection of
  Light Dark Matter and Solar Neutrinos via Color Center Production in
  Crystals}},  {\em Phys. Lett.} {\bf B782} (2018) 242--250,
  [\href{http://xxx.lanl.gov/abs/1705.03016}{{\tt arXiv:1705.03016}}].

\bibitem{Riedel:2012ur}
C.~J. Riedel, {\it {Direct detection of classically undetectable dark matter
  through quantum decoherence}},  {\em Phys. Rev.} {\bf D88} (2013), no.~11
  116005, [\href{http://xxx.lanl.gov/abs/1212.3061}{{\tt arXiv:1212.3061}}].

\bibitem{Bateman:2014lia}
J.~Bateman, I.~McHardy, A.~Merle, T.~R. Morris, and H.~Ulbricht, {\it {On the
  Existence of Low-Mass Dark Matter and its Direct Detection}},  {\em Sci.
  Rep.} {\bf 5} (2015) 8058, [\href{http://xxx.lanl.gov/abs/1405.5536}{{\tt
  arXiv:1405.5536}}].

\bibitem{Riedel:2016acj}
C.~J. Riedel and I.~Yavin, {\it {Decoherence as a way to measure extremely soft
  collisions with dark matter}},  {\em Phys. Rev.} {\bf D96} (2017), no.~2
  023007, [\href{http://xxx.lanl.gov/abs/1609.04145}{{\tt arXiv:1609.04145}}].

\bibitem{Derevianko:2013oaa}
A.~Derevianko and M.~Pospelov, {\it {Hunting for topological dark matter with
  atomic clocks}},  {\em Nature Phys.} {\bf 10} (2014) 933,
  [\href{http://xxx.lanl.gov/abs/1311.1244}{{\tt arXiv:1311.1244}}].

\bibitem{Arvanitaki:2014faa}
A.~Arvanitaki, J.~Huang, and K.~Van~Tilburg, {\it {Searching for dilaton dark
  matter with atomic clocks}},  {\em Phys. Rev.} {\bf D91} (2015), no.~1
  015015, [\href{http://xxx.lanl.gov/abs/1405.2925}{{\tt arXiv:1405.2925}}].

\bibitem{Stadnik:2015upa}
Y.~V. Stadnik and V.~V. Flambaum, {\it {Manifestations of dark matter and
  variations of fundamental constants in atoms and astrophysical phenomena}},
\newblock 2015.
\newblock \href{http://xxx.lanl.gov/abs/1509.00966}{{\tt arXiv:1509.00966}}.

\bibitem{Hees:2016gop}
A.~Hees, J.~Gu{\'e}na, M.~Abgrall, S.~Bize, and P.~Wolf, {\it {Searching for an
  oscillating massive scalar field as a dark matter candidate using atomic
  hyperfine frequency comparisons}},  {\em Phys. Rev. Lett.} {\bf 117} (2016),
  no.~6 061301, [\href{http://xxx.lanl.gov/abs/1604.08514}{{\tt
  arXiv:1604.08514}}].

\bibitem{VanTilburg:2015oza}
K.~Van~Tilburg, N.~Leefer, L.~Bougas, and D.~Budker, {\it {Search for
  ultralight scalar dark matter with atomic spectroscopy}},  {\em Phys. Rev.
  Lett.} {\bf 115} (2015), no.~1 011802,
  [\href{http://xxx.lanl.gov/abs/1503.06886}{{\tt arXiv:1503.06886}}].

\bibitem{Yang:2016odu}
Q.~Yang and H.~Di, {\it {Sub-MeV Bosonic Dark Matter, Misalignment Mechanism
  and Galactic Dark Matter Halo Luminosities}},  {\em Phys. Rev.} {\bf D95}
  (2017), no.~7 075032, [\href{http://xxx.lanl.gov/abs/1610.08378}{{\tt
  arXiv:1610.08378}}].

\bibitem{Dev:2016hxv}
P.~S.~B. Dev, M.~Lindner, and S.~Ohmer, {\it {Gravitational waves as a new
  probe of Bose?Einstein condensate Dark Matter}},  {\em Phys. Lett.} {\bf
  B773} (2017) 219--224, [\href{http://xxx.lanl.gov/abs/1609.03939}{{\tt
  arXiv:1609.03939}}].

\bibitem{Garcon:2017ixh}
A.~{Garcon}, D.~{Aybas}, J.~W. {Blanchard}, G.~{Centers}, N.~L. {Figueroa},
  P.~W. {Graham}, D.~F.~J. {Kimball}, S.~{Rajendran}, M.~{Gil Sendra}, A.~O.
  {Sushkov}, L.~{Trahms}, T.~{Wang}, A.~{Wickenbrock}, T.~{Wu}, and
  D.~{Budker}, {\it {The cosmic axion spin precession experiment (CASPEr): a
  dark-matter search with nuclear magnetic resonance}},  {\em Quantum Science
  and Technology} {\bf 3} (Jan., 2018) 014008,
  [\href{http://xxx.lanl.gov/abs/1707.05312}{{\tt arXiv:1707.05312}}].

\bibitem{Blas:2016ddr}
D.~Blas, D.~L. Nacir, and S.~Sibiryakov, {\it {Ultralight Dark Matter Resonates
  with Binary Pulsars}},  {\em Phys. Rev. Lett.} {\bf 118} (2017), no.~26
  261102, [\href{http://xxx.lanl.gov/abs/1612.06789}{{\tt arXiv:1612.06789}}].

\bibitem{Delaunay:2016brc}
C.~Delaunay, R.~Ozeri, G.~Perez, and Y.~Soreq, {\it {Probing Atomic Higgs-like
  Forces at the Precision Frontier}},  {\em Phys. Rev.} {\bf D96} (2017), no.~9
  093001, [\href{http://xxx.lanl.gov/abs/1601.05087}{{\tt arXiv:1601.05087}}].

\bibitem{Wcislo:2016qng}
P.~Wcislo, P.~Morzynski, M.~Bober, A.~Cygan, D.~Lisak, R.~Ciurylo, and
  M.~Zawada, {\it {Searching for dark matter with optical atomic clocks}},
  \href{http://xxx.lanl.gov/abs/1605.05763}{{\tt arXiv:1605.05763}}.

\bibitem{Wolf:2018xlz}
P.~Wolf, R.~Alonso, and D.~Blas, {\it {Scattering of light dark matter in
  atomic clocks}},  \href{http://xxx.lanl.gov/abs/1810.01632}{{\tt
  arXiv:1810.01632}}.

\bibitem{Brown:2010dt}
J.~M. Brown, S.~J. Smullin, T.~W. Kornack, and M.~V. Romalis, {\it {New limit
  on Lorentz and CPT-violating neutron spin interactions}},  {\em Phys. Rev.
  Lett.} {\bf 105} (2010) 151604,
  [\href{http://xxx.lanl.gov/abs/1006.5425}{{\tt arXiv:1006.5425}}].

\bibitem{Allmendinger:2013eya}
F.~Allmendinger, W.~Heil, S.~Karpuk, W.~Kilian, A.~Scharth, U.~Schmidt,
  A.~Schnabel, {\relax Yu}.~Sobolev, and K.~Tullney, {\it {New Limit on
  Lorentz-Invariance- and CPT-Violating Neutron Spin Interactions Using a
  Free-Spin-Precession $^3$He - $^{129}$Xe Comagnetometer}},  {\em Phys. Rev.
  Lett.} {\bf 112} (2014), no.~11 110801,
  [\href{http://xxx.lanl.gov/abs/1312.3225}{{\tt arXiv:1312.3225}}].

\bibitem{Graham:2017ivz}
P.~W. Graham, D.~E. Kaplan, J.~Mardon, S.~Rajendran, W.~A. Terrano, L.~Trahms,
  and T.~Wilkason, {\it {Spin Precession Experiments for Light Axionic Dark
  Matter}},  {\em Phys. Rev.} {\bf D97} (2018), no.~5 055006,
  [\href{http://xxx.lanl.gov/abs/1709.07852}{{\tt arXiv:1709.07852}}].

\bibitem{Stadnik}
Y.~V. Stadnik, {\it {PhD thesis, University of New South Wales (2017)}}, .

\bibitem{Becker:2007sv}
J.~K. Becker, {\it {High-energy neutrinos in the context of multimessenger
  physics}},  {\em Phys. Rept.} {\bf 458} (2008) 173--246,
  [\href{http://xxx.lanl.gov/abs/0710.1557}{{\tt arXiv:0710.1557}}].

\bibitem{Formaggio:2013kya}
J.~A. Formaggio and G.~P. Zeller, {\it {From eV to EeV: Neutrino Cross Sections
  Across Energy Scales}},  {\em Rev. Mod. Phys.} {\bf 84} (2012) 1307--1341,
  [\href{http://xxx.lanl.gov/abs/1305.7513}{{\tt arXiv:1305.7513}}].

\bibitem{Ringwald:2009bg}
A.~Ringwald, {\it {Prospects for the direct detection of the cosmic neutrino
  background}},  {\em Nucl. Phys.} {\bf A827} (2009) 501C--506C,
  [\href{http://xxx.lanl.gov/abs/0901.1529}{{\tt arXiv:0901.1529}}].

\bibitem{Baracchini:2018wwj}
{\bf PTOLEMY} Collaboration, E.~Baracchini et~al., {\it {PTOLEMY: A Proposal
  for Thermal Relic Detection of Massive Neutrinos and Directional Detection of
  MeV Dark Matter}},  \href{http://xxx.lanl.gov/abs/1808.01892}{{\tt
  arXiv:1808.01892}}.

\bibitem{Bishara:2017nnn}
F.~Bishara, J.~Brod, B.~Grinstein, and J.~Zupan, {\it {DirectDM: a tool for
  dark matter direct detection}},
  \href{http://xxx.lanl.gov/abs/1708.02678}{{\tt arXiv:1708.02678}}.

\bibitem{Preskill:1990fr}
J.~Preskill, {\it {Gauge anomalies in an effective field theory}},  {\em Annals
  Phys.} {\bf 210} (1991) 323--379.

\bibitem{Dror:2017ehi}
J.~A. Dror, R.~Lasenby, and M.~Pospelov, {\it {New constraints on light vectors
  coupled to anomalous currents}},  {\em Phys. Rev. Lett.} {\bf 119} (2017),
  no.~14 141803, [\href{http://xxx.lanl.gov/abs/1705.06726}{{\tt
  arXiv:1705.06726}}].

\bibitem{Dror:2017nsg}
J.~A. Dror, R.~Lasenby, and M.~Pospelov, {\it {Dark forces coupled to
  nonconserved currents}},  {\em Phys. Rev.} {\bf D96} (2017), no.~7 075036,
  [\href{http://xxx.lanl.gov/abs/1707.01503}{{\tt arXiv:1707.01503}}].

\bibitem{vanier1989quantum}
J.~Vanier and C.~Audoin, {\em The Quantum Physics of Atomic Frequency
  Standards: vols. I and II}.
\newblock A. Hilger, 1989.

\bibitem{Stone:2005rzh}
N.~J. Stone, {\it {Table of nuclear magnetic dipole and electric quadrupole
  moments}},  {\em Atom. Data Nucl. Data Tabl.} {\bf 90} (2005) 75--176.

\bibitem{goldberger2004collision}
M.~Goldberger and K.~Watson, {\em Collision Theory}.
\newblock Dover books on physics. Dover Publications, 2004.

\bibitem{Stadnik:2014xja}
Y.~V. Stadnik and V.~V. Flambaum, {\it {Nuclear spin-dependent interactions:
  Searches for WIMP, Axion and Topological Defect Dark Matter, and Tests of
  Fundamental Symmetries}},  {\em Eur. Phys. J.} {\bf C75} (2015), no.~3 110,
  [\href{http://xxx.lanl.gov/abs/1408.2184}{{\tt arXiv:1408.2184}}].

\bibitem{Brown:2011exm}
J.~Brown, {\it {Ph.D. Thesis, Princeton U. (2011)}}, .

\bibitem{Chakrabarty:2017fkd}
S.~S. Chakrabarty, S.~Enomoto, Y.~Han, P.~Sikivie, and E.~M. Todarello, {\it
  {Gravitational self-interactions of a degenerate quantum scalar field}},
  {\em Phys. Rev.} {\bf D97} (2018), no.~4 043531,
  [\href{http://xxx.lanl.gov/abs/1710.02195}{{\tt arXiv:1710.02195}}].

\bibitem{Hertzberg:2016tal}
M.~P. Hertzberg, {\it {Quantum and Classical Behavior in Interacting Bosonic
  Systems}},  {\em JCAP} {\bf 1611} (2016), no.~11 037,
  [\href{http://xxx.lanl.gov/abs/1609.01342}{{\tt arXiv:1609.01342}}].

\bibitem{Dvali:2017ruz}
G.~Dvali and S.~Zell, {\it {Classicality and Quantum Break-Time for Cosmic
  Axions}},  {\em JCAP} {\bf 1807} (2018), no.~07 064,
  [\href{http://xxx.lanl.gov/abs/1710.00835}{{\tt arXiv:1710.00835}}].

\bibitem{Preskill:1982cy}
J.~Preskill, M.~B. Wise, and F.~Wilczek, {\it {Cosmology of the Invisible
  Axion}},  {\em Phys. Lett.} {\bf 120B} (1983) 127--132.

\bibitem{Abbott:1982af}
L.~F. Abbott and P.~Sikivie, {\it {A Cosmological Bound on the Invisible
  Axion}},  {\em Phys. Lett.} {\bf 120B} (1983) 133--136.

\bibitem{Nelson:2011sf}
A.~E. Nelson and J.~Scholtz, {\it {Dark Light, Dark Matter and the Misalignment
  Mechanism}},  {\em Phys. Rev.} {\bf D84} (2011) 103501,
  [\href{http://xxx.lanl.gov/abs/1105.2812}{{\tt arXiv:1105.2812}}].

\bibitem{Li:2013nal}
B.~Li, T.~Rindler-Daller, and P.~R. Shapiro, {\it {Cosmological Constraints on
  Bose-Einstein-Condensed Scalar Field Dark Matter}},  {\em Phys. Rev.} {\bf
  D89} (2014), no.~8 083536, [\href{http://xxx.lanl.gov/abs/1310.6061}{{\tt
  arXiv:1310.6061}}].

\bibitem{Graham:2015ifn}
P.~W. Graham, D.~E. Kaplan, J.~Mardon, S.~Rajendran, and W.~A. Terrano, {\it
  {Dark Matter Direct Detection with Accelerometers}},  {\em Phys. Rev.} {\bf
  D93} (2016), no.~7 075029, [\href{http://xxx.lanl.gov/abs/1512.06165}{{\tt
  arXiv:1512.06165}}].

\bibitem{Arias:2012az}
P.~Arias, D.~Cadamuro, M.~Goodsell, J.~Jaeckel, J.~Redondo, and A.~Ringwald,
  {\it {WISPy Cold Dark Matter}},  {\em JCAP} {\bf 1206} (2012) 013,
  [\href{http://xxx.lanl.gov/abs/1201.5902}{{\tt arXiv:1201.5902}}].

\bibitem{Graham:2015rva}
P.~W. Graham, J.~Mardon, and S.~Rajendran, {\it {Vector Dark Matter from
  Inflationary Fluctuations}},  {\em Phys. Rev.} {\bf D93} (2016), no.~10
  103520, [\href{http://xxx.lanl.gov/abs/1504.02102}{{\tt arXiv:1504.02102}}].

\bibitem{Sikivie:2009qn}
P.~Sikivie and Q.~Yang, {\it {Bose-Einstein Condensation of Dark Matter
  Axions}},  {\em Phys. Rev. Lett.} {\bf 103} (2009) 111301,
  [\href{http://xxx.lanl.gov/abs/0901.1106}{{\tt arXiv:0901.1106}}].

\bibitem{Guth:2014hsa}
A.~H. Guth, M.~P. Hertzberg, and C.~Prescod-Weinstein, {\it {Do Dark Matter
  Axions Form a Condensate with Long-Range Correlation?}},  {\em Phys. Rev.}
  {\bf D92} (2015), no.~10 103513,
  [\href{http://xxx.lanl.gov/abs/1412.5930}{{\tt arXiv:1412.5930}}].

\bibitem{Schive:2014hza}
H.-Y. Schive, M.-H. Liao, T.-P. Woo, S.-K. Wong, T.~Chiueh, T.~Broadhurst, and
  W.~Y.~P. Hwang, {\it {Understanding the Core-Halo Relation of Quantum Wave
  Dark Matter from 3D Simulations}},  {\em Phys. Rev. Lett.} {\bf 113} (2014),
  no.~26 261302, [\href{http://xxx.lanl.gov/abs/1407.7762}{{\tt
  arXiv:1407.7762}}].

\bibitem{Schwabe:2016rze}
B.~Schwabe, J.~C. Niemeyer, and J.~F. Engels, {\it {Simulations of solitonic
  core mergers in ultralight axion dark matter cosmologies}},  {\em Phys. Rev.}
  {\bf D94} (2016), no.~4 043513,
  [\href{http://xxx.lanl.gov/abs/1606.05151}{{\tt arXiv:1606.05151}}].

\bibitem{Levkov:2018kau}
D.~G. Levkov, A.~G. Panin, and I.~I. Tkachev, {\it {Bose Condensation by
  Gravitational Interactions}},  \href{http://xxx.lanl.gov/abs/1804.05857}{{\tt
  arXiv:1804.05857}}.

\bibitem{Widdicombe:2018oeo}
J.~Y. Widdicombe, T.~Helfer, D.~J.~E. Marsh, and E.~A. Lim, {\it {Formation of
  Relativistic Axion Stars}},  \href{http://xxx.lanl.gov/abs/1806.09367}{{\tt
  arXiv:1806.09367}}.

\bibitem{Arvanitaki:2017nhi}
A.~Arvanitaki, S.~Dimopoulos, and K.~Van~Tilburg, {\it {Resonant absorption of
  bosonic dark matter in molecules}},
  \href{http://xxx.lanl.gov/abs/1709.05354}{{\tt arXiv:1709.05354}}.

\bibitem{Foster:2017hbq}
J.~W. Foster, N.~L. Rodd, and B.~R. Safdi, {\it {Revealing the Dark Matter Halo
  with Axion Direct Detection}},  {\em Phys. Rev.} {\bf D97} (2018), no.~12
  123006, [\href{http://xxx.lanl.gov/abs/1711.10489}{{\tt arXiv:1711.10489}}].

\bibitem{2012arXiv1204.3621G}
J.~{Gu{\'e}na}, M.~{Abgrall}, D.~{Rovera}, P.~{Laurent}, B.~{Chupin},
  M.~{Lours}, G.~{Santarelli}, P.~{Rosenbusch}, M.~E. {Tobar}, R.~{Li},
  K.~{Gibble}, A.~{Clairon}, and S.~{Bize}, {\it {Progress in Atomic Fountains
  at LNE-SYRTE}},  {\em IEEE Transactions on Ultrasonics, Ferroelectrics, and
  Frequency Control} {\bf 59} (2012), no.~3, 391--409,
  [\href{http://xxx.lanl.gov/abs/1204.3621}{{\tt arXiv:1204.3621}}].

\bibitem{2013PhRvL.110r0802G}
K.~{Gibble}, {\it {Scattering of Cold-Atom Coherences by Hot Atoms: Frequency
  Shifts from Background-Gas Collisions}},  {\em Physical Review Letters} {\bf
  110} (May, 2013) 180802, [\href{http://xxx.lanl.gov/abs/1304.3486}{{\tt
  arXiv:1304.3486}}].

\bibitem{2015RvMP...87..637L}
A.~D. {Ludlow}, M.~M. {Boyd}, J.~{Ye}, E.~{Peik}, and P.~O. {Schmidt}, {\it
  {Optical atomic clocks}},  {\em Reviews of Modern Physics} {\bf 87} (Apr.,
  2015) 637--701, [\href{http://xxx.lanl.gov/abs/1407.3493}{{\tt
  arXiv:1407.3493}}].

\bibitem{weinberg2013lectures}
S.~Weinberg, {\em Lectures on Quantum Mechanics}.
\newblock Cambridge University Press, 2013.

\bibitem{2014Metro..51..108G}
J.~{Gu{\'e}na}, M.~{Abgrall}, A.~{Clairon}, and S.~{Bize}, {\it {Contributing
  to TAI with a secondary representation of the SI second}},  {\em Metrologia}
  {\bf 51} (Feb., 2014) 108, [\href{http://xxx.lanl.gov/abs/1401.7976}{{\tt
  arXiv:1401.7976}}].

\bibitem{Erickcek:2007jv}
A.~L. Erickcek, P.~J. Steinhardt, D.~McCammon, and P.~C. McGuire, {\it
  {Constraints on the Interactions between Dark Matter and Baryons from the
  X-ray Quantum Calorimetry Experiment}},  {\em Phys. Rev.} {\bf D76} (2007)
  042007, [\href{http://xxx.lanl.gov/abs/0704.0794}{{\tt arXiv:0704.0794}}].

\bibitem{Kavanagh:2016pyr}
B.~J. Kavanagh, R.~Catena, and C.~Kouvaris, {\it {Signatures of
  Earth-scattering in the direct detection of Dark Matter}},  {\em JCAP} {\bf
  1701} (2017), no.~01 012, [\href{http://xxx.lanl.gov/abs/1611.05453}{{\tt
  arXiv:1611.05453}}].

\bibitem{Emken:2017qmp}
T.~Emken and C.~Kouvaris, {\it {DaMaSCUS: The Impact of Underground Scatterings
  on Direct Detection of Light Dark Matter}},  {\em JCAP} {\bf 1710} (2017),
  no.~10 031, [\href{http://xxx.lanl.gov/abs/1706.02249}{{\tt
  arXiv:1706.02249}}].

\bibitem{LAURENT2015540}
P.~Laurent, D.~Massonnet, L.~Cacciapuoti, and C.~Salomon, {\it The aces/pharao
  space mission},  {\em Comptes Rendus Physique} {\bf 16} (2015), no.~5 540 --
  552. The measurement of time / La mesure du temps.

\bibitem{2017arXiv170903256L}
L.~{Liu}, D.~{L{\"u}}, W.~{Chen}, T.~{Li}, Q.~{Qu}, B.~{Wang}, L.~{Li},
  W.~{Ren}, Z.~{Dong}, J.~{Zhao}, W.~{Xia}, X.~{Zhao}, J.~{Ji}, M.~{Ye},
  Y.~{Sun}, Y.~{Yao}, D.~{Song}, Z.~{Liang}, S.~{Hu}, D.~{Yu}, X.~{Hou},
  W.~{Shi}, H.~{Zang}, J.~{Xiang}, X.~{Peng}, and Y.~{Wang}, {\it {Tests of
  Cold Atom Clock in Orbit}},  \href{http://xxx.lanl.gov/abs/1709.03256}{{\tt
  arXiv:1709.03256}}.

\bibitem{Gemmel:2010ft}
C.~Gemmel et~al., {\it {Limit on Lorentz and CPT violation of the bound Neutron
  Using a Free Precession 3He/129Xe co-magnetometer}},  {\em Phys. Rev.} {\bf
  D82} (2010) 111901, [\href{http://xxx.lanl.gov/abs/1011.2143}{{\tt
  arXiv:1011.2143}}].

\bibitem{Safronova:2017xyt}
M.~S. Safronova, D.~Budker, D.~DeMille, D.~F.~J. Kimball, A.~Derevianko, and
  C.~W. Clark, {\it {Search for New Physics with Atoms and Molecules}},  {\em
  Rev. Mod. Phys.} {\bf 90} (2018), no.~2 025008,
  [\href{http://xxx.lanl.gov/abs/1710.01833}{{\tt arXiv:1710.01833}}].

\bibitem{Terrano:2015sna}
W.~A. Terrano, E.~G. Adelberger, J.~G. Lee, and B.~R. Heckel, {\it {Short-range
  spin-dependent interactions of electrons: a probe for exotic pseudo-Goldstone
  bosons}},  {\em Phys. Rev. Lett.} {\bf 115} (2015), no.~20 201801,
  [\href{http://xxx.lanl.gov/abs/1508.02463}{{\tt arXiv:1508.02463}}].

\bibitem{Vasilakis:2008yn}
G.~Vasilakis, J.~M. Brown, T.~W. Kornack, and M.~V. Romalis, {\it {Limits on
  new long range nuclear spin-dependent forces set with a K - He-3
  co-magnetometer}},  {\em Phys. Rev. Lett.} {\bf 103} (2009) 261801,
  [\href{http://xxx.lanl.gov/abs/0809.4700}{{\tt arXiv:0809.4700}}].

\bibitem{Raffelt:2012sp}
G.~Raffelt, {\it {Limits on a CP-violating scalar axion-nucleon interaction}},
  {\em Phys. Rev.} {\bf D86} (2012) 015001,
  [\href{http://xxx.lanl.gov/abs/1205.1776}{{\tt arXiv:1205.1776}}].

\bibitem{Abel:2017rtm}
C.~Abel et~al., {\it {Search for Axionlike Dark Matter through Nuclear Spin
  Precession in Electric and Magnetic Fields}},  {\em Phys. Rev.} {\bf X7}
  (2017), no.~4 041034, [\href{http://xxx.lanl.gov/abs/1708.06367}{{\tt
  arXiv:1708.06367}}].

\bibitem{Tulin:2017ara}
S.~Tulin and H.-B. Yu, {\it {Dark Matter Self-interactions and Small Scale
  Structure}},  {\em Phys. Rept.} {\bf 730} (2018) 1--57,
  [\href{http://xxx.lanl.gov/abs/1705.02358}{{\tt arXiv:1705.02358}}].

\bibitem{Pollack:2014rja}
J.~Pollack, D.~N. Spergel, and P.~J. Steinhardt, {\it {Supermassive Black Holes
  from Ultra-Strongly Self-Interacting Dark Matter}},  {\em Astrophys. J.} {\bf
  804} (2015), no.~2 131, [\href{http://xxx.lanl.gov/abs/1501.00017}{{\tt
  arXiv:1501.00017}}].

\bibitem{Davidson:2000hf}
S.~Davidson, S.~Hannestad, and G.~Raffelt, {\it {Updated bounds on millicharged
  particles}},  {\em JHEP} {\bf 05} (2000) 003,
  [\href{http://xxx.lanl.gov/abs/hep-ph/0001179}{{\tt hep-ph/0001179}}].

\bibitem{Raffelt:1996wa}
G.~G. Raffelt, {\em {Stars as laboratories for fundamental physics}}.
\newblock Chicago, USA: Univ. Pr. (1996) 664 p, 1996.

\bibitem{Raffelt:1999tx}
G.~G. Raffelt, {\it {Particle physics from stars}},  {\em Ann. Rev. Nucl. Part.
  Sci.} {\bf 49} (1999) 163--216,
  [\href{http://xxx.lanl.gov/abs/hep-ph/9903472}{{\tt hep-ph/9903472}}].

\bibitem{1998BrJPh..28...72E}
R.~{Ejnisman} and N.~P. {Bigelow}, {\it {Is it possible to do experimental
  cosmology using cold atoms?}},  {\em Brazilian Journal of Physics} {\bf 28}
  (Mar., 1998) 72--76.

\bibitem{Cocco:2007za}
A.~G. Cocco, G.~Mangano, and M.~Messina, {\it {Probing low energy neutrino
  backgrounds with neutrino capture on beta decaying nuclei}},  {\em JCAP} {\bf
  0706} (2007) 015, [\href{http://xxx.lanl.gov/abs/hep-ph/0703075}{{\tt
  hep-ph/0703075}}].

\bibitem{Stodolsky:1974aq}
L.~Stodolsky, {\it {Speculations on Detection of the Neutrino Sea}},  {\em
  Phys. Rev. Lett.} {\bf 34} (1975) 110. [Erratum: Phys. Rev.
  Lett.34,508(1975)].

\bibitem{Langacker:1982ih}
P.~Langacker, J.~P. Leveille, and J.~Sheiman, {\it {On the Detection of
  Cosmological Neutrinos by Coherent Scattering}},  {\em Phys. Rev.} {\bf D27}
  (1983) 1228.

\bibitem{Ringwald:2004np}
A.~Ringwald and Y.~Y.~Y. Wong, {\it {Gravitational clustering of relic
  neutrinos and implications for their detection}},  {\em JCAP} {\bf 0412}
  (2004) 005, [\href{http://xxx.lanl.gov/abs/hep-ph/0408241}{{\tt
  hep-ph/0408241}}].

\bibitem{Domcke:2017aqj}
V.~Domcke and M.~Spinrath, {\it {Detection prospects for the Cosmic Neutrino
  Background using laser interferometers}},  {\em JCAP} {\bf 1706} (2017),
  no.~06 055, [\href{http://xxx.lanl.gov/abs/1703.08629}{{\tt
  arXiv:1703.08629}}].

\bibitem{Akhmedov:2018wlf}
E.~Akhmedov, G.~Arcadi, M.~Lindner, and S.~Vogl, {\it {Coherent scattering and
  macroscopic coherence: Implications for neutrino, dark matter and axion
  detection}},  \href{http://xxx.lanl.gov/abs/1806.10962}{{\tt
  arXiv:1806.10962}}.

\bibitem{Green:2017ybv}
D.~Green and S.~Rajendran, {\it {The Cosmology of Sub-MeV Dark Matter}},  {\em
  JHEP} {\bf 10} (2017) 013, [\href{http://xxx.lanl.gov/abs/1701.08750}{{\tt
  arXiv:1701.08750}}].

\bibitem{Kostelecky:1999mr}
V.~A. Kostelecky and C.~D. Lane, {\it {Constraints on Lorentz violation from
  clock comparison experiments}},  {\em Phys. Rev.} {\bf D60} (1999) 116010,
  [\href{http://xxx.lanl.gov/abs/hep-ph/9908504}{{\tt hep-ph/9908504}}].

\end{thebibliography}\endgroup
\bibliographystyle{JHEP}

\end{document}